\documentclass{article}

\usepackage[utf8]{inputenc}
\usepackage{listings}
\usepackage{color}
\usepackage{fullpage}
\usepackage{amsfonts}
\usepackage{epsfig,endnotes}
\usepackage{subfig}
\usepackage{comment}
\usepackage{amsmath}
\usepackage{graphicx}
\usepackage{array}
\usepackage{multirow}
\usepackage{color}
\usepackage{booktabs}
\usepackage{algorithm}
\usepackage{caption}
\usepackage{comment}
\usepackage{filecontents}
\usepackage{soul}
\usepackage[noend]{algpseudocode}
\usepackage{balance}
\usepackage{url}

\makeatletter
\newcommand\footnoteref[1]{\protected@xdef\@thefnmark{\ref{#1}}\@footnotemark}
\makeatother

\algrenewcommand\algorithmicindent{1.0em}
\newcommand{\correct}[1]{{#1}}
\newcommand{\PROTNAME}{AV}
\newcommand{\PROTFULLNAME}{Aggregated Victims}

\date{}
\title{Lightweight Robust Size Aware Cache Management}

\author{
	Gil Einziger\\
	Computer Science Department\\
	Ben Gurion University\\
	\texttt{gilga1983@gmail.com}
\and	
	Ohad Eytan\\
	Computer Science Department\\
	Technion\\
	\texttt{ohadey@cs.technion.ac.il}
\and
	Roy Friedman\\
	Computer Science Department\\
	Technion\\
	\texttt{roy@cs.technion.ac.il}
\and
	Benjamin Manes\\
	Independent\\
	\texttt{ben.manes@gmail.com}
}

\begin{document}
\maketitle

\begin{abstract}
Modern key-value stores, object stores, Internet proxy caches, as well as Content Delivery Networks (CDN) often manage objects of diverse sizes, e.g., blobs, video files of different lengths, images with varying resolution, and small documents.  
In such workloads, size-aware cache policies outperform size-oblivious algorithms.  
Unfortunately, existing size-aware algorithms tend to be overly complicated and computationally~expensive.  

Our work follows a more approachable pattern; we extend the prevalent (size-oblivious) TinyLFU cache admission policy to handle variable sized items.
Implementing our approach inside two popular caching libraries only requires minor~changes. 
We show that our algorithms yield competitive or better hit-ratios and byte hit-ratios compared to the state of the art size-aware algorithms such as AdaptSize, LHD, LRB, and GDSF.
Further, a runtime comparison indicates that our implementation is faster by up to x3 compared to the best alternative, i.e., it imposes much lower CPU~overhead.

\end{abstract}

\section{Introduction}

Caches improve the performance of diverse computing and networked systems by storing portions of the data closer to the applications and users.
Caching is widely used in many databases and storage systems in which DRAM memory serves as a cache for slower secondary storage~\cite{lirs}, Web caching~\cite{UsageStudy}, and even in hardware fast SRAM memory serves as a cache to slower DRAM memory.
As another example, \emph{Content Delivery Networks} (CDN) reduce the data access latency by storing popular content closer (in terms of Internet topology) to its consumers~\cite{CDN1}.
CDNs deploy multiple distributed content caching servers, also known as \emph{Points of Presence} (PoP), in multiple ASs.
PoPs serve popular content faster than fetching the data from the original server and accomplish three goals:
reducing Internet traffic~\cite{netflix-corona}, reducing the load on the content providers' servers, and reducing users' perceived response times, thereby improving the user~experience.

Given that the cache is smaller than the entire storage whose content it stores, a \emph{cache management policy} decides which items to store in the cache at any given point.
Due to the diversity of workloads and systems, a plethora of such policies have been devised, e.g.,~\cite{2Q,ARC,AdaptSize,AdaptiveCaches,AdaptiveTinyLFU,CAR,LHD,LRB,ClockPro,Cloud1,Cloud2, MiniSim}.

The vast majority of research on cache management policies ignore the size of cached items.
The logic behind this simplification is that many databases and storage systems use block devices as secondary storage, so caches only need to work at block boundary.
The sizes of all blocks or pages within a given system are the same or very similar.

However, in modern databases and data-stores, the reality is more complicated.
Specifically, there are many important scenarios, such as imaging and video objects, CDN's caches, cell content in column-family data models, blobs, and application caches, where the size of cache-able items vary, emphasizing the need for size-aware cache policies.
Further, much of today's Internet content volume is consumed by video streaming, which is encoded and transmitted using variable bit rates.
Therefore, the video is partitioned into chunks (each corresponding to a short video duration) whose size varies considerably~\cite{AViC}.
In fact, even the same sub-stream may be encoded multiple times at different bit rates, each of which is transmitted to different clients based on their temporal line quality.
In such workloads, size-aware cache policies often outperform size oblivious cache policies.
Indeed, a modern analysis has recently shown that sophisticated variable sized objects caching indeed has room for significant system performance improvements~\cite{PracticalBounds}.

Alas, size-aware cache management policies are often more complex than unit sized policies. 
For example, size-aware policies may divide the cache into \emph{slabs}, where each slab is devoted to objects of similar size and is managed independently of other slabs~\cite{redis,LHD}.
Such an approach leads to sub-optimal cache utilization as objects of a certain slab may be more popular than objects of other slabs. 
Further, the popularity of objects varies, so there is no single best static slabbing partitioning.  
Some works mitigate these complications by periodically re-balancing the slab allocation~\cite{LHD}.

When objects' sizes vary, we distinguish between \emph{hit-ratio} and \emph{byte hit-ratio}.
Hit-ratio is the ratio between the number of accesses that are a hit in the cache and the total number of accesses.
Hit-ratio is important to improve the systems' perceived access latency, and thereby its user experience.
In contrast, byte hit-ratio is the accumulated number of bytes served by hits in the cache vs. the total number of bytes returned by all accesses.
This indicates the cache's effectiveness in reducing network bandwidth~consumption.

The seminal GDSF policy~\cite{GDSF} is very effective in terms of both hit ratio and byte hit-ratio.
But, GDSF suffers from a poor reputation of being computationally expensive and it operates in logarithmic complexity.
Three other prominent size-aware cache management policies include AdaptSize~\cite{AdaptSize}, LHD~\cite{LHD} and the recent LRB~\cite{LRB}.
AdaptSize involves parameter tuning through non-trivial stochastic models.
LHD and LRB both apply a heavy machine learning process
to learn from previous accesses to predict future ones.
Here we are looking for a lightweight, yet effective, size aware caching~policy.


\subsection*{Our Contributions}
This paper explores the benefits of simple size-aware cache policies that generalize the W-TinyLFU policy~\cite{TinyLFU} to handle varying sized objects\footnote{W-TinyLFU is used, directly or through caching libraries, in Cassandra, Accumulo, HBase, Apache Solr, Infinispan, OpenWhisk, Corfu, Finagle, Spring, Akka, Neo4j, DGraph, Druid, and many others as listed in~\cite{CaffeineProject}.}.
In a gist, W-TinyLFU decides which objects to hold in the cache for a long duration according to their access frequency compared to the cache victim. 
Intuitively, variable-sized items imply that we may need to compare a newly arriving item against multiple cache victims to make enough room for it in the cache.

Our work evaluates three such possible extensions.
The first two, nicknamed \emph{Implicit Victims} (IV) and \emph{Queue of Victims} (QV), are implemented in production extensions of W-TinyLFU to variable-sized objects (our Caffeine~\cite{CaffeineProject} and Ristretto~\cite{RistrettoProject}), but were never described outside their code, nor analyzed and compared in any manner.
The third is our newly proposed approach, nicknamed \emph{Aggregated Victims} (AV).
We demonstrate that our proposed method AV outperforms the two existing solutions IV and QV in hit-ratio.
When it comes to byte hit-ratio, QV has a slight advantage in the CDN traces.
Next, we show that our new policy AV achieves competitive hit-ratio and byte hit-ratio compared to the state-of-the-art policies GDSF~\cite{GDSF}, AdaptSize~\cite{AdaptSize}, LHD~\cite{LHD} and LRB~\cite{LRB} over a wide variety of real-world traces.
While none of the evaluated policies is best in all the cases, our policy consistently attains high hit ratios and faster constant time operation and is also very simple to~implement.

Interestingly, our measurements revealed the following inherent limitation of cache management approaches, such as AdaptSize~\cite{AdaptSize}, that admit objects with a probability that is inversely proportional to the object's size.
The problem is that such policies fail to utilize the entire cache area when operating over large caches, as found in modern CDN PoPs~\cite{CDN1}.
We explain this problem in the evaluation~section.


\paragraph*{Paper road-map:} The rest of this paper is structured as follows:
We survey related work in Section~\ref{sec:related}.
We present the (size oblivious) W-TinyLFU approach in Section~\ref{sec:w-tinylfu}, and the extensions to variable size objects in Section~\ref{sec:algorithms}.
Section~\ref{sec:eval} presents the performance evaluation of the hit-ratio, byte hit-ratio, and throughput for the various approaches.
Finally, we conclude with a discussion in Section~\ref{sec:discussion}.

\section{Related Work}
\label{sec:related}

As indicated above, the majority of cache management policies assume that all objects are of similar or equal size.
Such policies include LFU and its variants~\cite{WLFU,LFUIMPl,LFUAGING} as well as LRU~\cite{LRU} and its performance-driven approximations~\cite{Tanenbaum01,ClockPro}.
More modern approaches improve on LRU by adding frequency aspects to it.
These include, e.g., LRU-K~\cite{LRUK}, 2Q~\cite{2Q}, ARC~\cite{ARC}, CAR~\cite{CAR}, LIRS~\cite{lirs}, DLIRS~\cite{dlirs}, FRD~\cite{FRD} and Hyperbolic~\cite{Hyperbolic}. Mini\nobreakdash-Sim~\cite{MiniSim} efficiently simulates multiple policies and configurations in order to choose the best one.
W\nobreakdash-TinyLFU proposes having two cache segments with an approximated LFU admission filter for transferring items between these segments~\cite{AdaptiveTinyLFU,TinyLFU}.

The seminal Greedy-Dual work was one of the first to incorporate size into the replacement decision~\cite{GD}.
The Greedy-Dual was later extended to accommodate for recency, frequency, cost, and size in the GDSF scheme~\cite{GDSF}.
GDSF gives a dynamic score to each item by combining the above four criteria into a unifying formula.
To make room for a newly arriving item $x$, GDSF collects cached items whose dynamic score is minimal, until finding enough items whose combined size is larger than the size of $x$ and then evicts all these found items.
To date, GDSF remains a tough performance benchmark to beat as far as hit ratio and byte hit ratio are concerned.
Yet, its computational complexity is largely considered too high for most production caching systems.

The popular Redis system~\cite{redis} divides the total cache area into multiple slabs, each dedicated to similarly-sized objects.
Every slab is maintained using either sampled LRU or sampled LFU for eviction coupled with blind~admission.

Three promising recent size aware caching schemes are AdaptSize~\cite{AdaptSize}, LHD~\cite{LHD} and LRB~\cite{LRB}.
AdaptSize implements a stochastic sized-based admission policy where large items have a smaller probability of being admitted compared with small items. 
Specifically, AdaptSize utilizes Markov models to optimize the admission policy. 

In contrast, LHD~\cite{LHD} introduced the concept of \emph{hit density} to capture the combination of frequency and size.
It also considers recency through machine learning based prediction of when each item would be reaccessed.
These parameters are placed in a unifying formula, and the cache victims are those whose score is minimal.
Yet, to keep the computational cost practical, LHD randomly samples 64 uniformly selected items, computes the scores only for those items, and evicts the minimal among them.
LHD also employs slabbing to avoid having to evict too many items at once, as well as slabs re-balancing to improve the overall cache utilization.

LRB~\cite{LRB} is a recent sophisticated cache replacement policy whose design goal is to maximize byte hit ratio in order to minimize bandwidth consumption in CDNs.
LRB employs machine learning in order to approximate the behavior of a \emph{relaxed Belady} algorithm.
The classical offline Belady algorithm~\cite{Belady} evicts from the cache the object whose future access is the furthest, which is optimal.
Yet, as the authors of~\cite{LRB} have discovered, learning to approximate Belady is very hard, but is also usually not needed.
Instead, they proposed a variant called relaxed Belady, in which the algorithm evicts some object whose next access is beyond what they referred to as the \emph{Belady boundary}.
The LRB policy utilizes machine learning in order to mimic the relaxed Belady algorithm, which is feasible and indeed provides very good byte hit ratio results.

\section{An Overview of W-TinyLFU}
\label{sec:w-tinylfu}

\begin{figure}[h]
	\begin{center}
		\includegraphics[width=0.5\textwidth]{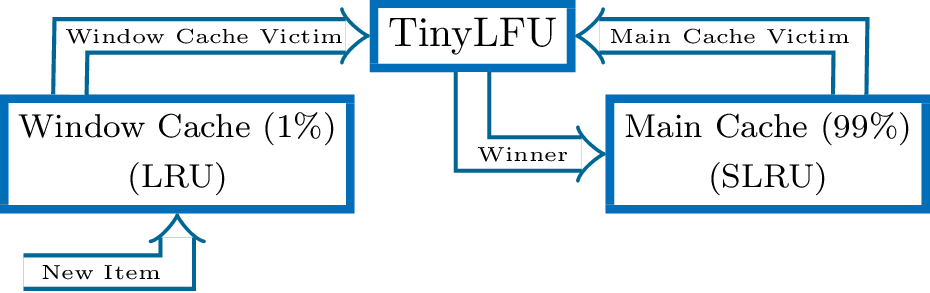}
	\end{center}	
	\caption{\normalfont{W-TinyLFU scheme: Items are first always admitted to the Window cache. The Window cache victim is offered to the Main cache, which employs TinyLFU as its admission filter.
			We preserve here the terminology of~\cite{TinyLFU}.}}
	\label{fig:wtinylfu}
\end{figure}

In this section, we give a brief overview of the (size oblivious) W-TinyLFU scheme. 
For more details, see~\cite{AdaptiveTinyLFU,TinyLFU}.
As shown in Figure~\ref{fig:wtinylfu}, W-TinyLFU relies on three modules:
the \emph{Main cache}, the \emph{Window cache} and an approximated LFU based \emph{admission filter} called \emph{TinyLFU}.
On a cache miss, the accessed item is inserted into the Window cache.
When this results in another item being evicted from Window cache, aka \emph{Window victim}, the Window victim is introduced to the TinyLFU filter.
The filter provides approximate frequency estimations of the Window victim, and the would-be victim of the Main cache, aka \emph{Main victim}.
If the Window victim's estimated frequency is higher than that of the Main victim, then the Window victim is admitted to the Main cache at the expense of the Main victim.
Else, the Main victim remains in the Main cache, and the Window victim gets~evicted.

In principle, this is a generic architecture where one can use any cache management scheme for the Main and Window caches, and the admission filter of the Main cache could employ any utility comparison function.
Yet, the authors of~\cite{TinyLFU} recommended deploying LRU as the eviction policy of the Window cache, SLRU as the eviction scheme of the Main cache, and TinyLFU as the admission filter.
As for the relative sizes of the two cache regions, in~\cite{TinyLFU} is was recommended that the Window cache be allocated $1\%$ of the overall cache size, whereas a followup work~\cite{AdaptiveTinyLFU} showed how to adapt these region sizes dynamically.

The TinyLFU admission filter is implemented through a sketch such as a \emph{minimal increment counting Bloom filter}~\cite{SpectralBloom}, or a \emph{count min sketch}~\cite{CMSketch}.
All sketch' counters are halved for aging purposes every $ S $ accesses, where $ S $ is a parameter.
Given a cache of size $C$ items, these counters are also capped by $S/C$ and therefore require $O(\log{S/C})$ bits.
Typically, $S$ is at least one order of magnitude larger than $C$.
The sketch counters corresponding to $x$ are updated for every occurrence of $x$, even if it is not in the cache. 
Such an approach eliminates the need for maintaining ghost entries, as is common in other schemes~\cite{ARC,LIRS-Gil,FRD}.

\section{Size-aware Admission Policy}
\label{sec:algorithms}
This section discusses how to extend the W-TinyLFU policy to variable-sized objects. 
There are three main differences between the similar size case and variable size scenarios:
First, a new item might be larger than the Window cache.
In this case, the newly arriving item skips the Window cache and is immediately submitted for the TinyLFU filter to be determined if it can enter the Main~cache.

\begin{figure}[t]
	\begin{center}
		\includegraphics[width=0.7\textwidth]{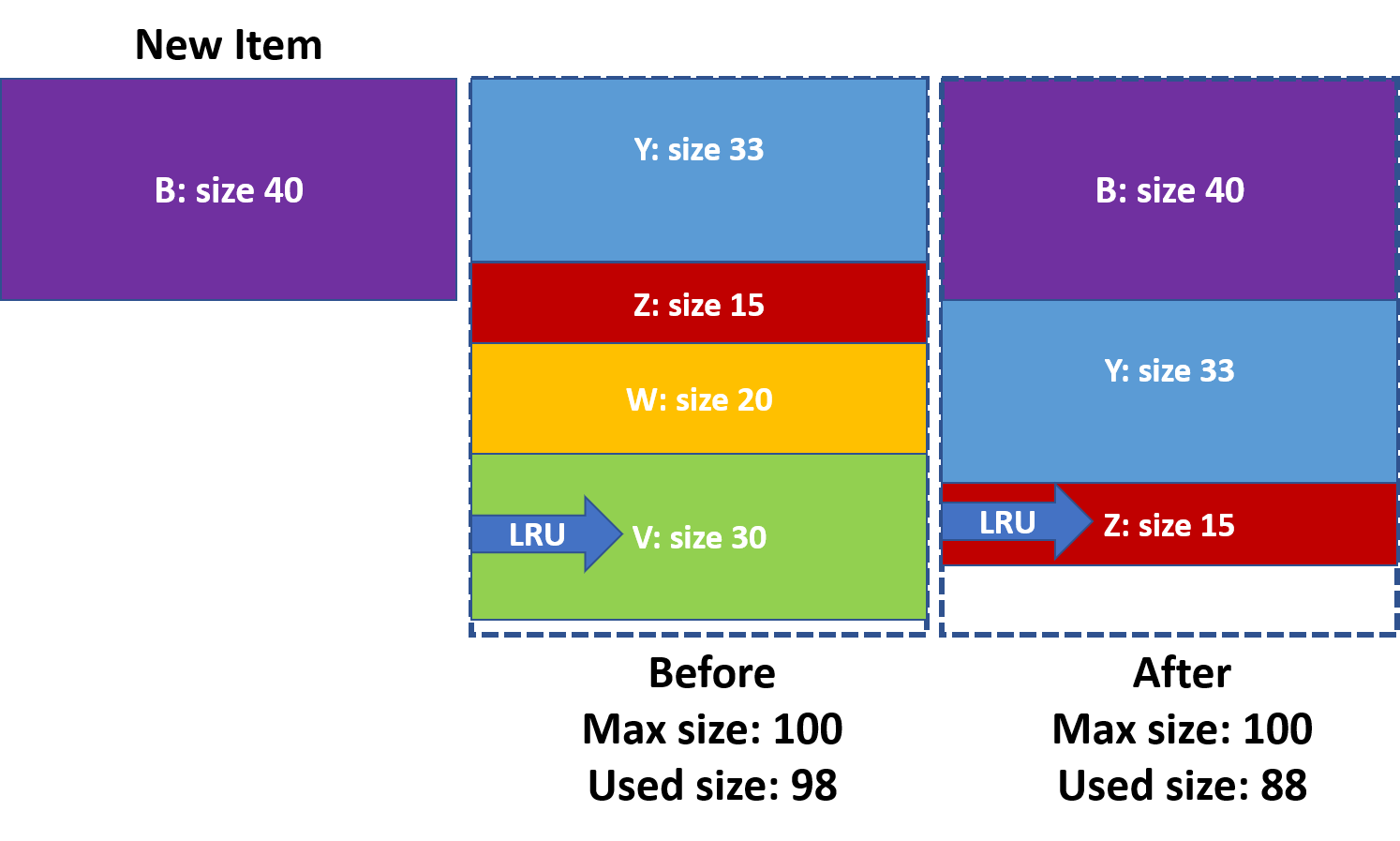}
	\end{center}	
	\caption{\normalfont{An illustration of the admission of a new item (B) to the Window cache, in this example, evicting the LRU item (V), does not make enough room for B, and we need to evict also the 2nd LRU item (W).}
		\label{fig:ex1}}
\end{figure}

Second, we may need to consider multiple potential Window victims whose aggregated size is large enough to make room for the newly arriving item. 
Figure~\ref{fig:ex1} exemplifies this point.
In this example, B is the new item, while W and V are the Window cache~victims. 

Third, in W-TinyLFU, the Window cache victim is tested against the Main cache victim. 
Here, we may have multiple Window cache victims, each tested against (possibly multiple) Main cache victims.
There are several options on how to pick these potential Main cache victims and how to compare them to the candidate item(s).
Specifically, as mentioned before, in this work, we consider three options.
These include Implicit Victims (IV), Queue of Victims (QV), and \PROTFULLNAME{} (\PROTNAME).
The Caffeine Java-based project~\cite{CaffeineProject} employs IV, the Go-based Ristretto project~\cite{ RistrettoProject} implements QV, and \PROTNAME{} is our newly designed scheme.
While IV and QV were adapted into popular libraries, we are the first to evaluate them~systematically. 
\begin{figure}[t]
	\begin{center}
		\includegraphics[width=0.5\textwidth]{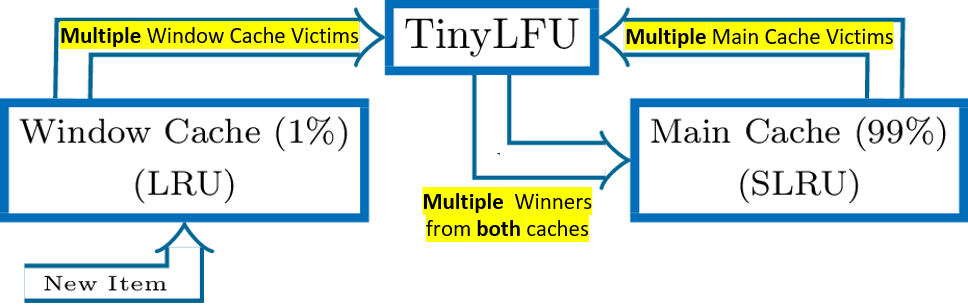}
	\end{center}	
	\caption{\normalfont{We are extending W-TinyLFU to size-awareness.
			The main change to the (size oblivious) W-TinyLFU policy is that adding the new item to the Window cache may result in multiple victims (in yellow).
We use TinyLFU to decide for each one of these Window cache victims.
Similarly, admitting a Window cache victim to the Main cache may result in multiple Main cache victims. 
Thus, in the end, the set of victims may come from both caches.}
		\label{fig:sizewtinylfu}}
\end{figure}

\begin{algorithm}[t]
	\caption{W-TinyLFU Becomes Size Aware} \label{alg:wtinylfu}
	\small
	\begin{algorithmic}[1]
		\Procedure{OnMiss}{$newItem$}
		\If {$newItem$ size $>$ cache size}
		\State reject $newItem$
		\State return
		\EndIf
		\State $candidates$ $\gets$ emptyList() 
		\If {$newItem$ size $>$ maximal window size} \color{blue}\Comment{The new item is too large for the Window cache}\color{black}
		\State $candidates$.append($newItem$) \color{blue}\Comment{Skip Window cache $\rightarrow$ a candidate for Main}\color{black}
		\Else
		\State insert $newItem$ to window
		\While{window size $>$ maximal window size} \color{blue}\Comment{Evict items from Window to make room}\color{black}
		\State $candidate$ $\gets$ evict from window \color{blue}\Comment{Each eviction from Window becomes}\color{black}
		\State $candidates$.append($candidate$) \color{blue}\Comment{a candidate for Main}\color{black}
		\EndWhile
		\EndIf
		\For{$candidate$ in $candidates$}
		\State EvictOrAdmit($candidate$) \color{blue}\Comment{This function is where IV, QV and \PROTNAME{} differ}\color{black}
		\EndFor
		\EndProcedure
	\end{algorithmic}
	\normalsize
\end{algorithm}

Figure~\ref{fig:sizewtinylfu} outlines the size-aware W-TinyLFU policies. 
The differences with the basic W-TinyLFU policy are marked in yellow.
In our measurements, we use a Window cache whose size is 1\% of the total cache size (and 99\% is allocated to the Main cache), as suggested in~\cite{TinyLFU}.
The eviction policy for the Window cache is LRU, while for the Main cache, we experiment with few alternatives, as described below.
Algorithm~\ref{alg:wtinylfu} provides pseudocode for miss handling by the size-aware adaptation of W-TinyLFU. 

\subsection{Implicit Victims (IV)}

\begin{figure}[t]
	\begin{center}
		\includegraphics[width=0.7\textwidth]{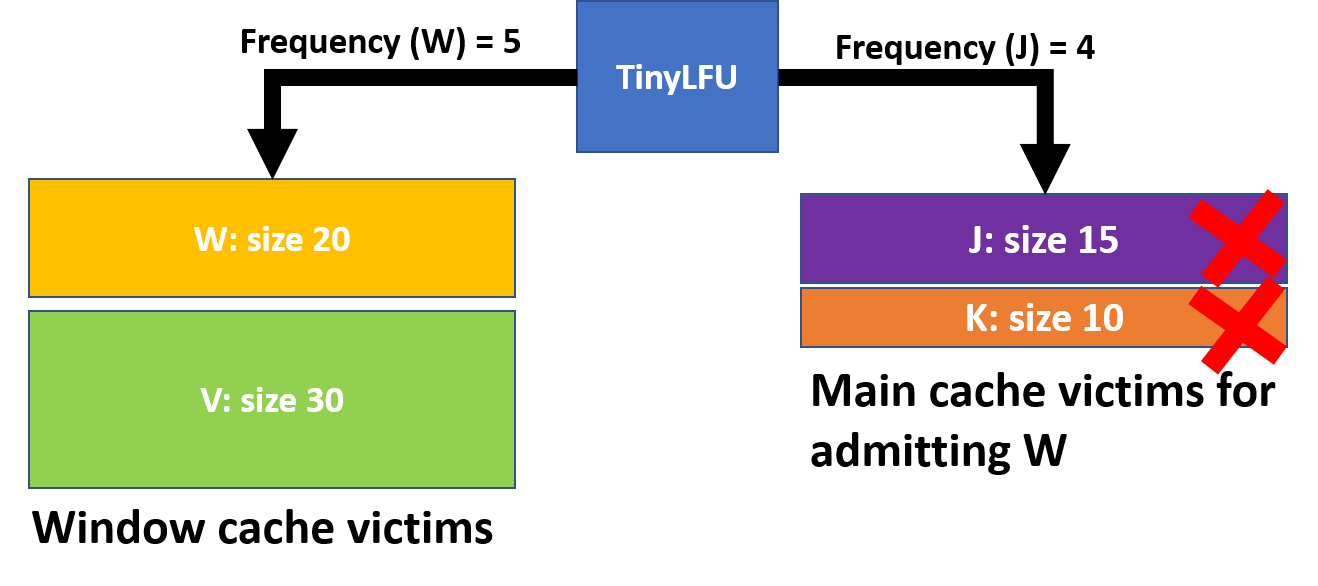}
	\end{center}	
	\caption{\normalfont{An example of the Implicit Victims (IV) approach: We compare the frequency of the Window cache victim (W) to that of the \emph{first} Main cache victim (J). Since W is more frequent, we admit W and evict J and K.}
		\label{fig:ex2}}
\end{figure}

The IV scheme deals with variable sized objects by following the simplistic approach in Algorithm~\ref{alg:iv-alg}:
IV compares the candidate item's frequency vs. the frequency of the would-be victim, as reported by the approximate filter (Line~\ref{line:iv:compare}).
If the candidate is more frequent, IV blindly evicts enough victims, one after the other, until there is enough free space in the cache for the candidate to be admitted (Lines~\ref{line:iv:collect}--\ref{line:iv:evict}).
The method is exemplified in Figure~\ref{fig:ex2}.  

Notice that IV only cares about the first victim's frequency and ignores the frequency of subsequent victims that may be evicted to admit the candidate.
In case the victim is more frequent, IV promotes it (Line~\ref{line:iv:promote}) and rejects the candidate (Line~\ref{line:iv:reject}).

This approach is straightforward and computationally efficient to implement.
As mentioned before, we developed IV and implemented it in Caffeine~\cite{CaffeineProject}, but it was never explained before beyond the source code of Caffeine nor studied~systematically.
\begin{algorithm}[t]
	\caption{Implicit Victims (IV)}\label{alg:iv-alg}
	\small
	\begin{algorithmic}[1]
		\Procedure{EvictOrAdmit}{$candidate$}
		\State $victim\gets$ getVictim()
		\If{frequency($candidate$) $\ge$ frequency($victim$)} \label{line:iv:compare}
			\While{space is needed} \label{line:iv:collect}
				\State $victim\gets$ getVictim()
				\State evict $victim$ \label{line:iv:evict}
			\EndWhile
			\State admit $candidate$ \label{line:iv:admit}
		\Else
			\State promote $victim$ \label{line:iv:promote}
			\State reject $candidate$ \label{line:iv:reject}
		\EndIf
		\EndProcedure
	\end{algorithmic}
	\normalsize
\end{algorithm}

\subsection{Queue of Victims (QV)}

QV takes a somewhat different approach to handle variable-sized objects, as outlined in Algorithm~\ref{alg:qv-alg}.
Specifically, QV repeatedly seeks for victims.
For each victim, if the frequency of the candidate is higher than the victim's frequency, the victim is evicted (Lines~\ref{line:qv:compare}--\ref{line:qv:evict}).
Otherwise, the victim is promoted in the cache, as if it was accessed, and the iterative process stops (Lines~\ref{line:qv:promote}--\ref{line:qv:break}).
At the end of the iterative process, if there is enough vacant space in the cache for the candidate, then the candidate is admitted (Line~\ref{line:qv:admit}).

\begin{figure}[t]
	\begin{center}
		\includegraphics[width=0.7\textwidth]{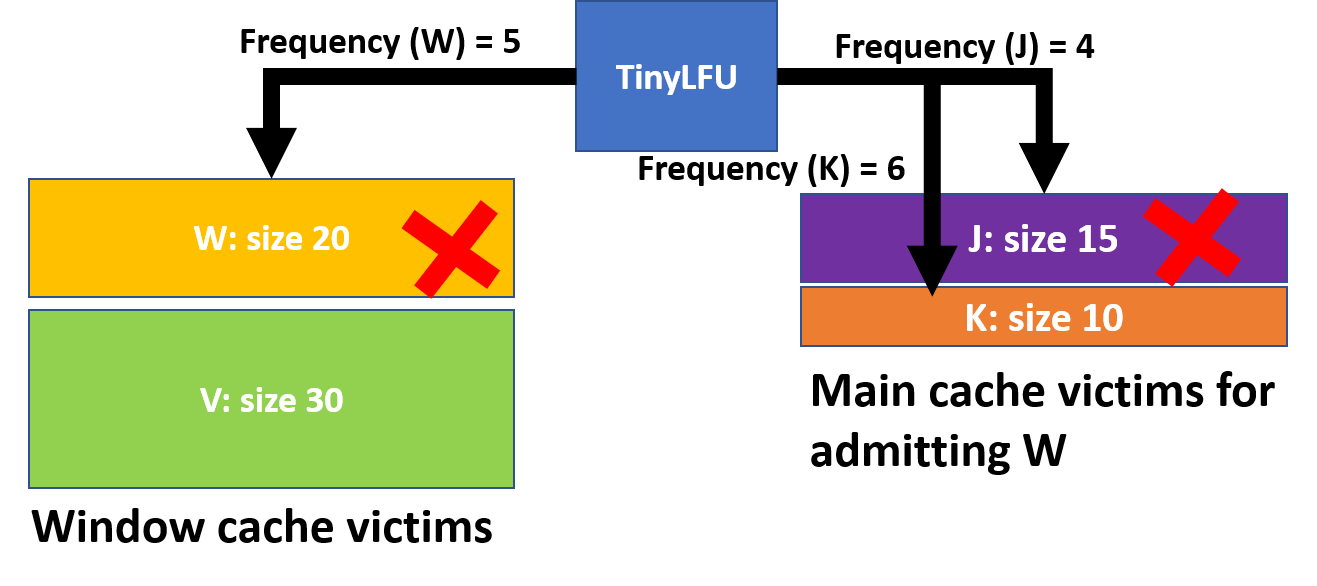}
	\end{center}	
	\caption{\normalfont{The Queue of Victims (QV) approach: Here, we compare the frequency of the Window cache victim (W) to that of the Main cache victim (J). Since W is more frequent, evict J and move to the next victim (K). In this case, K is more frequent than W, so we do not admit W to the Main cache.}
		\label{fig:ex3}}
\end{figure}

Figure~\ref{fig:ex3} illustrates this process. 
In this example, we evict an item (J) from the Main cache without admitting an item.
The lack of admission is because the evicted item was not large enough to make room for the candidate, while the next victim was more frequent than the candidate.

\begin{algorithm}[t]
	\caption{Queue of Victims (QV)} \label{alg:qv-alg}
	\small
	\begin{algorithmic}[1]
		\Procedure{EvictOrAdmit}{$candidate$}
		\While{space is needed}
			\State $victim\gets$ getVictim()
			\If {frequency($candidate$) $\ge$ frequency($victim$)} \label{line:qv:compare}
				\State evict $victim$  \label{line:qv:evict}
			\Else
				\State promote $victim$ \label{line:qv:promote}
				\State break \label{line:qv:break}
			\EndIf
		\EndWhile
		\If{there is enough space}
			\State admit $candidate$ \label{line:qv:admit}
		\Else
			\State reject $candidate$ \label{line:qv:reject}
		\EndIf
		\EndProcedure
	\end{algorithmic}
	\normalsize
\end{algorithm}

The motivation behind QV, compared to IV, is two-fold:
($i$) avoid evicting items that are more frequent than the candidate, which could happen with IV,
($ii$) yet, if a victim is less frequent than the candidate, evict the victim even if eventually the candidate is not admitted.
The rationale behind ($ii$) is that if a victim is less frequent than a candidate, it is polluting the cache, and so it should be evicted in any case.

\subsection{\PROTFULLNAME{} (\PROTNAME)}
We now present our \PROTFULLNAME{} (\PROTNAME) approach. 
\PROTNAME{} compares a cache candidate's frequency to that of the entire group of potential victims. 
We admit the candidate if its frequency is higher than that of \emph{all} the potential cache victims. 
Intuitively, since admitting the candidate causes all cache victims' eviction, the admission policy asserts that such a transaction is beneficial to the cache as a~whole. 

\begin{algorithm}[t]
	\caption{\PROTFULLNAME{} (\PROTNAME)}\label{alg:our-alg}
	\small
	\begin{algorithmic}[1]
		\Procedure{EvictOrAdmit}{$candidate$}
		\State $victims\gets$ emptyList()
		\While{$victims$ size $<$ space needed} \label{line:our:collect}
		\State $victim\gets$ getVictim()
		\State $victims$.append($victim$) \label{line:our:append}
		\If {frequency($candidate$) $<$ frequency($victims$)} \label{line:our:early}
		\State break \color{blue}\Comment{early pruning}\color{black} \label{alg:our-alg:pruning}
		\EndIf
		\EndWhile
		\If {frequency($candidate$) $\ge$ frequency($victims$)} \label{line:our:compare}
		\For{$victim$ in $victims$}  \label{line:our:evict}
		\State evict $victim$
		\EndFor
		\State admit $candidate$ \label{line:our:admit}	
		\Else
		\For{$victim$ in $victims$} \label{line:our:promoting}
		\State promote $victim$ \label{alg:our-alg:promote}
		\EndFor	
		\State reject $candidate$ \label{line:our:reject}
		\EndIf
		\EndProcedure
	\end{algorithmic}
	\normalsize
\end{algorithm}

\begin{figure}[t]
	\begin{center}
		\includegraphics[width=0.7\textwidth]{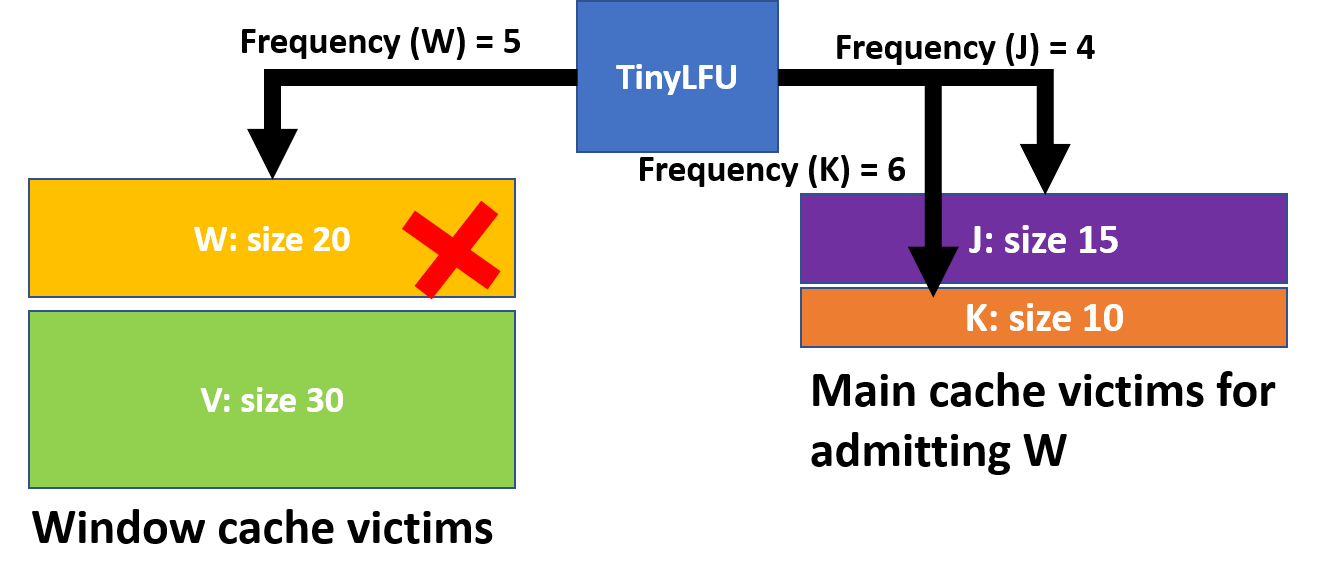}
	\end{center}	
	\caption{\normalfont{Illustrating the Aggregate Victims (AV) approach: In this example, we compare the frequency of the Window cache victim (W) to that of the the Main cache victims (J,K). The combined frequency of the Main cache victims (6+4=10) is larger than the frequency of W (5). Therefore, we do not admit W to the Main~cache. }
		\label{fig:ex4}}
\end{figure}

Algorithm~\ref{alg:our-alg} details \PROTNAME's behavior: 
Specifically, \PROTNAME{} gathers potential victims (according to the eviction policy) until their total size is sufficient to admit the candidate (Lines~\ref{line:our:collect}--\ref{line:our:append}).
\PROTNAME{} then uses TinyLFU to compare the candidates' frequency to the total frequency of all victims (Line~\ref{line:our:compare}).
If the candidate's frequency outweighs the total frequency of all victims, then it is admitted at the expense of the victims (Lines~\ref{line:our:evict}--\ref{line:our:admit}).
Otherwise, the candidate is discarded, and no entries are evicted from the cache (Lines~\ref{line:our:promoting}--\ref{line:our:reject}).
As before, if the candidate is rejected, we promote the potential victims (Line~\ref{alg:our-alg:promote}).
Figure~\ref{fig:ex4} exemplifies this process.
In the figure, W fails to enter the Main cache and is dropped from the cache entirely.

In most simple eviction policies (e.g., LRU), we also promote the victims as if they were accessed once after rejecting the candidate, to force the eviction policy to select a different set of victims for the next candidate. 
Yet, some eviction policies may not require this~step.  

\subsubsection{The Early Pruning Optimization}
The description of AV so far suffers from inefficiency. If the victims' cumulative frequencies surpass the candidate's frequency during its iterative process, we can abort the calculation since the candidate will not be admitted.
Hence, we can improve run-time by gradually gathering the victims and summing their frequency estimations until the total victims' frequency outweighs the candidate's frequency, or their total size is sufficient to admit the candidate.
This optimization is listed in (Lines 6--8) of Algorithm~\ref{alg:our-alg}; we name it \emph{early pruning}. 

\begin{figure}[t]
	\begin{center}
	\includegraphics[trim=0 0 20 0, clip, width=0.75\columnwidth]{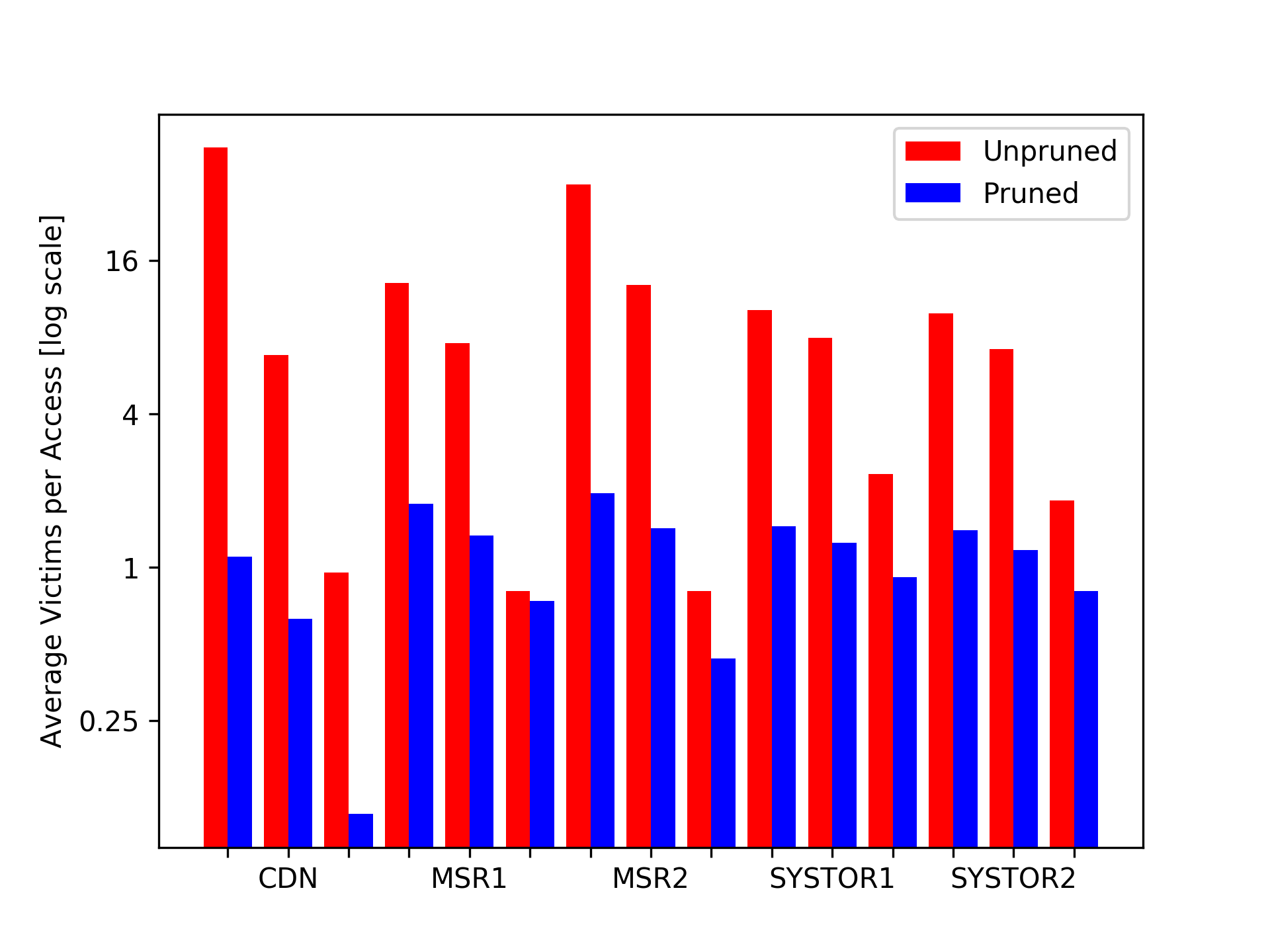}
	\end{center}
	\caption{The number of victims per access, with and without the pruning optimization (see algorithm~\ref{alg:our-alg} line~\ref{alg:our-alg:pruning}), for different traces with cache sizes of $10MB$, $1GB$, and $100GB$ for~each.}
	\label{fig:pruning}
\end{figure}

Early pruning improves the run time of \PROTNAME{} since we need to compare rejected candidates (which are often the vast majority~\cite{TinyCache}) with fewer victims. 
Figure~\ref{fig:pruning} demonstrates the effectiveness of this optimization. 
Notice that early pruning reduces victim comparisons by a factor of $x4$-$x16$. 

Let us note that potential victims that were eventually not evicted are promoted in all these algorithms, e.g., Line~\ref{alg:our-alg:promote} of Algorithm~\ref{alg:our-alg}.
Hence, early pruning may result in fewer objects being promoted than without this optimization.
Yet, we empirically found that its impact on the hit ratio is~negligible.


\section{Evaluation}
\label{sec:eval}


In this section, we present our conducted trace-based simulations results. 
First, we compare the three discussed approaches of the size-aware admission filter (IV, QV, and \PROTNAME), combining six different Main cache eviction policies for W-TinyLFU: \textbf{SLRU} as in Caffeine~\cite{CaffeineProject}, four versions of sampled policies and random eviction (\textbf{Random}). 
Each of the sampling policies samples five entries and selects the victim according to one of the following four rules respectively: lowest frequency (\textbf{Sampled Frequency}), largest size (\textbf{Sampled Size}), lowest frequency/size (\textbf{Sampled Frequency/Size}), and the closest match where the evicted item is the one yielding highest memory utilization (\textbf{Sampled Needed Size}), e.g., when admitting a 100k item into a cache with 50k free space, the victim is the one whose size is closest to 50k).
Our sampled configurations mimic Ristretto's SampledLFU configuration~\cite{RistrettoProject} that randomly samples five potential victims and evicts the one with the lowest frequency. 
Overall, we suggest eighteen combinations of admission and eviction policies in total.
 
Next, we picked our top versions, AV with SLRU for hit-ratio and QV with SLRU for byte-hit-ratio, and compared it to four other leading size-aware policies: GDSF~\cite{GDSF}, AdaptSize~\cite{AdaptSize}, LHD~\cite{LHD} and LRB~\cite{LRB}.
Following previous works, our primary measurement was hit-ratio, byte-hit-ratio, and run-time performance.
As mentioned before, hit-ratio improves latencies, resulting in better user experience and byte-hit-ratio indicates how much bandwidth is saved, meaning better network utilization, while run-time points to the CPU overhead for maintaining the cache itself. 
 
To perform the evaluation, we used our own Java implementation to measure all but AdaptSize, LHD, and LRB policies. 
For these latter three, we used their authors' C++ implementations available in~\cite{AdaptSizeSimulator},~\cite{LHDSimulator} and~\cite{LRBSimulator} respectively.
We used their default configurations, with minor modifications, mainly to address trace parsing issues and remove redundant stats measuring. 
To verify a baseline, we make sure that LRU hit-ratio and byte-hit-ratio are the same in all three frameworks.

\paragraph*{Traces}
\begin{figure*}[!h]

	\center{

		\subfloat[MSR1\label{fig:cdf:msr1}]{\includegraphics[trim=0 0 0 0, clip, height = 3.0cm]{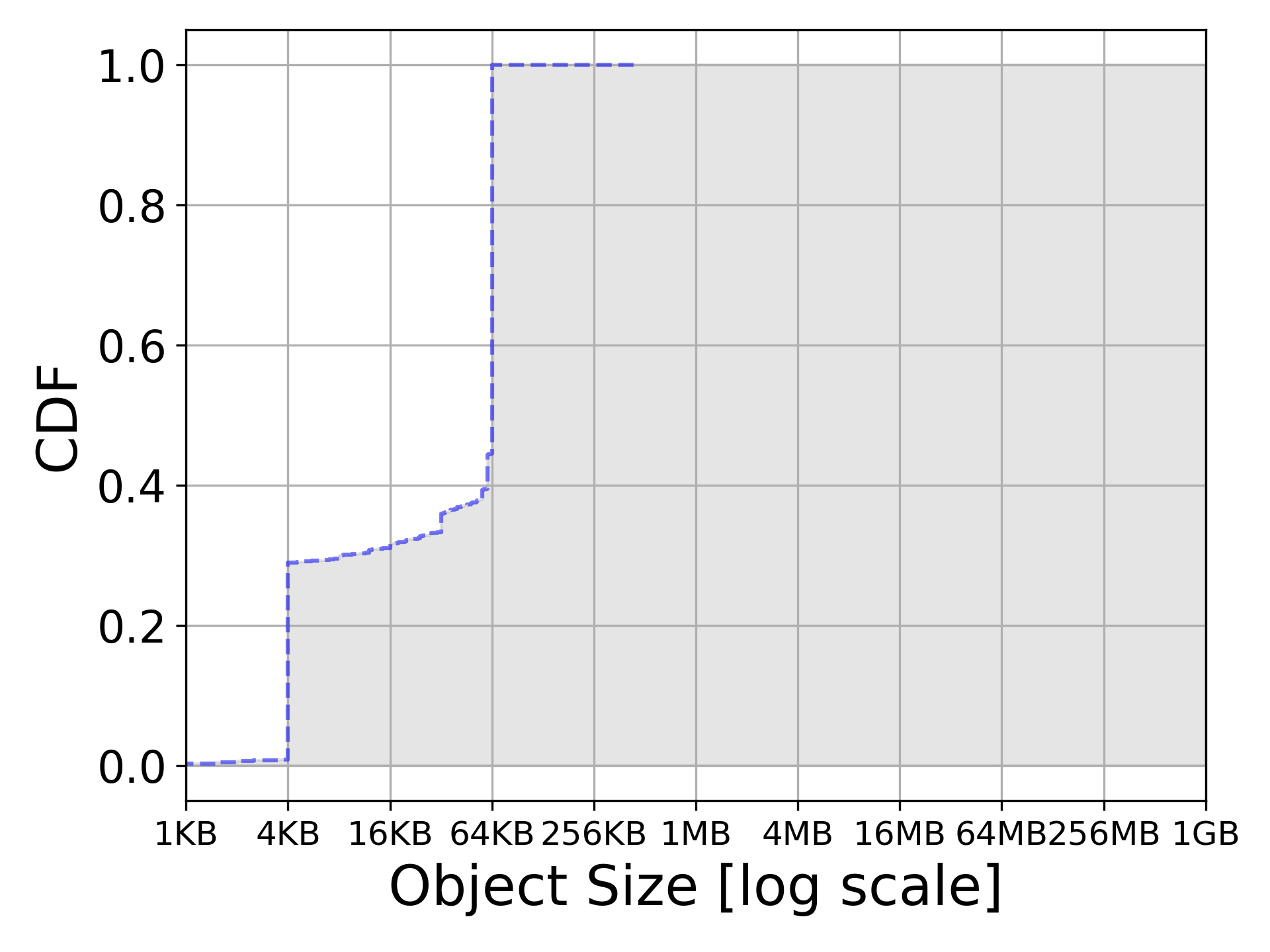}} 
		\subfloat[MSR2\label{fig:cdf:msr2}]{\includegraphics[trim=0 0 0 0, clip, height = 3.0cm]{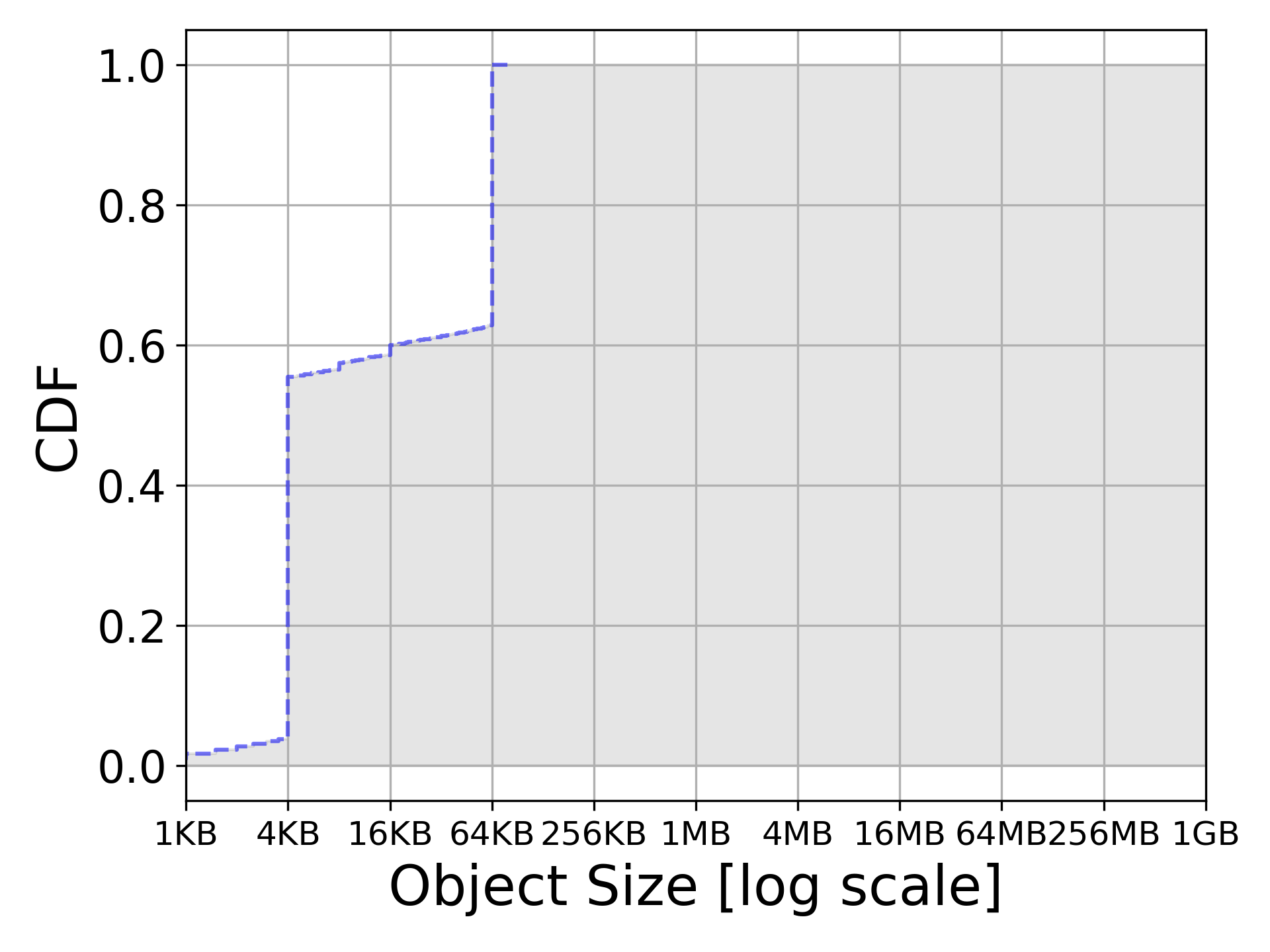}} 
		\subfloat[MSR3\label{fig:cdf:msr3}]{\includegraphics[trim=0 0 0 0, clip, height = 3.0cm]{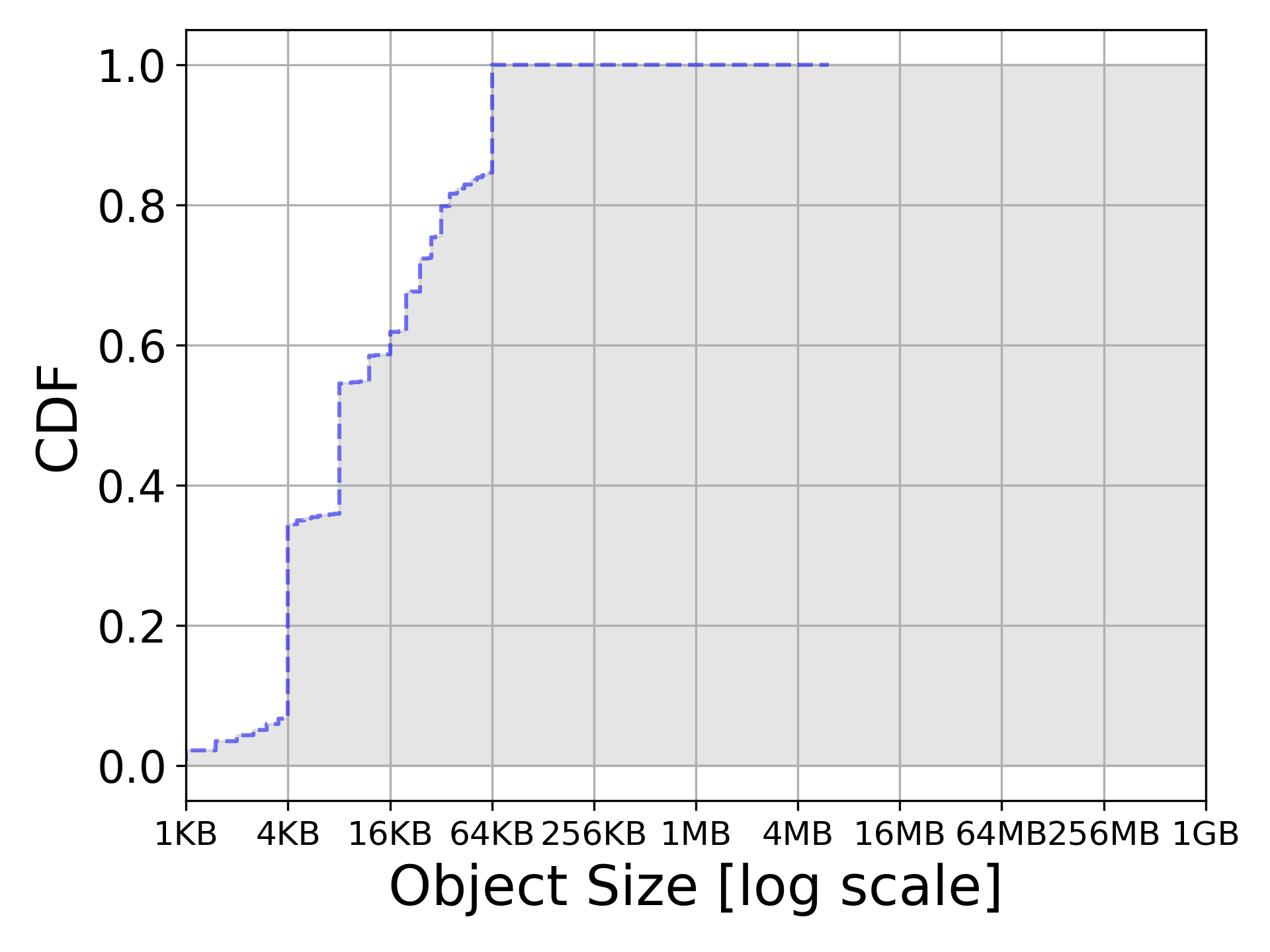}}
\\

		\subfloat[SYSTOR1\label{fig:cdf:systor1}]{\includegraphics[trim=0 0 0 0, clip, height = 3.0cm]{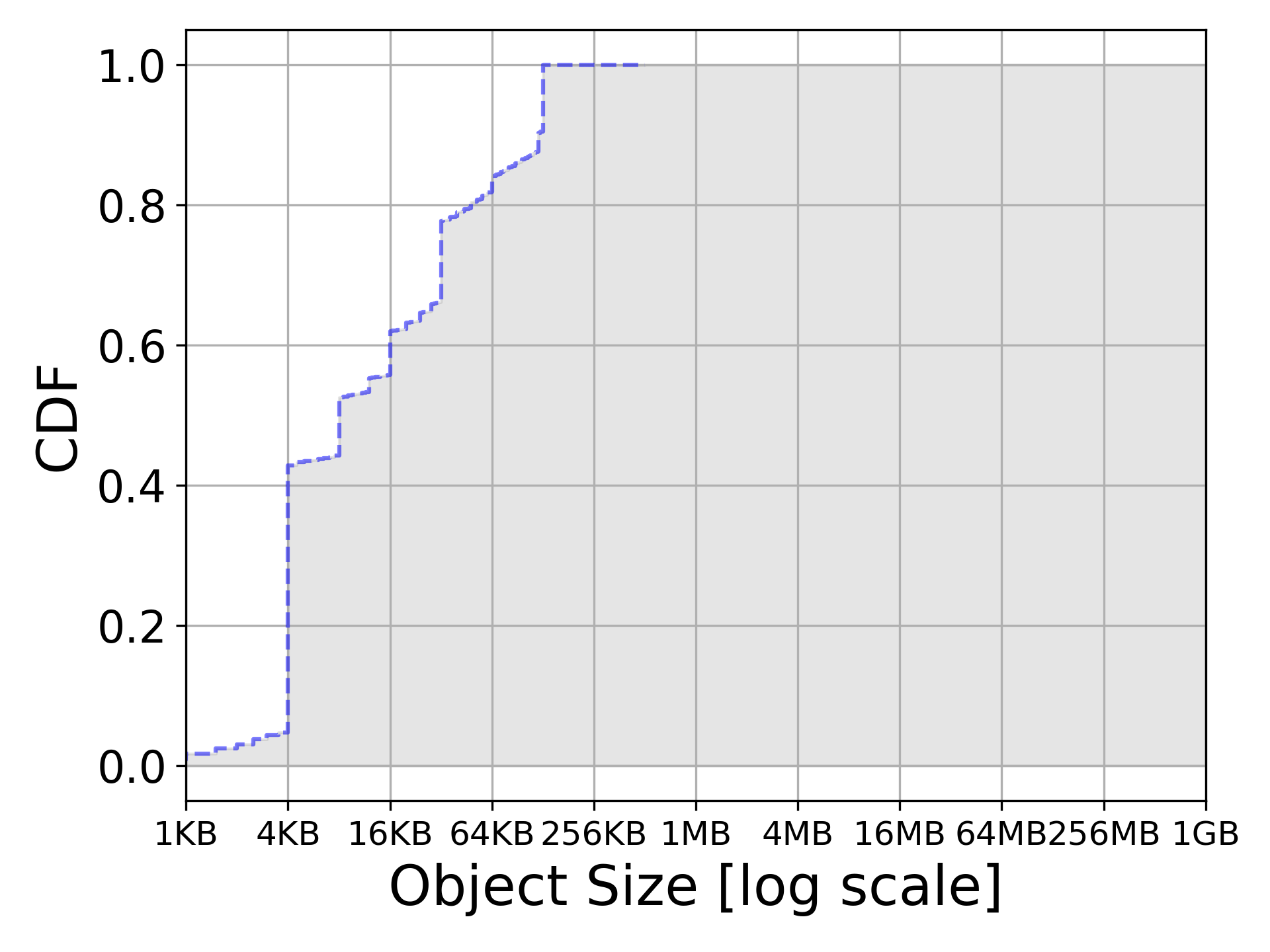}}
		\subfloat[SYSTOR2\label{fig:cdf:systor2}]{\includegraphics[trim=0 0 0 0, clip, height = 3.0cm]{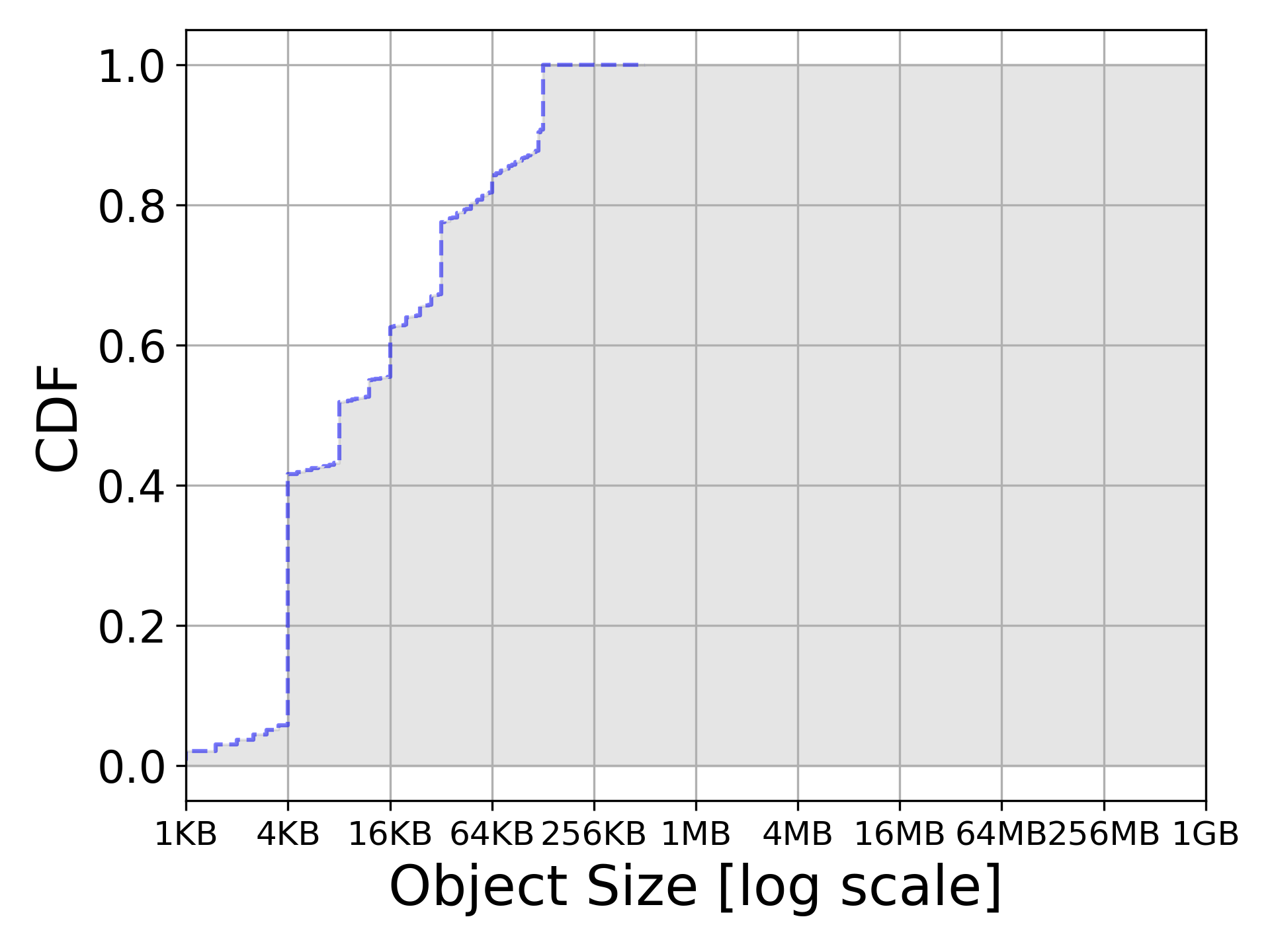}}
		\subfloat[SYSTOR3\label{fig:cdf:systor3}]{\includegraphics[trim=0 0 0 0, clip, height = 3.0cm]{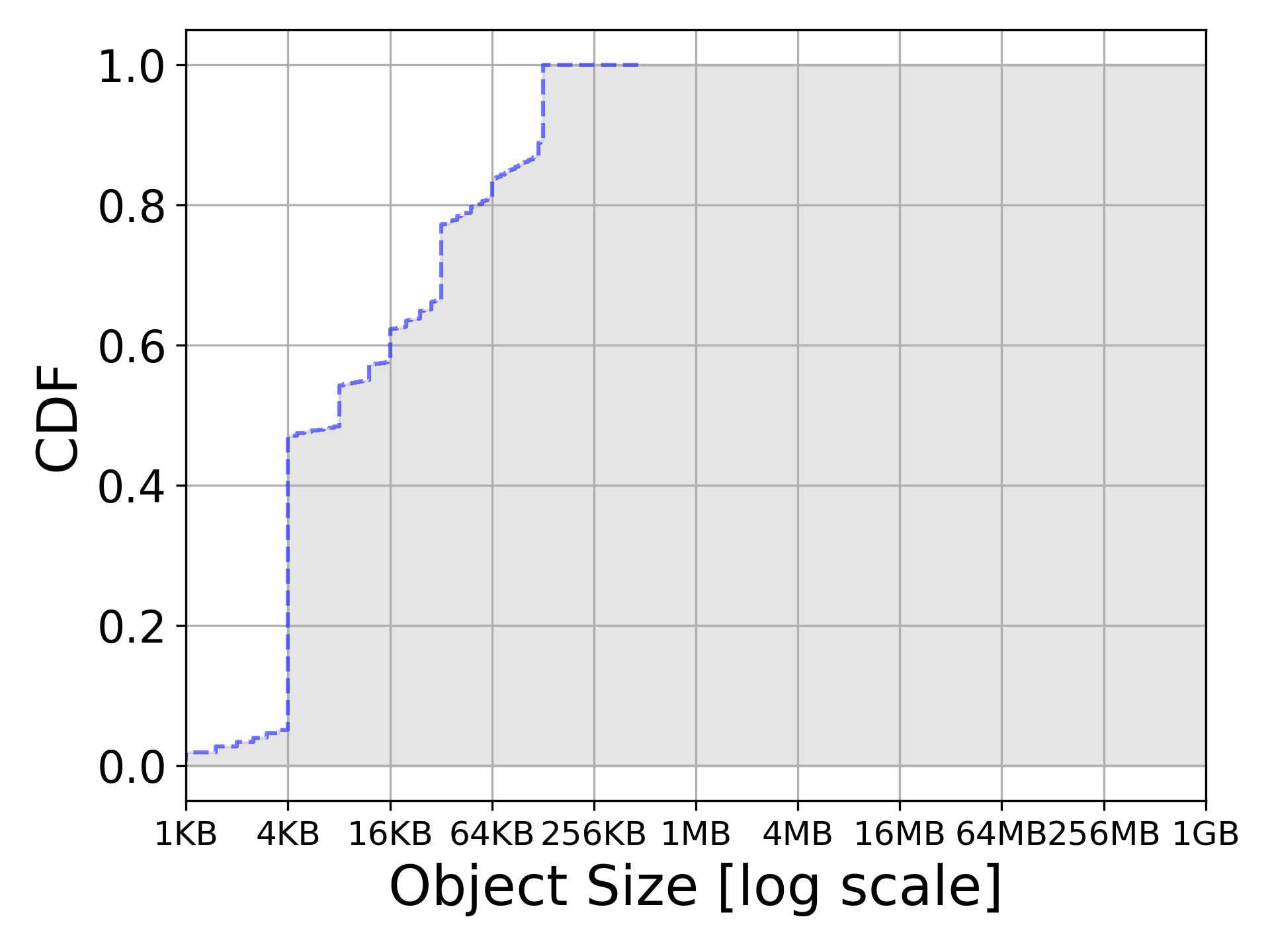}}
\\

		\subfloat[CDN1\label{fig:cdf:cdn1}]{\includegraphics[trim=0 0 0 0, clip, height = 3.0cm]{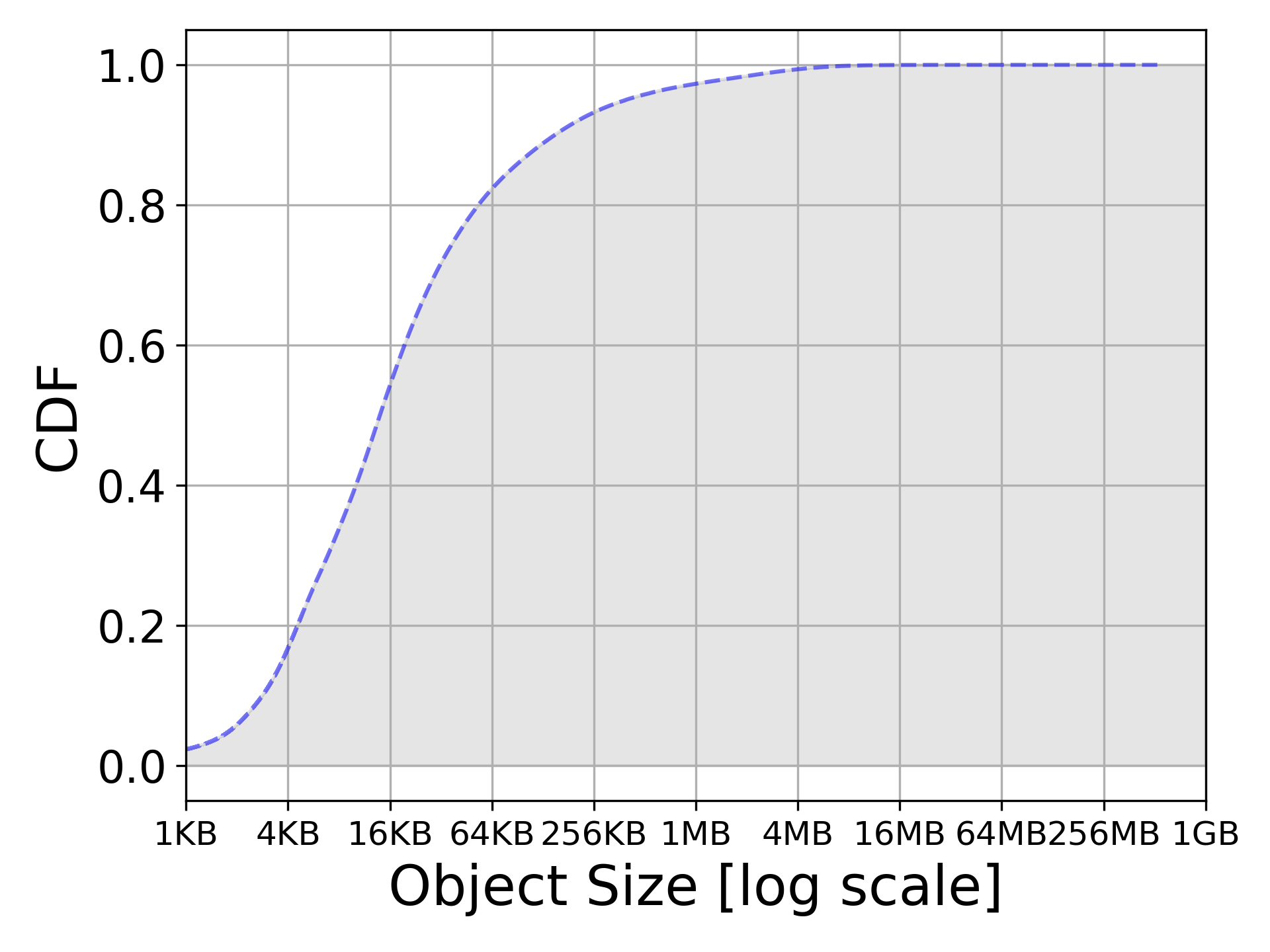}}
		\subfloat[CDN2\label{fig:cdf:cdn2}]{\includegraphics[trim=0 0 0 0, clip, height = 3.0cm]{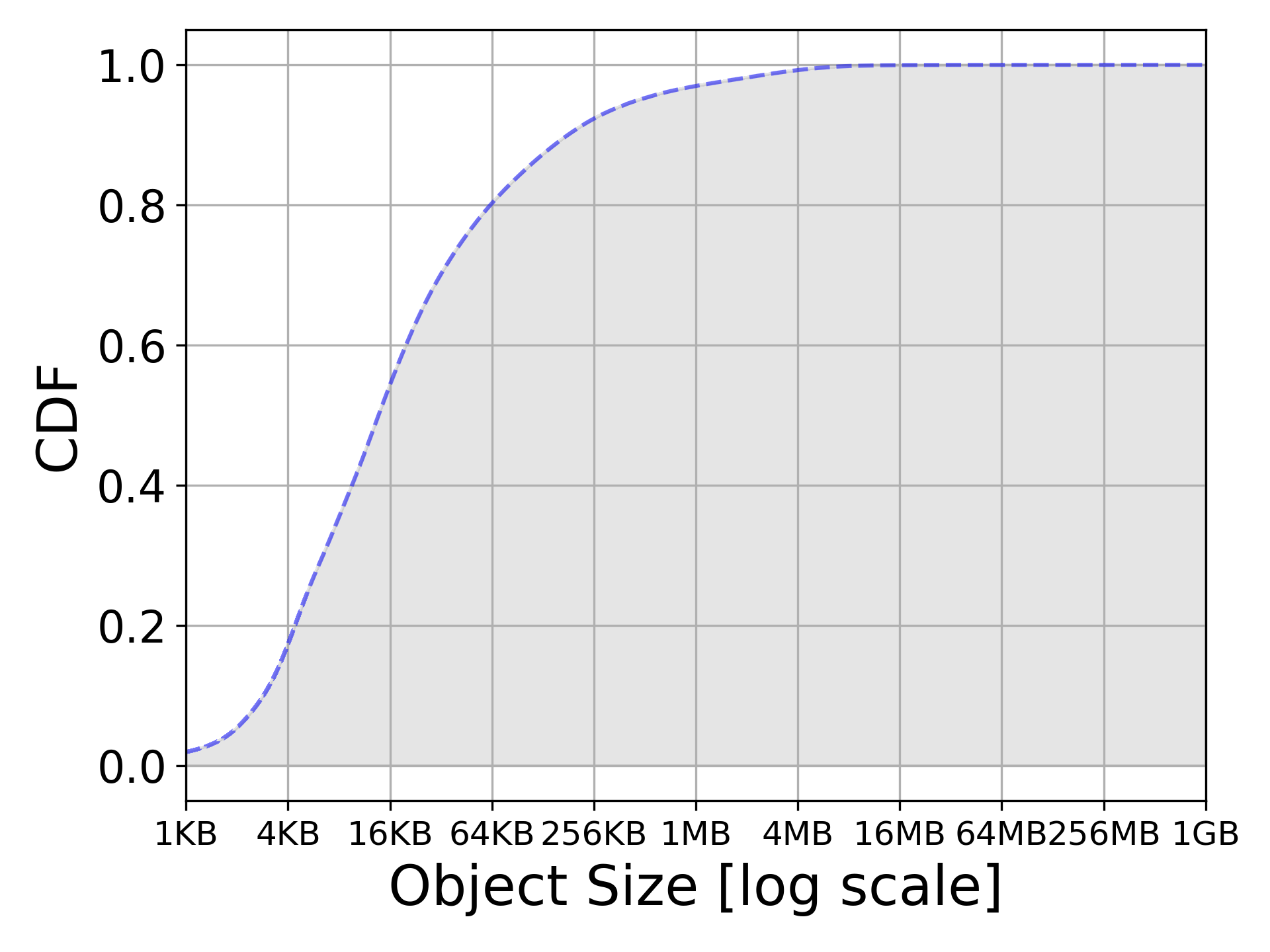}}
		\subfloat[CDN3\label{fig:cdf:cdn3}]{\includegraphics[trim=0 0 0 0, clip, height = 3.0cm]{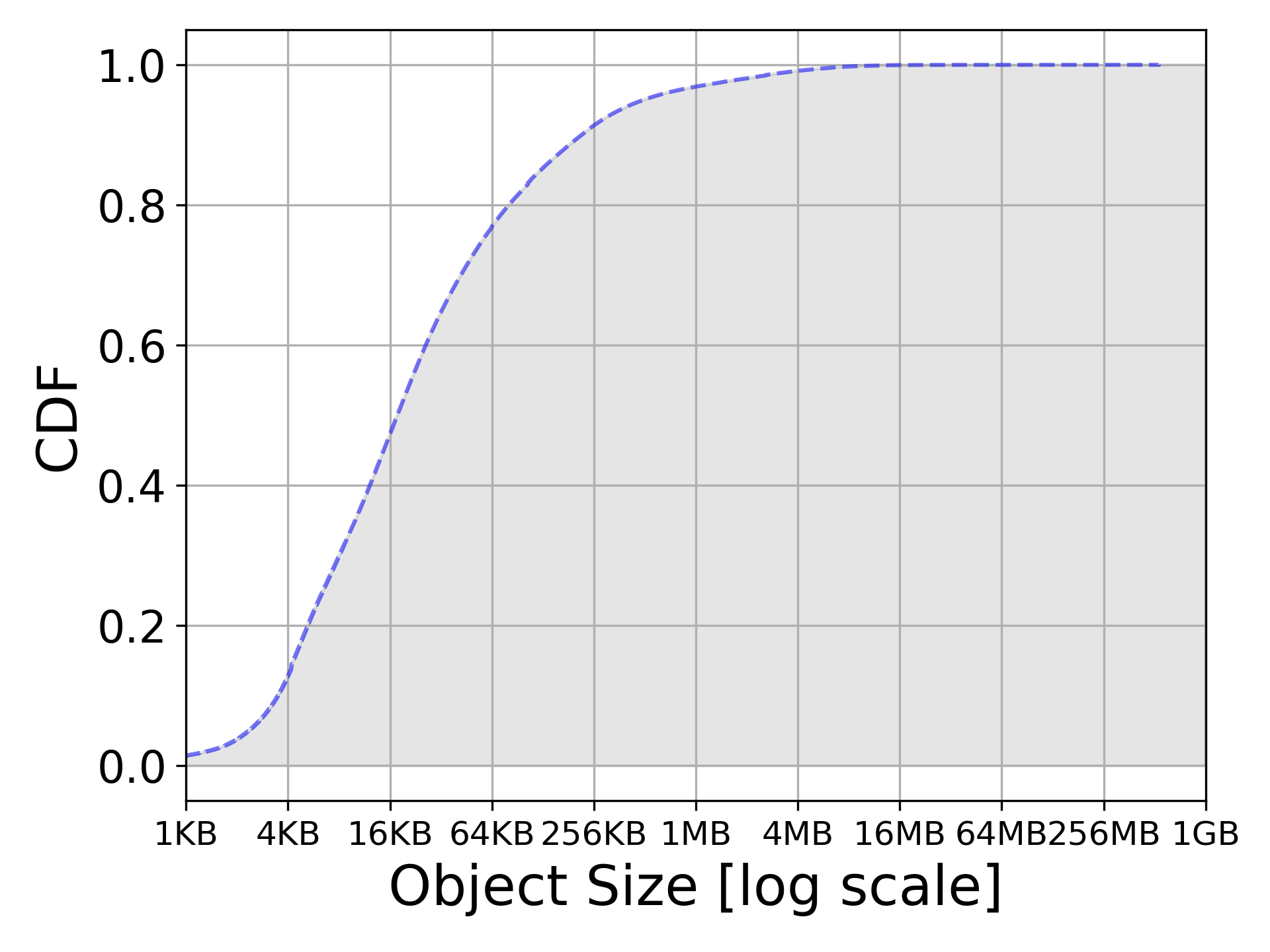}}
\\
		\subfloat[TENCENT1\label{fig:cdf:tencent1}]{\includegraphics[trim=0 0 0 0, clip, height = 3.0cm]{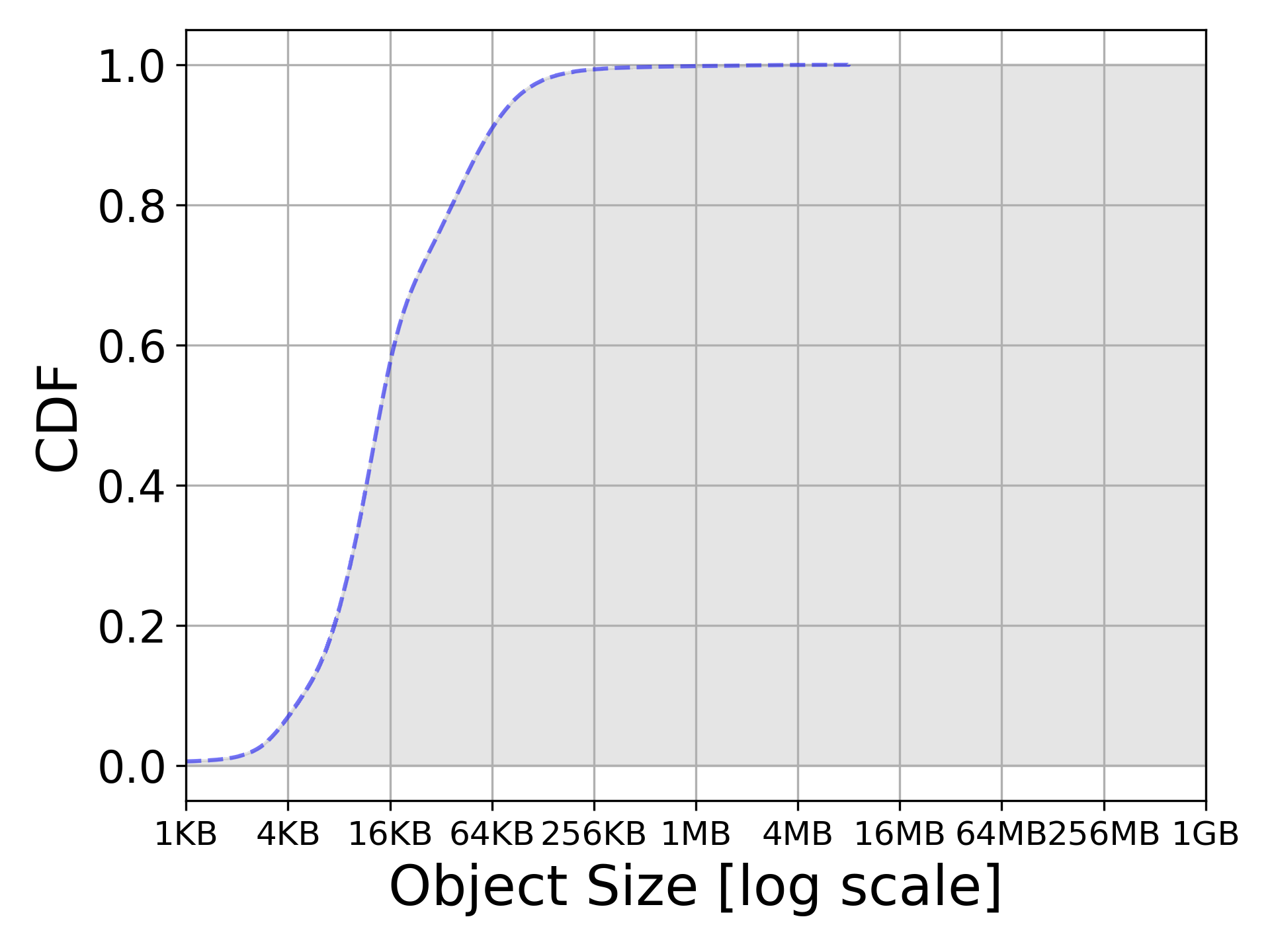}}

	}

	\caption{The cumulative distribution for objects sizes in each trace. The dotted blue lines indicate the actual span of objects sizes in the trace. Notice that in MSR1 and MSR2, most elements belong to a small number of tight object size ranges. In MSR3, TENCENT and the three SYSTOR traces, object sizes are spread further apart. Finally, in CDN traces objects sizes span the entire range.}

	\label{fig:cdfs}

\end{figure*}

\label{sec:traces}
Our measurements have used four types of real traces, taken from four different resources, as detailed in Table~\ref{tab:traces}.
MSR1, MSR2, and MSR3 are taken from an enterprise storage system~\cite{MSR};
SYSTOR1, SYSTOR2, and SYSTOR3 are taken from the storage of a VDI system~\cite{Systor};
CDN1~\cite{PracticalBounds} along with CDN2 and CDN3~\cite{LRB} are taken from CDN deployments;
TENCENT1~\cite{Tencent} is taken from a large-scale photo service.
Table~\ref{tab:traces} details for each trace the total number of accesses in the trace, the number of unique objects, and the total size of all unique objects. 

\begin{table}[h]
	\caption{Summary of Traces}
	\label{tab:traces}		
	\begin{center}
		\begin{tabular}{l|c|c|c} 
			\textbf{Name} & \textbf{Accesses} & \textbf{Objects} & \textbf{Total} \\
			
			 & $millions$ & $millions$ & \textbf{Objects Size} \\
			
			\hline
			
			MSR1 \cite{MSR} & 29 & 18 & 738 $GB$ \\
			MSR2 \cite{MSR} & 37 & 6 & 161 $GB$ \\
			MSR3 \cite{MSR} & 2.2 & 0.27 & 5.3 $GB$ \\
			
			SYSTOR1 \cite{Systor} & 77 & 52 & 1.5 $TB$ \\
			SYSTOR2 \cite{Systor} & 79 & 48 & 1.4 $TB$ \\
			SYSTOR3 \cite{Systor} & 78 & 53 & 1.5 $TB$ \\
			
			CDN1 \cite{PracticalBounds} & 500 & 18 & 2.3 $TB$ \\
			CDN2 \cite{LRB} & 2,800 & 37 & 5.5 $TB$ \\
			CDN3 \cite{LRB} & 2,655 & 50 & 8.3 $TB$ \\

			TENCENT1 \cite{Tencent} & 339 & 137 & 9.8 $TB$ \\
			
		\end{tabular}
		
		\vspace{0.4cm}
		Total number of accesses, unique objects and total size of unique objects for each trace.
	\end{center}
	\normalsize
\end{table}

\begin{figure*}[!h]
	\center{
		\subfloat[MSR2\label{fig:wtinylfu:base:hr:msr2}]{\includegraphics[trim=10 0 0 0, clip, height = 3.0cm]{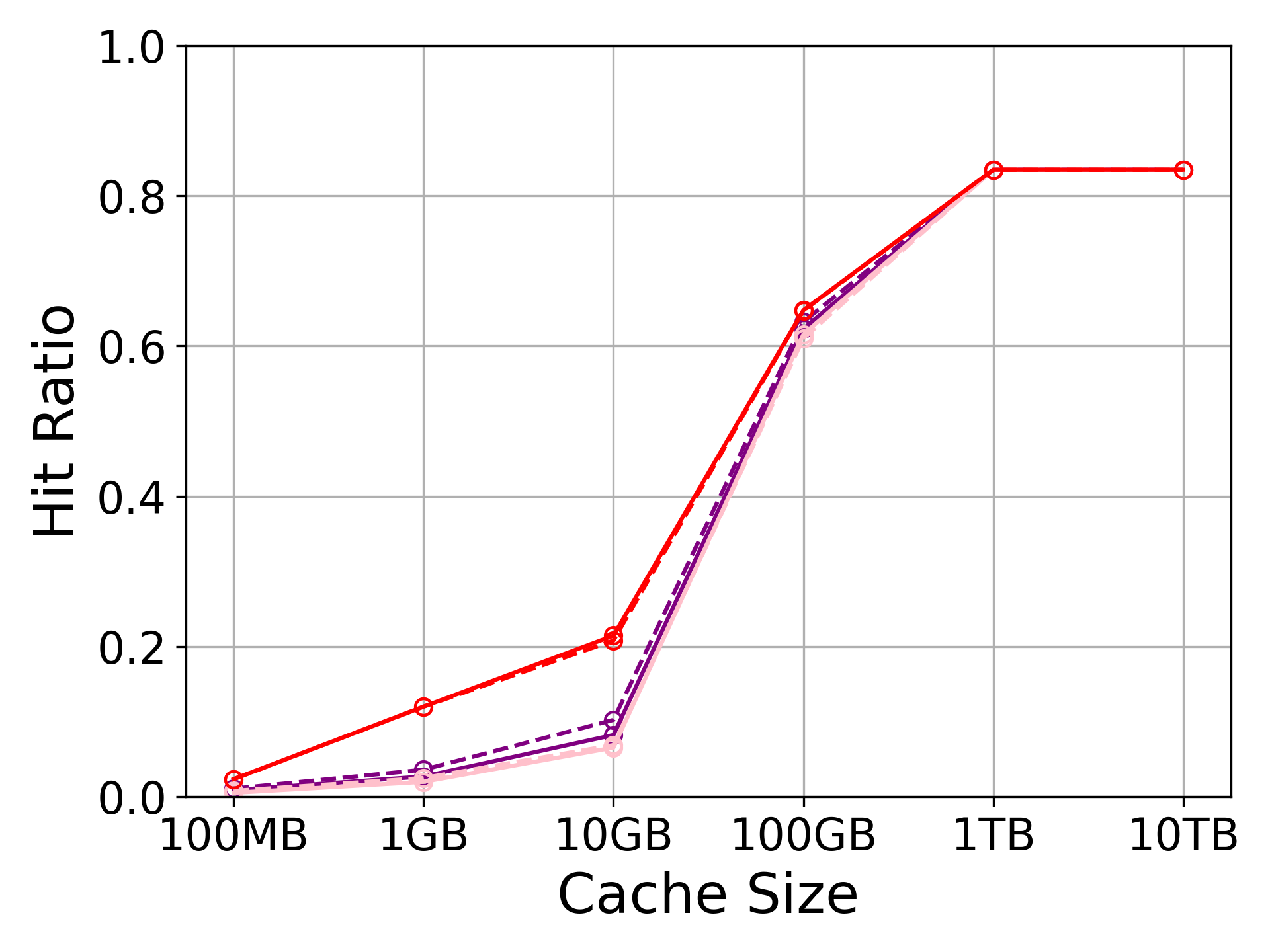}}
		\subfloat[SYSTOR2\label{fig:wtinylfu:base:hr:systor2}]{\includegraphics[trim=30 0 0 0, clip, height = 3.0cm]{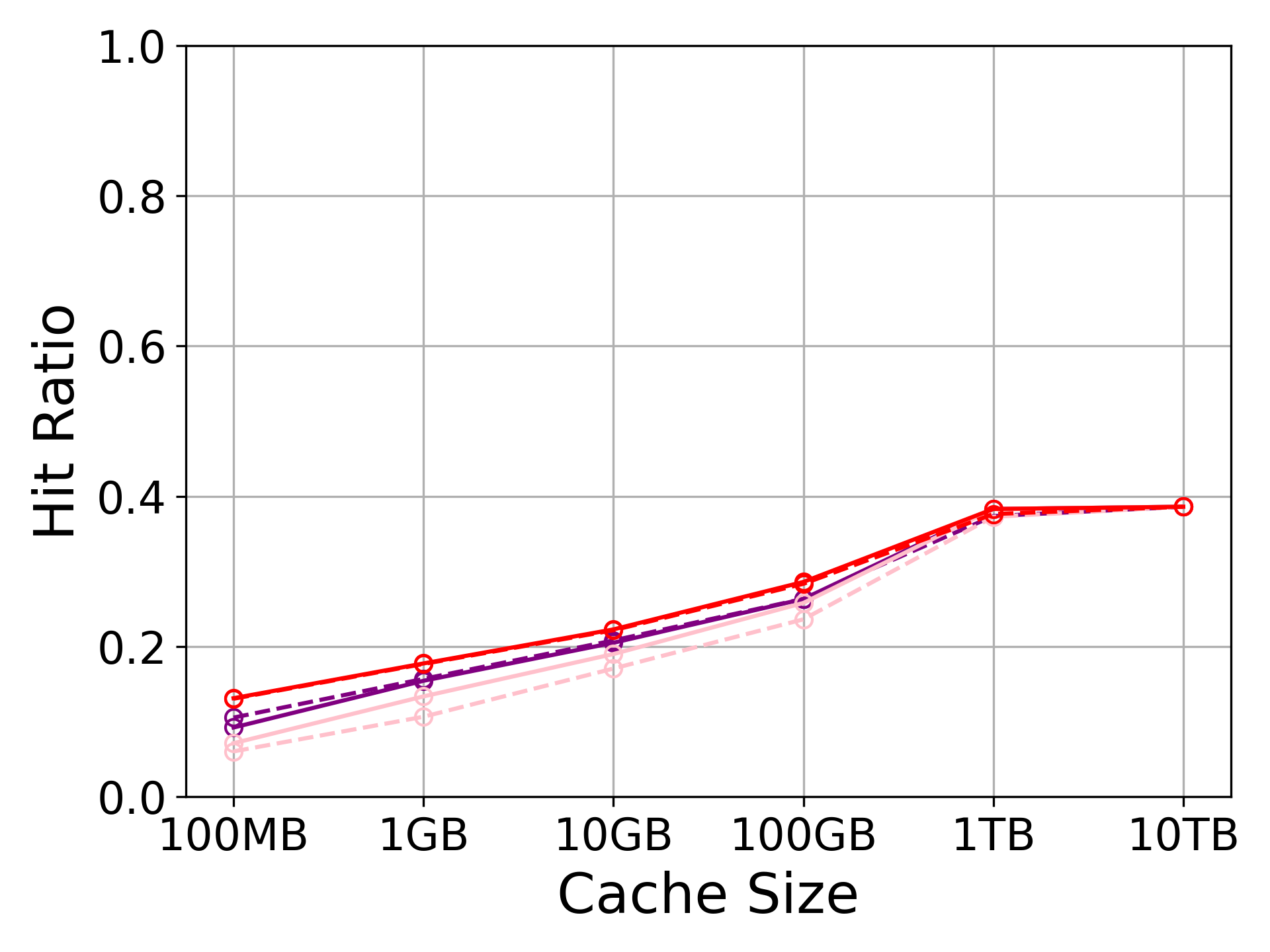}}
		\subfloat[CDN1\label{fig:wtinylfu:base:hr:cdn1}]{\includegraphics[trim=30 0 0 0, clip, height = 3.0cm]{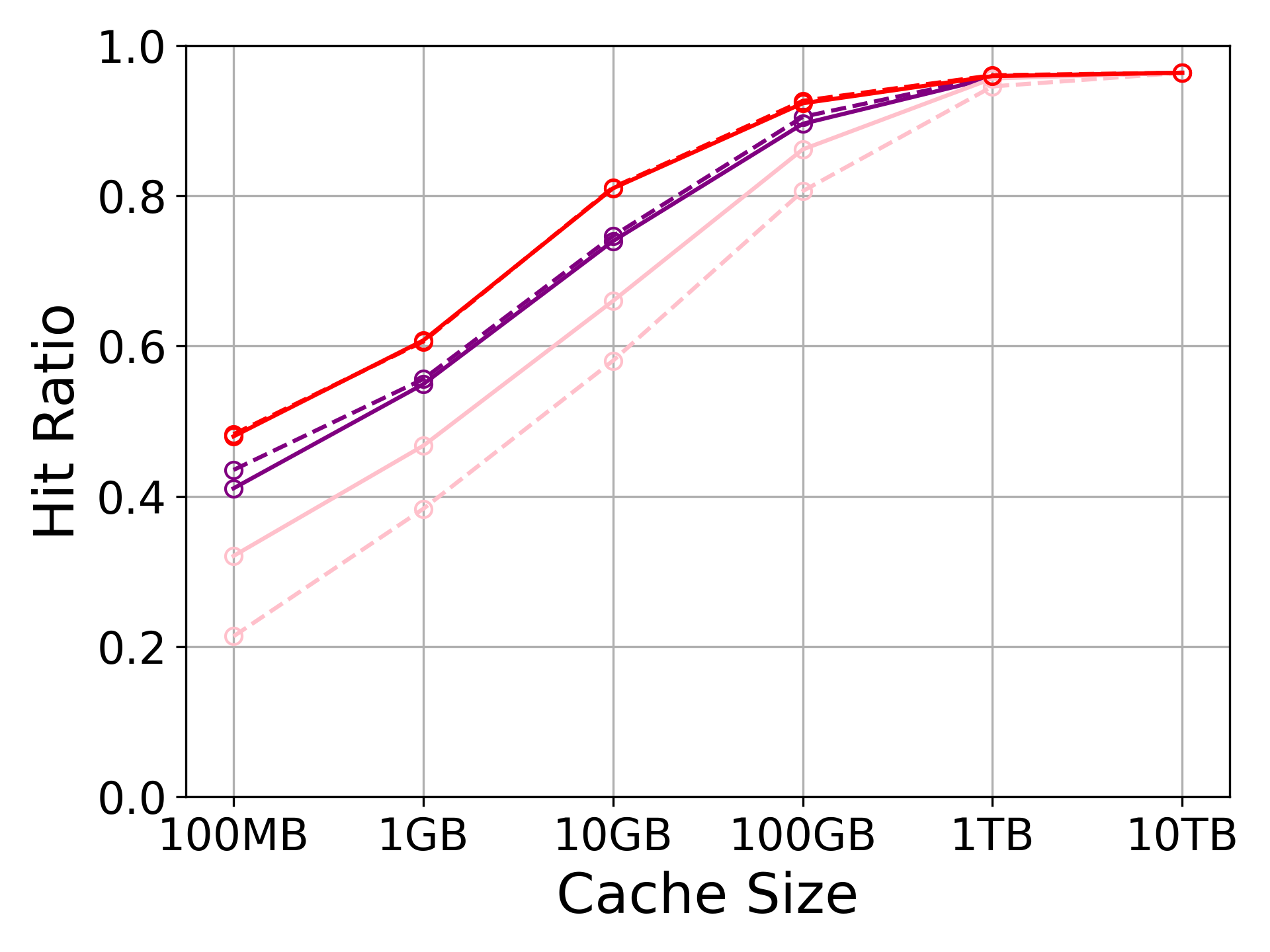}}
		\subfloat[TENCENT1\label{fig:wtinylfu:base:hr:tencent1}]{\includegraphics[trim=30 0 0 0, clip, height = 3.0cm]{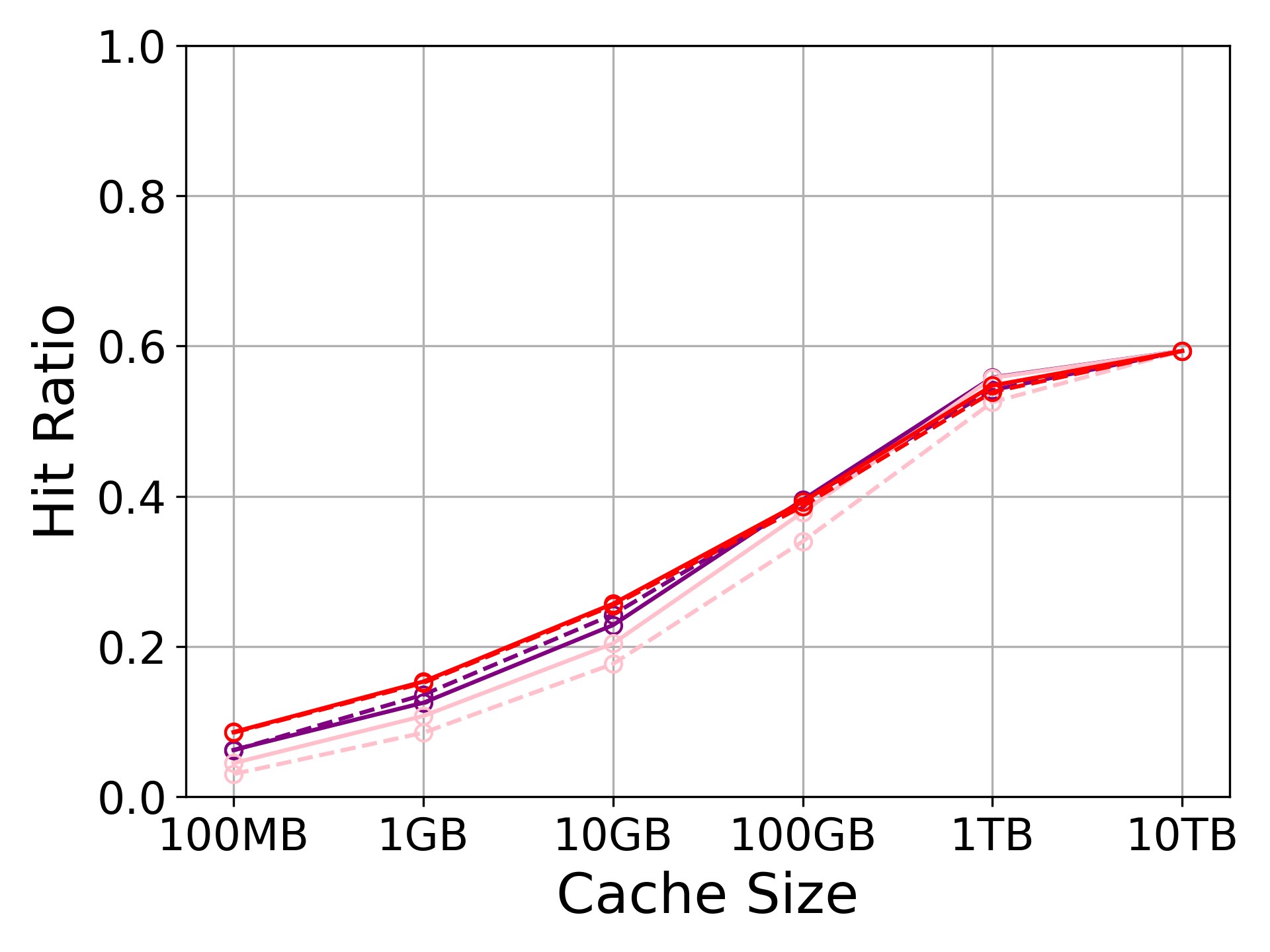}}
		\\
		\vspace{0.15cm}
		\includegraphics[trim=0 0 0 0, clip, height = 0.8cm]{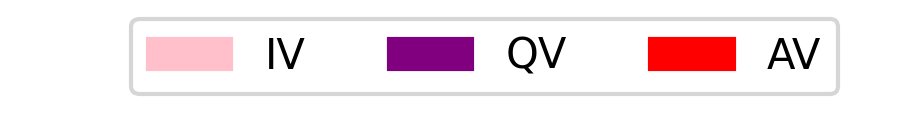}
		\includegraphics[trim=0 0 0 0, clip, height = 0.8cm]{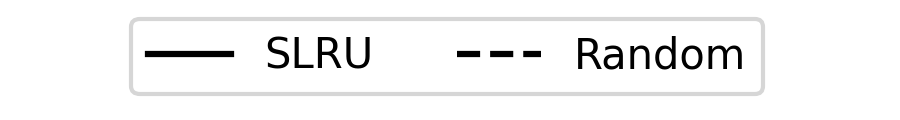}
		\\
		\subfloat[MSR2\label{fig:wtinylfu:freq:hr:msr2}]{\includegraphics[trim=10 0 0 0, clip, height = 3.0cm]{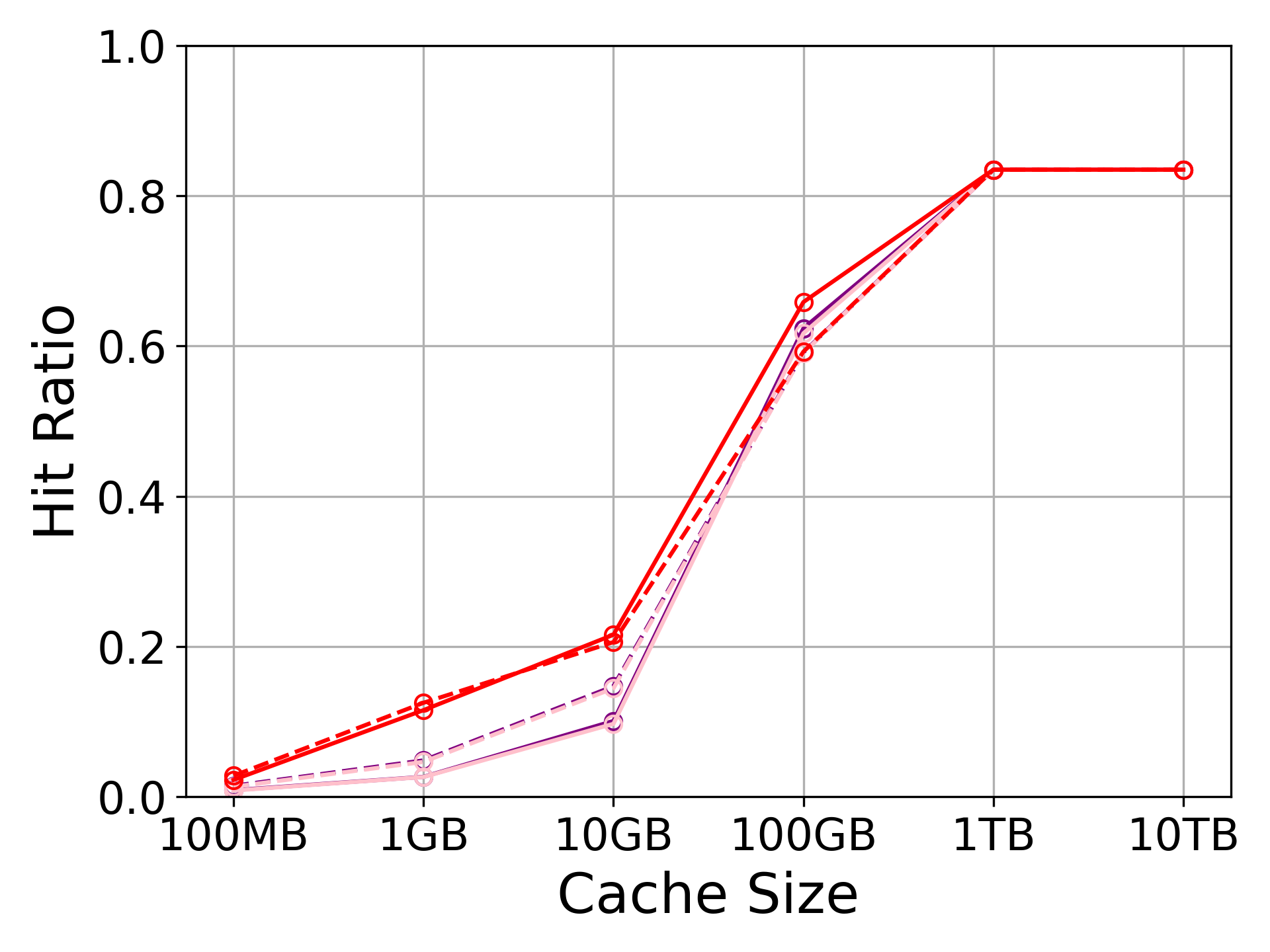}}
		\subfloat[SYSTOR2\label{fig:wtinylfu:freq:hr:systor2}]{\includegraphics[trim=30 0 0 0, clip, height = 3.0cm]{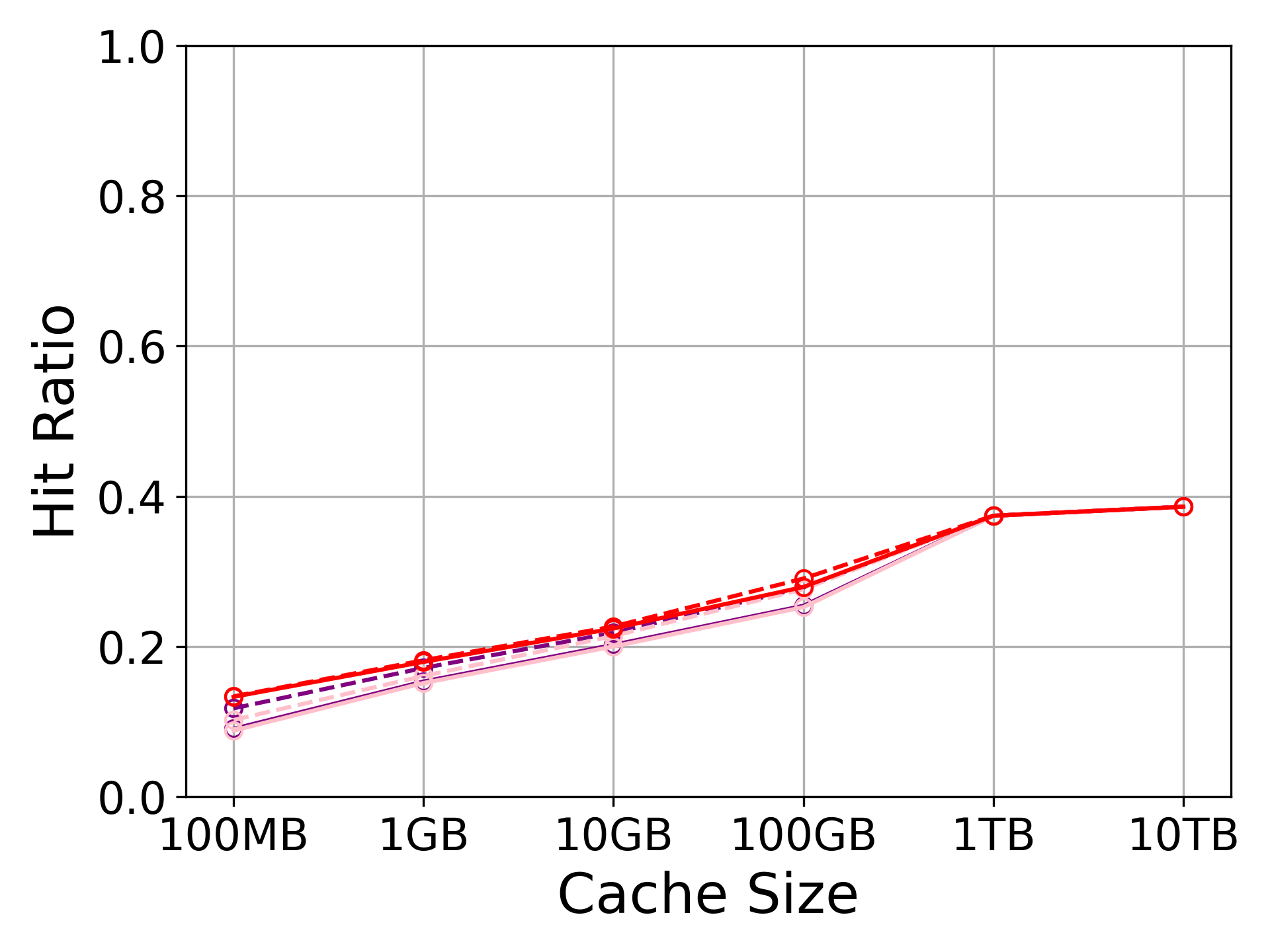}}
		\subfloat[CDN1\label{fig:wtinylfu:freq:hr:cdn1}]{\includegraphics[trim=30 0 0 0, clip, height = 3.0cm]{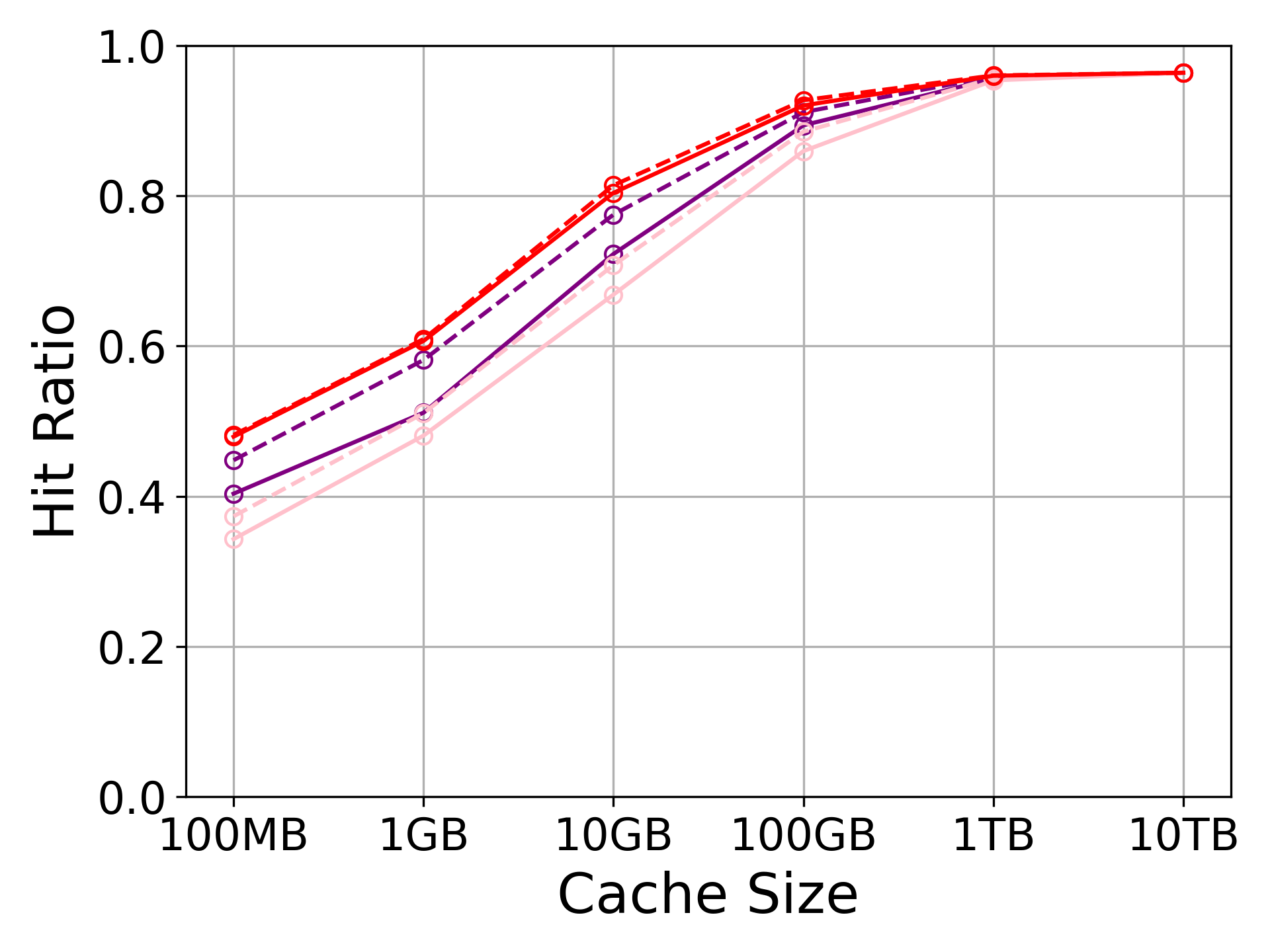}}
		\subfloat[TENCENT1\label{fig:wtinylfu:freq:hr:tencent1}]{\includegraphics[trim=30 0 0 0, clip, height = 3.0cm]{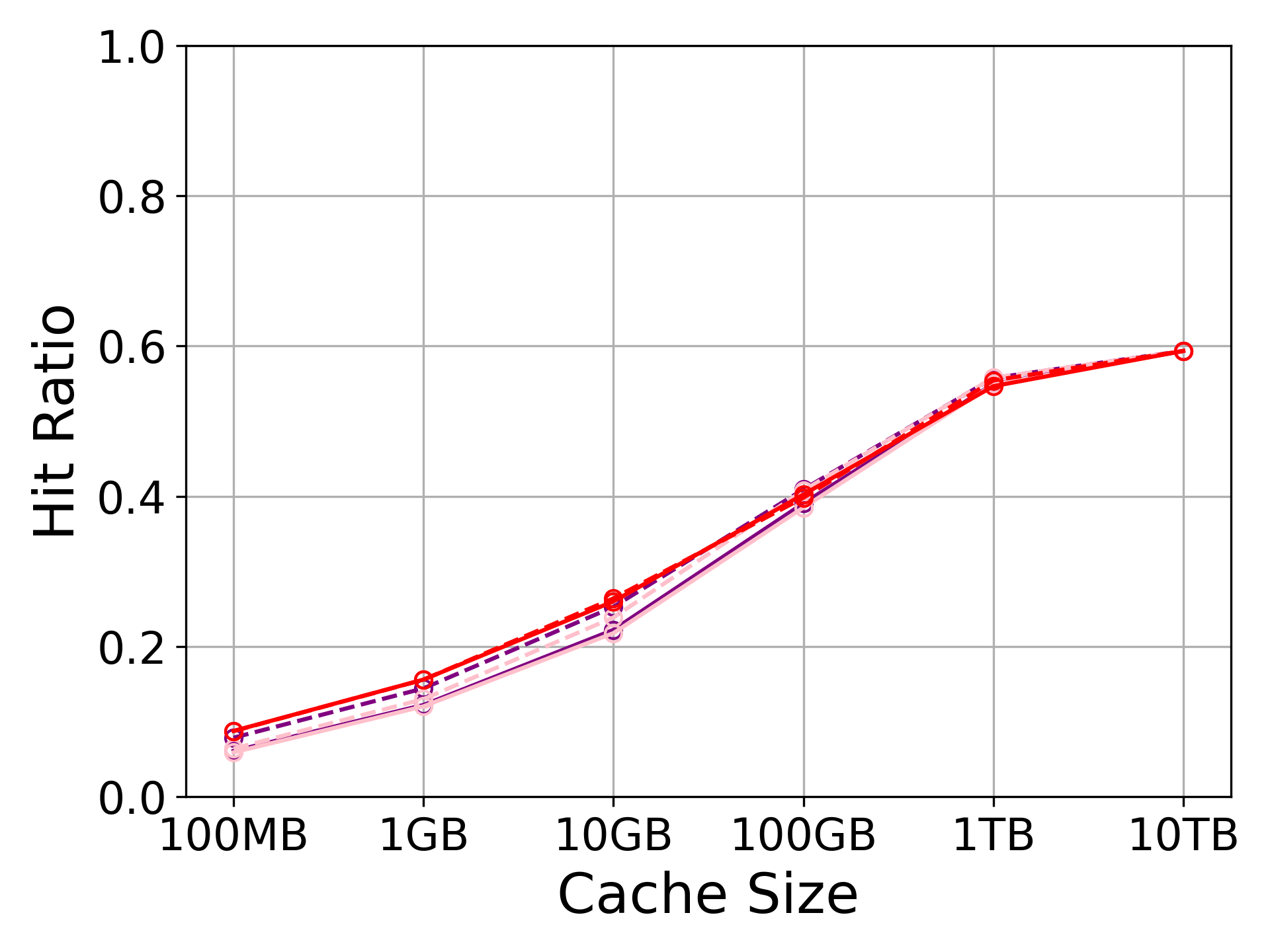}}
		\\
		\vspace{0.15cm}
		\includegraphics[trim=20 0 10 0, clip, height = 0.8cm]{Plots/wtinylfu/size-legend.png}
		\includegraphics[trim=20 0 20 0, clip, height = 0.8cm]{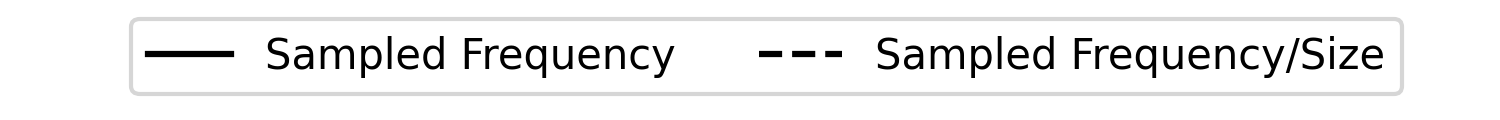}
		\\
		\subfloat[MSR2\label{fig:wtinylfu:size:hr:msr2}]{\includegraphics[trim=10 0 0 0, clip, height = 3.0cm]{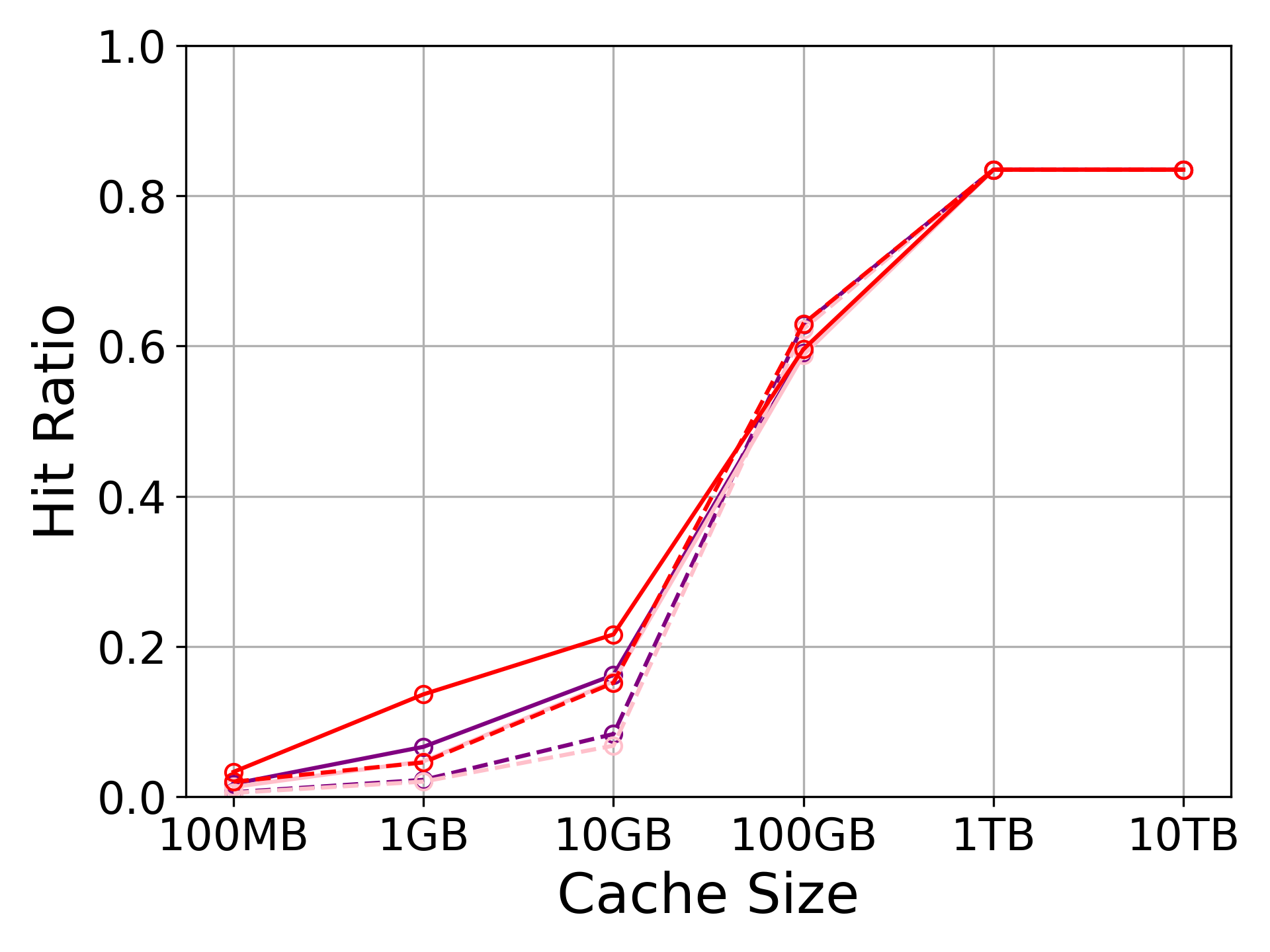}}
		\subfloat[SYSTOR2\label{fig:wtinylfu:size:hr:systor2}]{\includegraphics[trim=30 0 0 0, clip, height = 3.0cm]{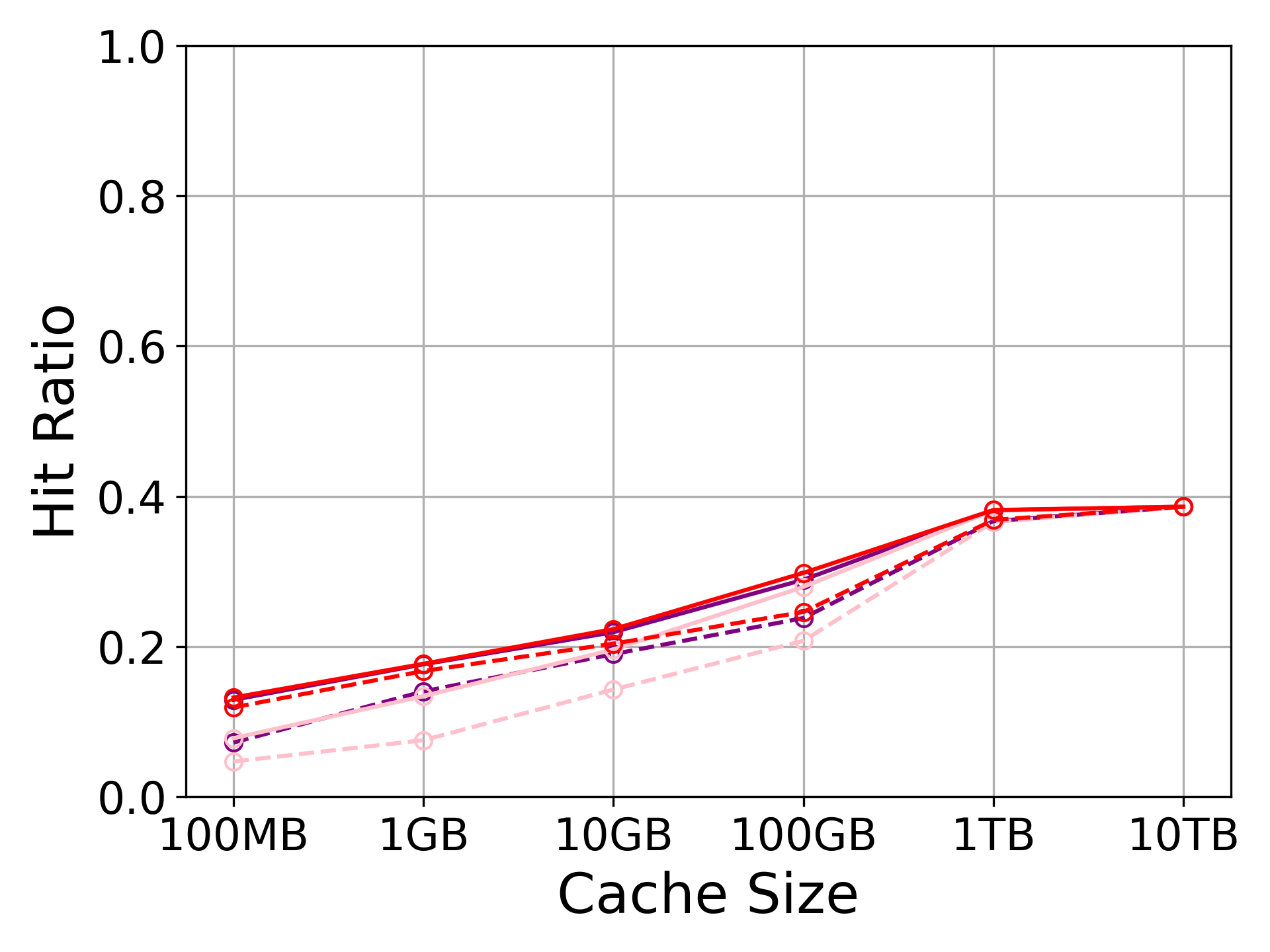}}
		\subfloat[CDN1\label{fig:wtinylfu:size:hr:cdn1}]{\includegraphics[trim=30 0 0 0, clip, height = 3.0cm]{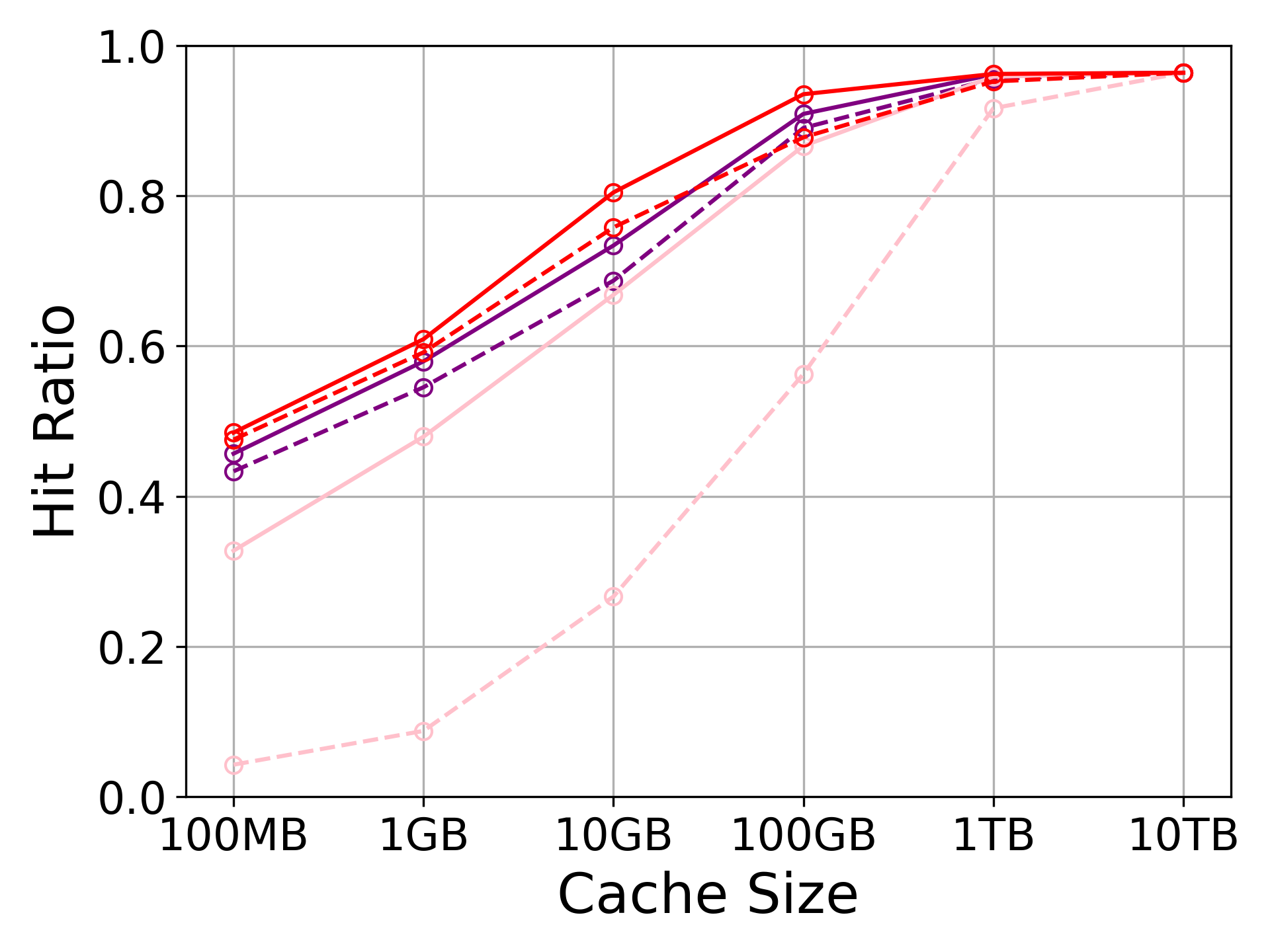}}
		\subfloat[TENCENT1\label{fig:wtinylfu:size:hr:tencent1}]{\includegraphics[trim=30 0 0 0, clip, height = 3.0cm]{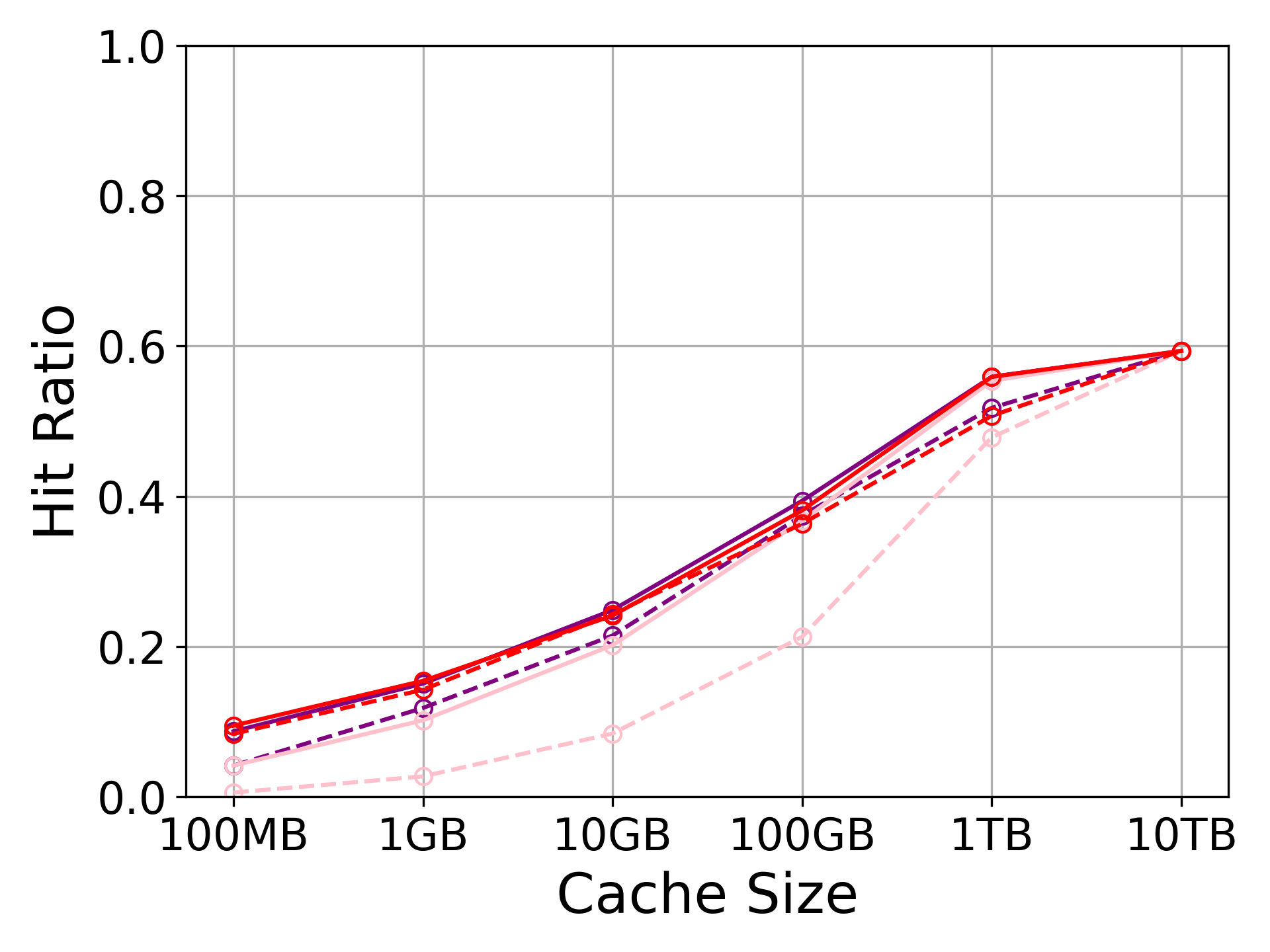}}
		\\
		\vspace{0.15cm}
		\includegraphics[trim=20 0 10 0, clip, height = 0.8cm]{Plots/wtinylfu/size-legend.png}
		\includegraphics[trim=20 0 20 0, clip, height = 0.8cm]{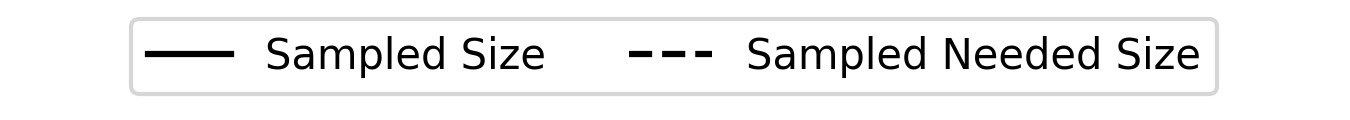}
	}
	\caption{Evaluation of hit-ratio for W-TinyLFU versions. For brevity, we present here only MSR2, SYSTOR2, TENCENT1, and CDN1 traces. Results for the rest of the traces are similar.}
	\label{fig:evaluation-between:hr}
\end{figure*}
\begin{figure*}[!h]
	\center{
		\subfloat[MSR2\label{fig:wtinylfu:base:bhr:msr2}]{\includegraphics[trim=10 0 0 0, clip, height = 3.0cm]{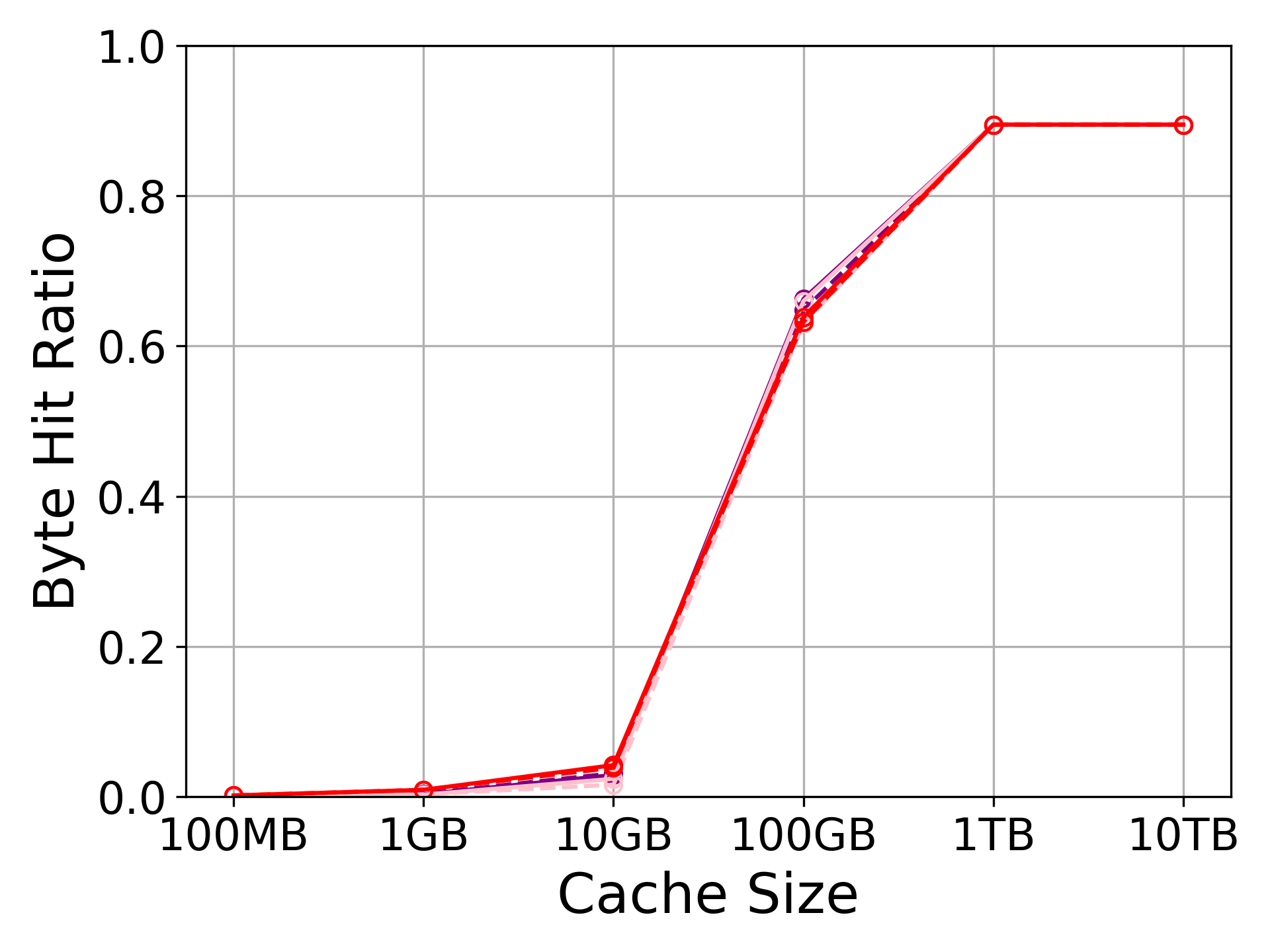}}
		\subfloat[SYSTOR2\label{fig:wtinylfu:base:bhr:systor2}]{\includegraphics[trim=30 0 0 0, clip, height = 3.0cm]{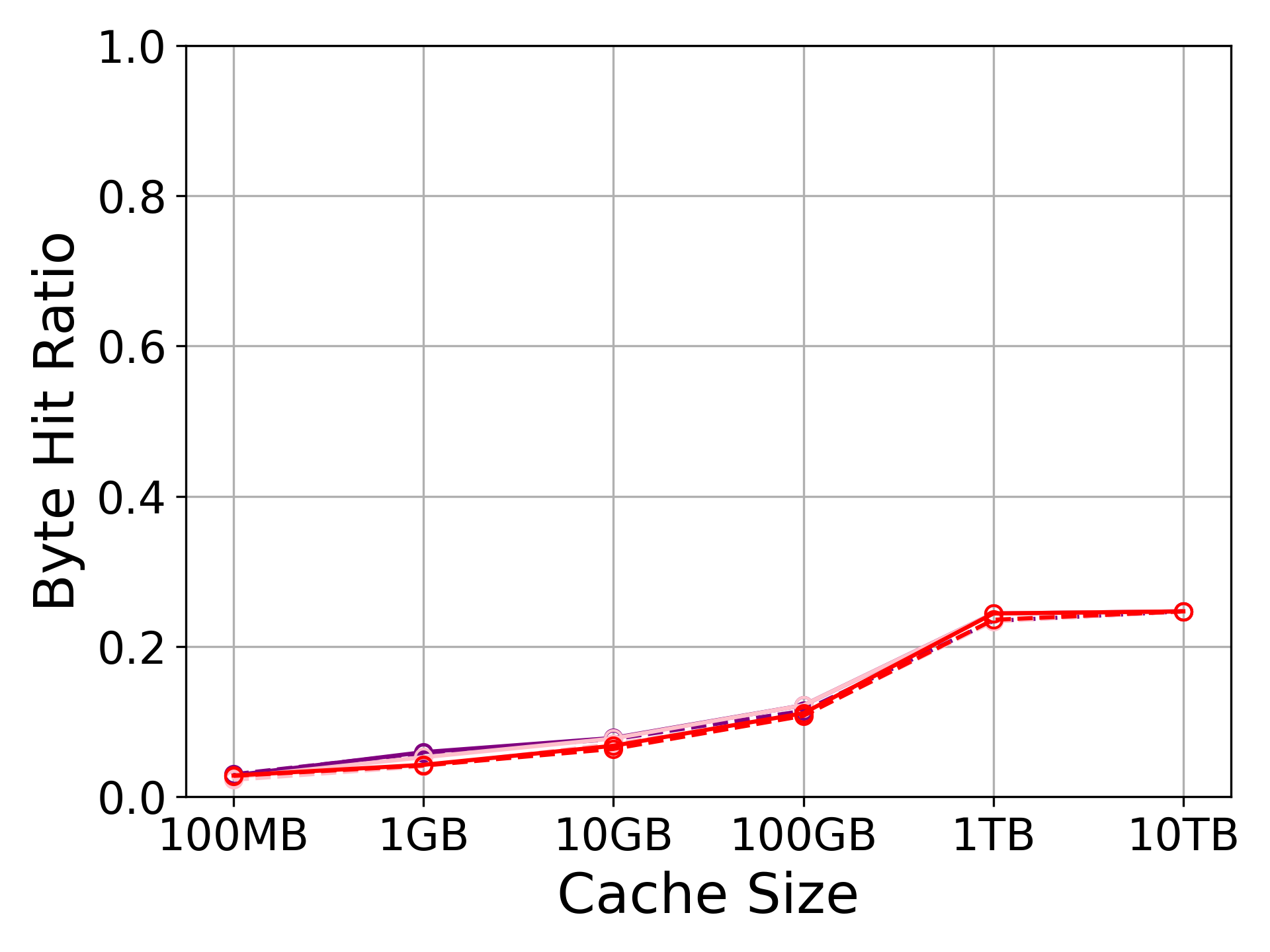}}
		\subfloat[CDN1\label{fig:wtinylfu:base:bhr:cdn1}]{\includegraphics[trim=30 0 0 0, clip, height = 3.0cm]{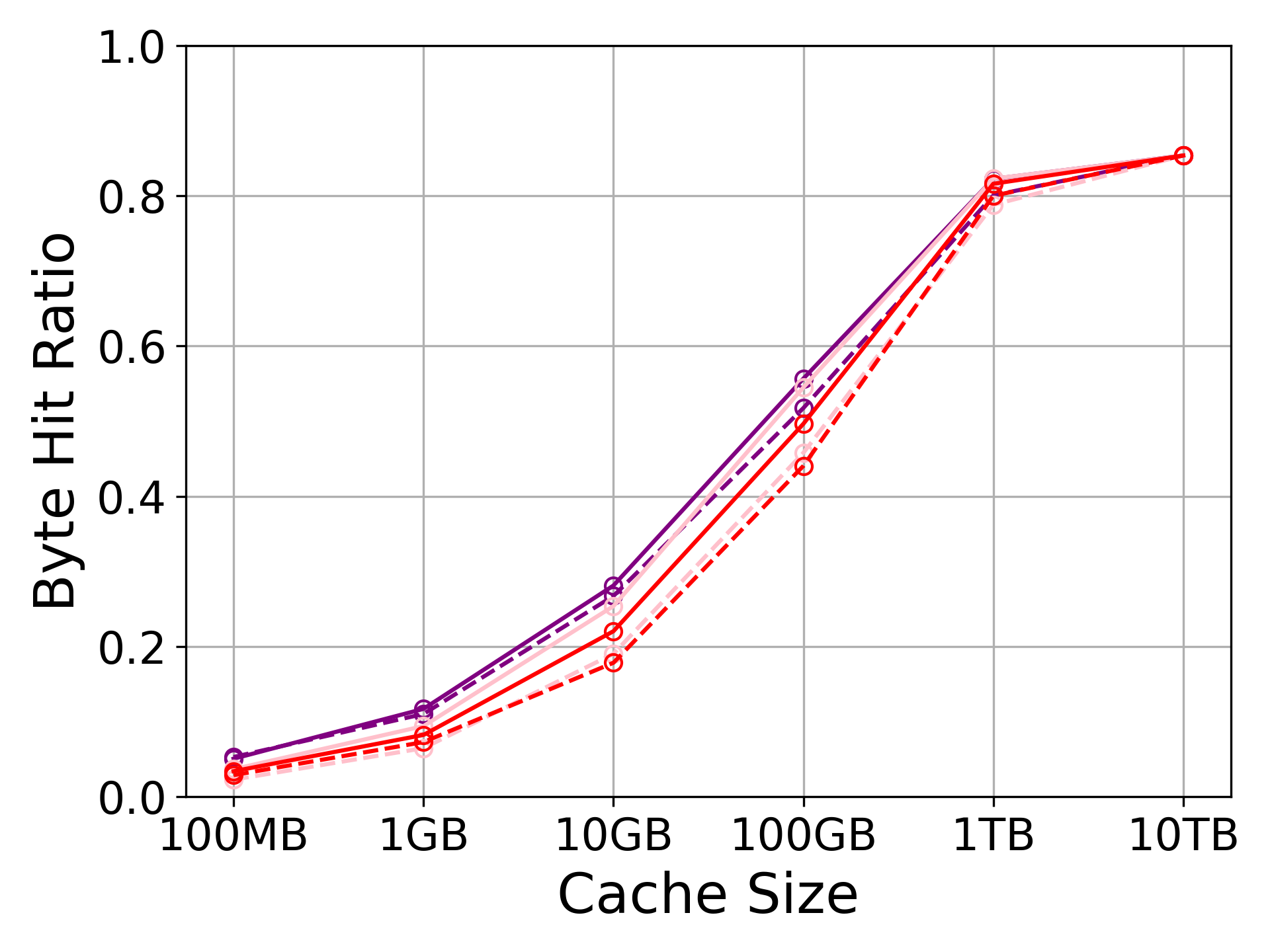}}
		\subfloat[TENCENT1\label{fig:wtinylfu:base:bhr:tencent1}]{\includegraphics[trim=30 0 0 0, clip, height = 3.0cm]{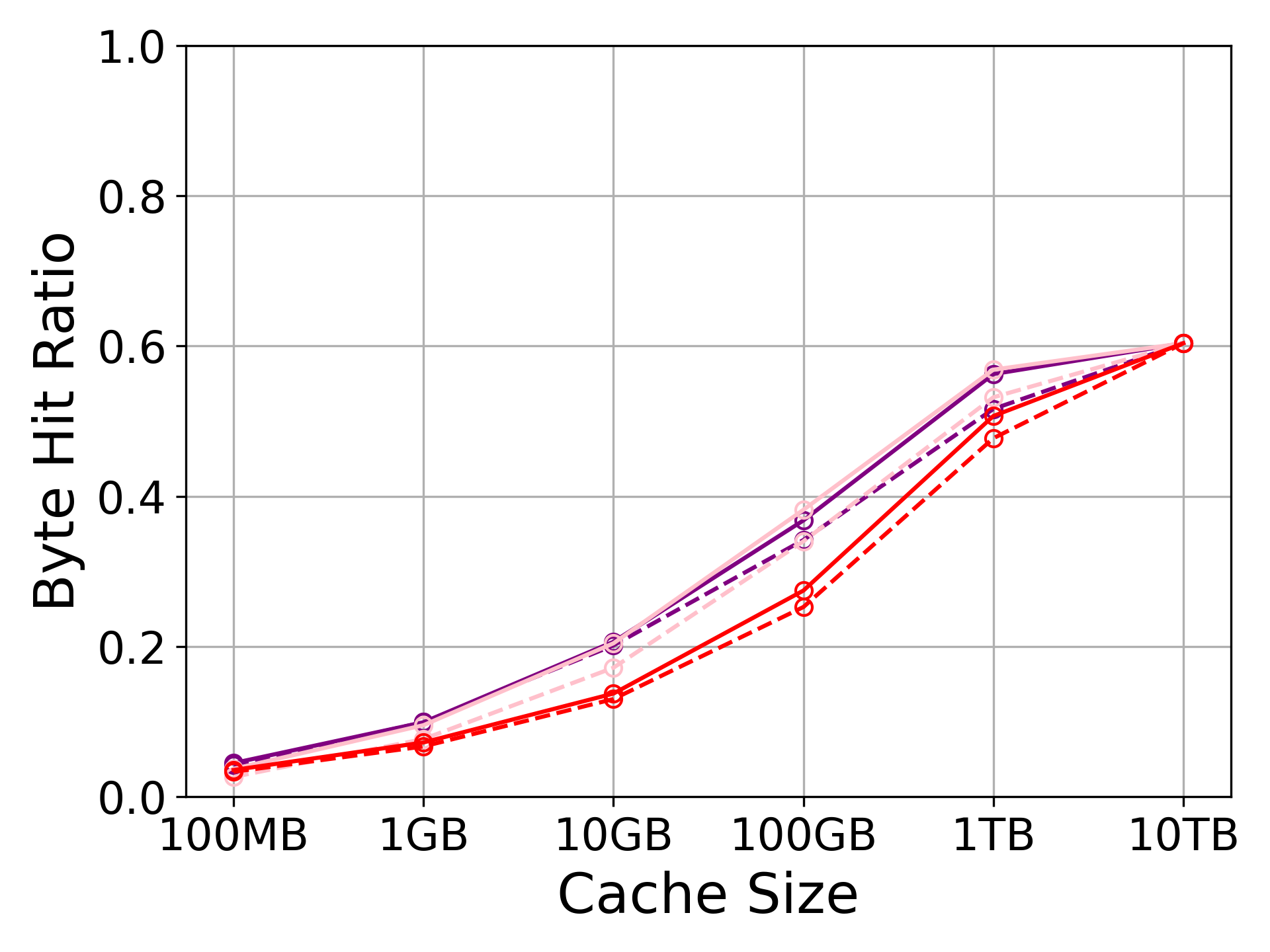}}
		\\
		\vspace{0.15cm}
		\includegraphics[trim=0 0 0 0, clip, height = 0.8cm]{Plots/wtinylfu/size-legend.png}
		\includegraphics[trim=0 0 0 0, clip, height = 0.8cm]{Plots/wtinylfu/base/legend.png}
		\\
		\subfloat[MSR2\label{fig:wtinylfu:freq:bhr:msr2}]{\includegraphics[trim=10 0 0 0, clip, height = 3.0cm]{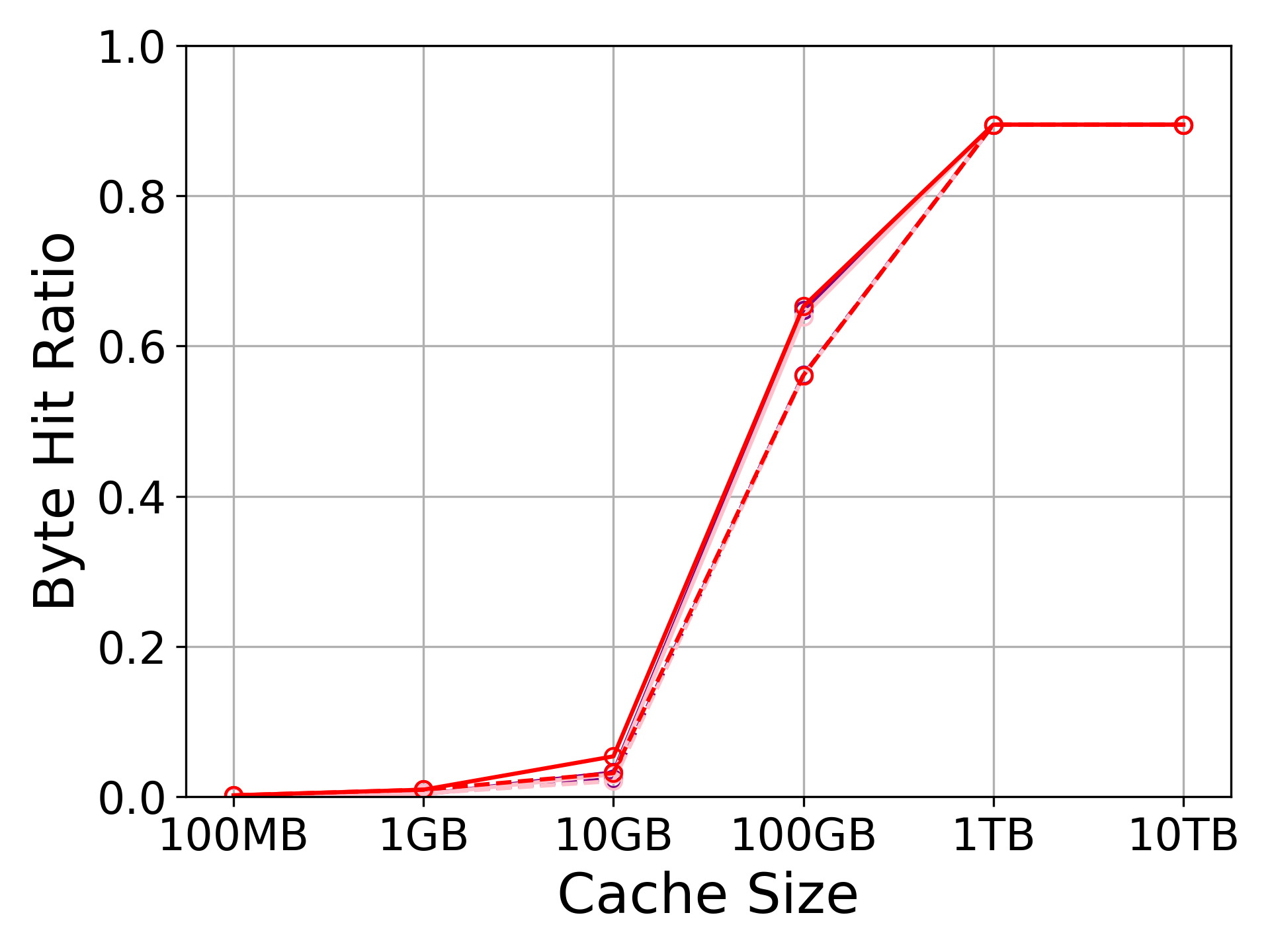}}
		\subfloat[SYSTOR2\label{fig:wtinylfu:freq:bhr:systor2}]{\includegraphics[trim=30 0 0 0, clip, height = 3.0cm]{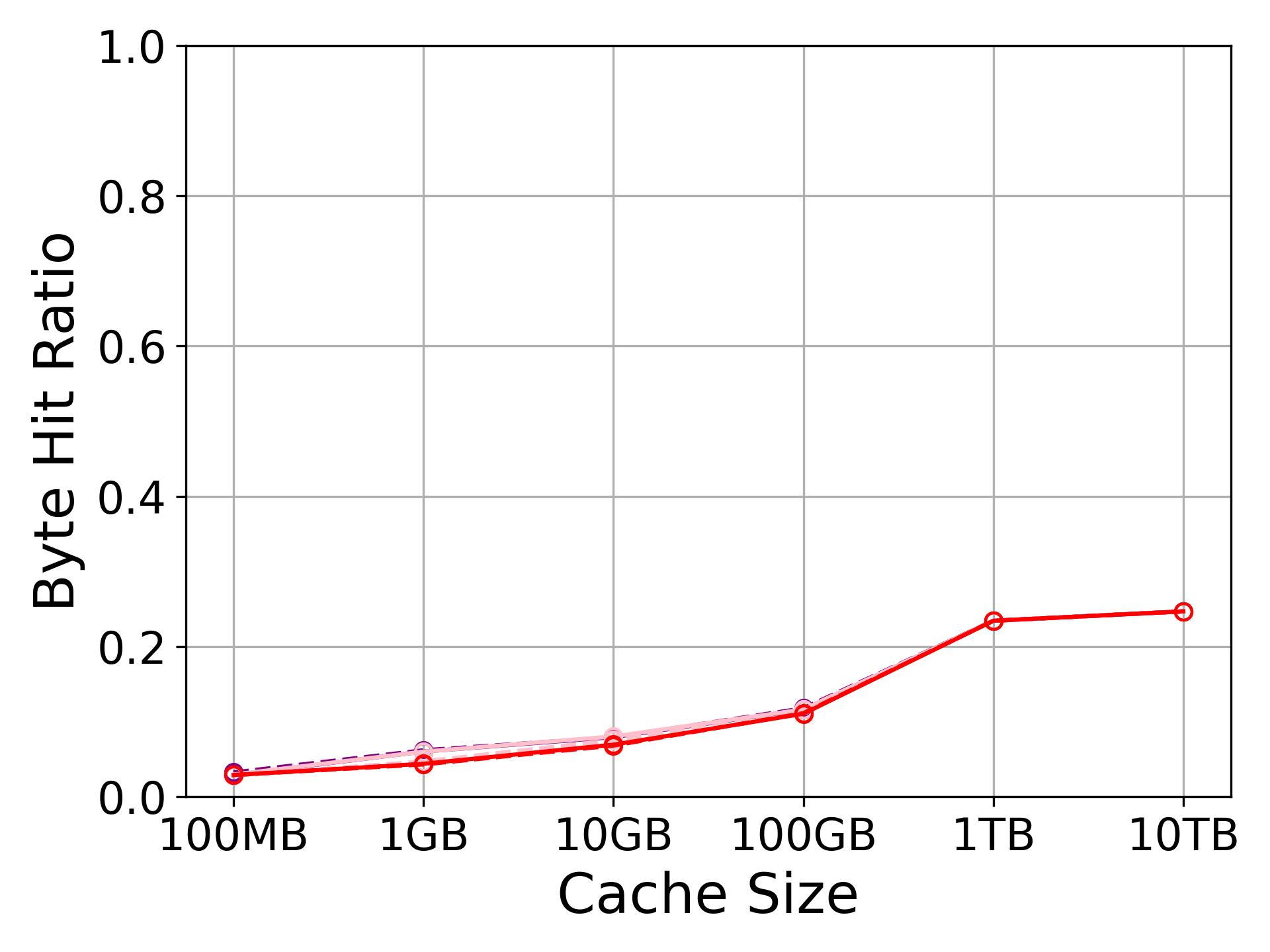}}
		\subfloat[CDN1\label{fig:wtinylfu:freq:bhr:cdn1}]{\includegraphics[trim=30 0 0 0, clip, height = 3.0cm]{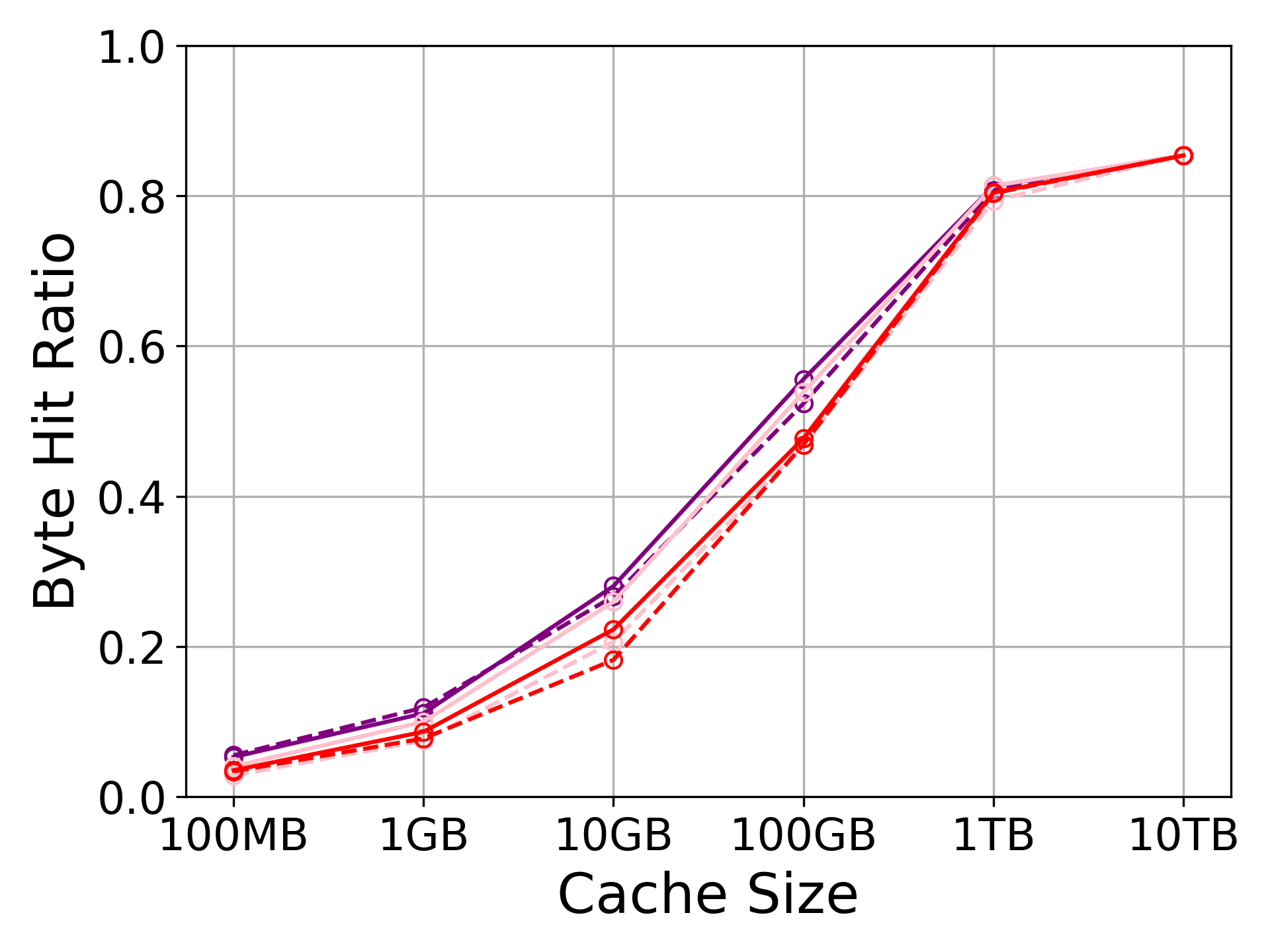}}
		\subfloat[TENCENT1\label{fig:wtinylfu:freq:bhr:tencent1}]{\includegraphics[trim=30 0 0 0, clip, height = 3.0cm]{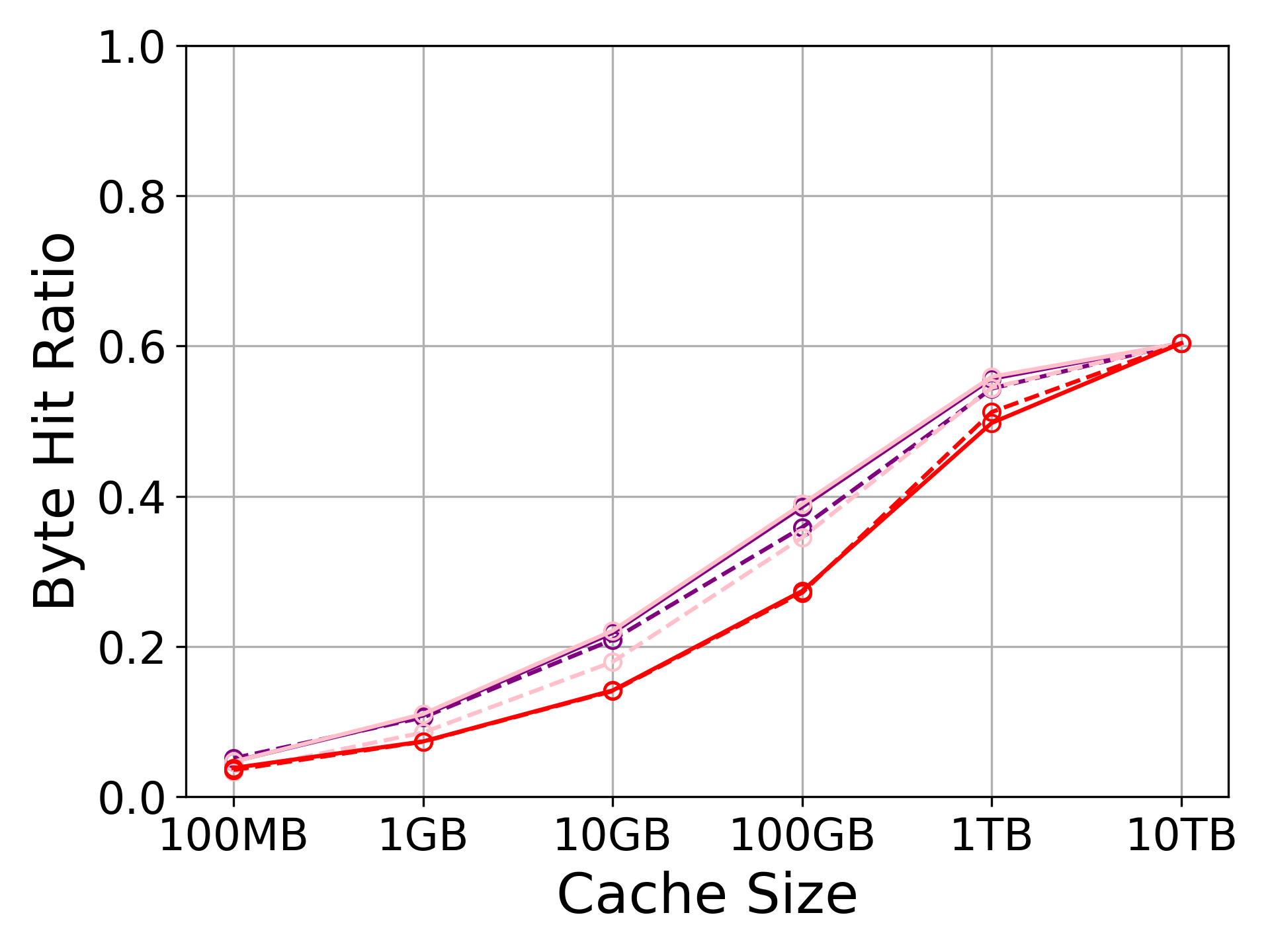}}
		\\
		\vspace{0.15cm}
		\includegraphics[trim=20 0 10 0, clip, height = 0.8cm]{Plots/wtinylfu/size-legend.png}
		\includegraphics[trim=20 0 20 0, clip, height = 0.8cm]{Plots/wtinylfu/freq/legend.png}
		\\
		\subfloat[MSR2\label{fig:wtinylfu:size:bhr:msr2}]{\includegraphics[trim=10 0 0 0, clip, height = 3.0cm]{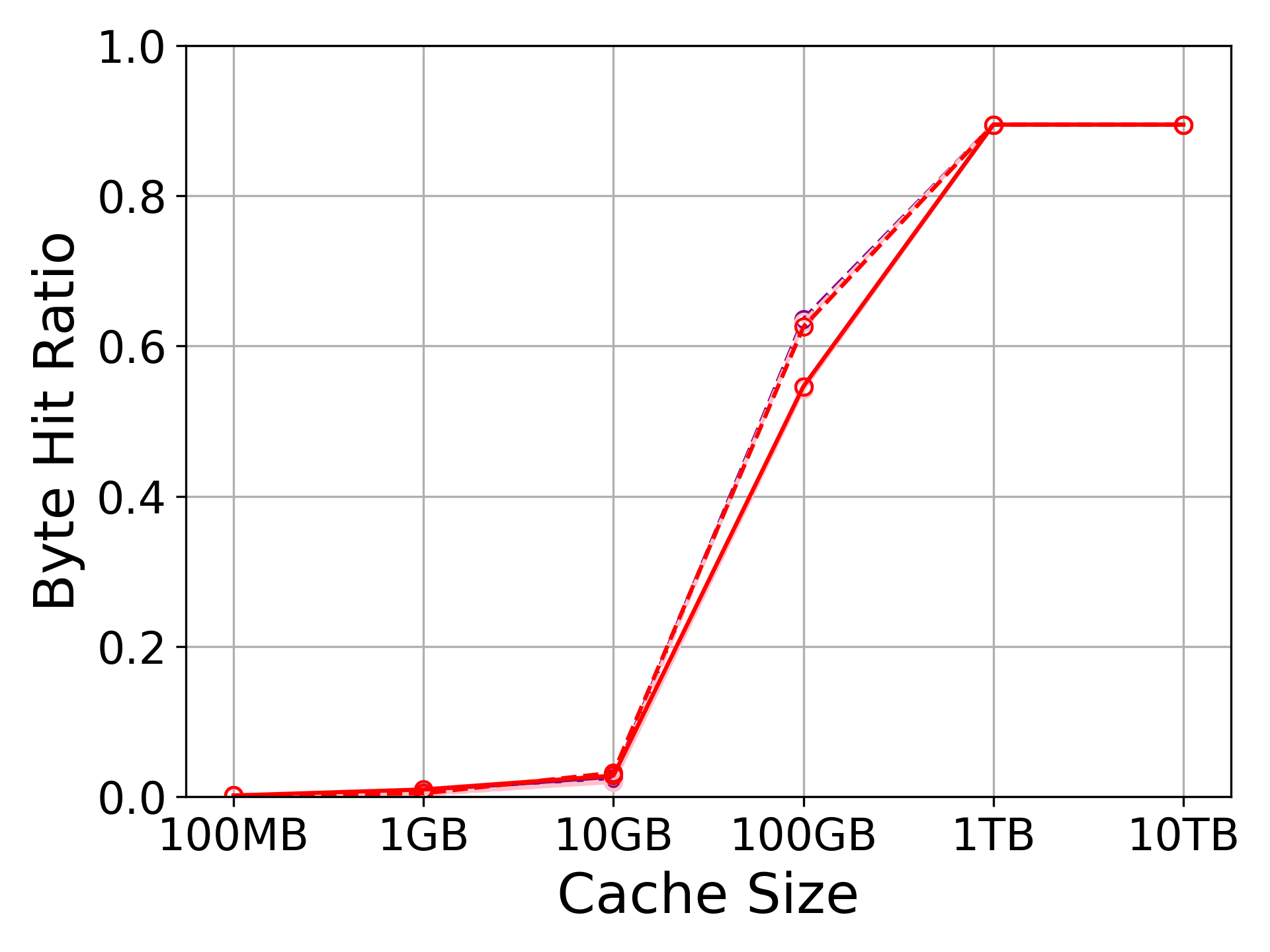}}
		\subfloat[SYSTOR2\label{fig:wtinylfu:size:bhr:systor2}]{\includegraphics[trim=30 0 0 0, clip, height = 3.0cm]{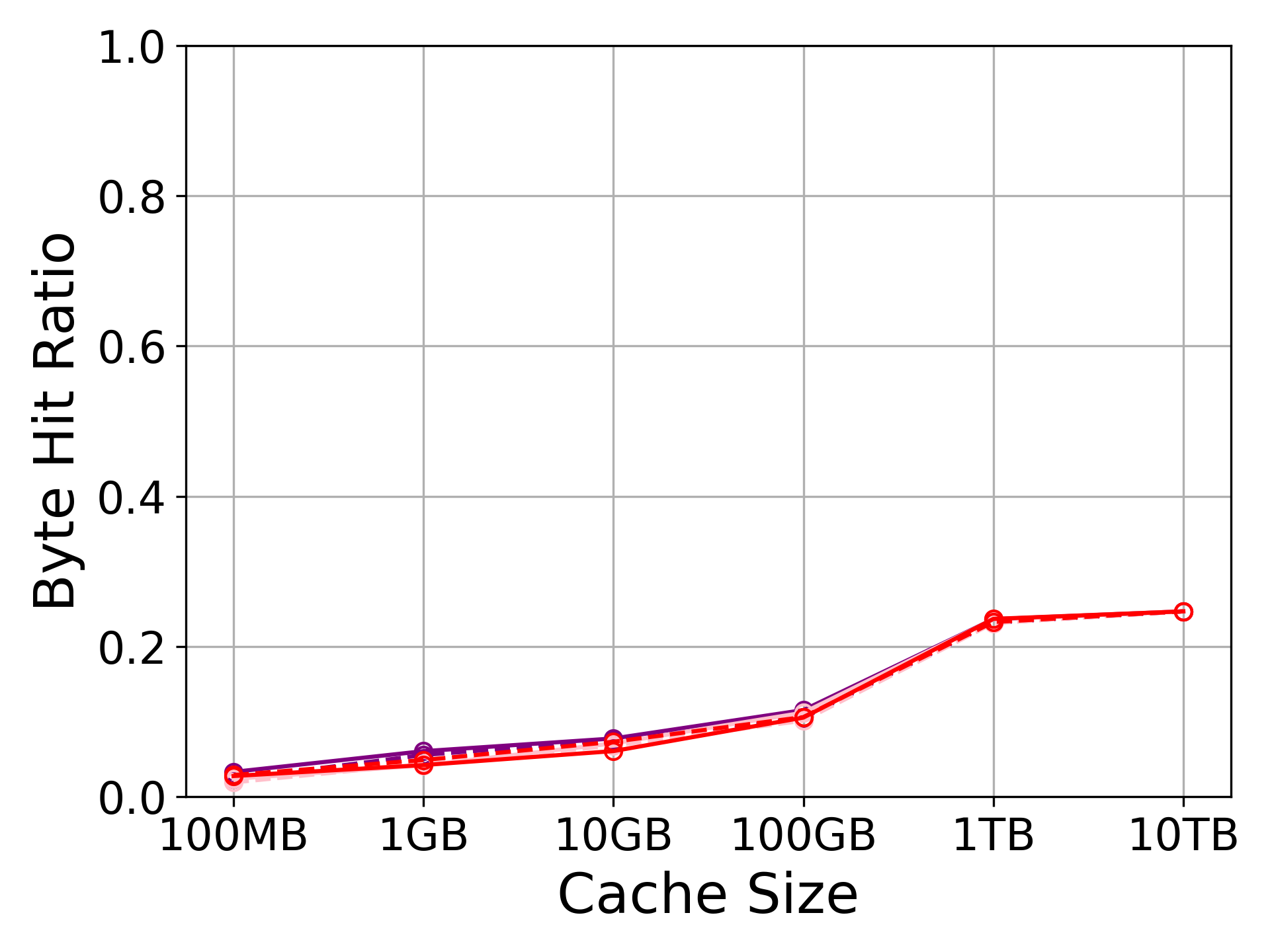}}
		\subfloat[CDN1\label{fig:wtinylfu:size:bhr:cdn1}]{\includegraphics[trim=30 0 0 0, clip, height = 3.0cm]{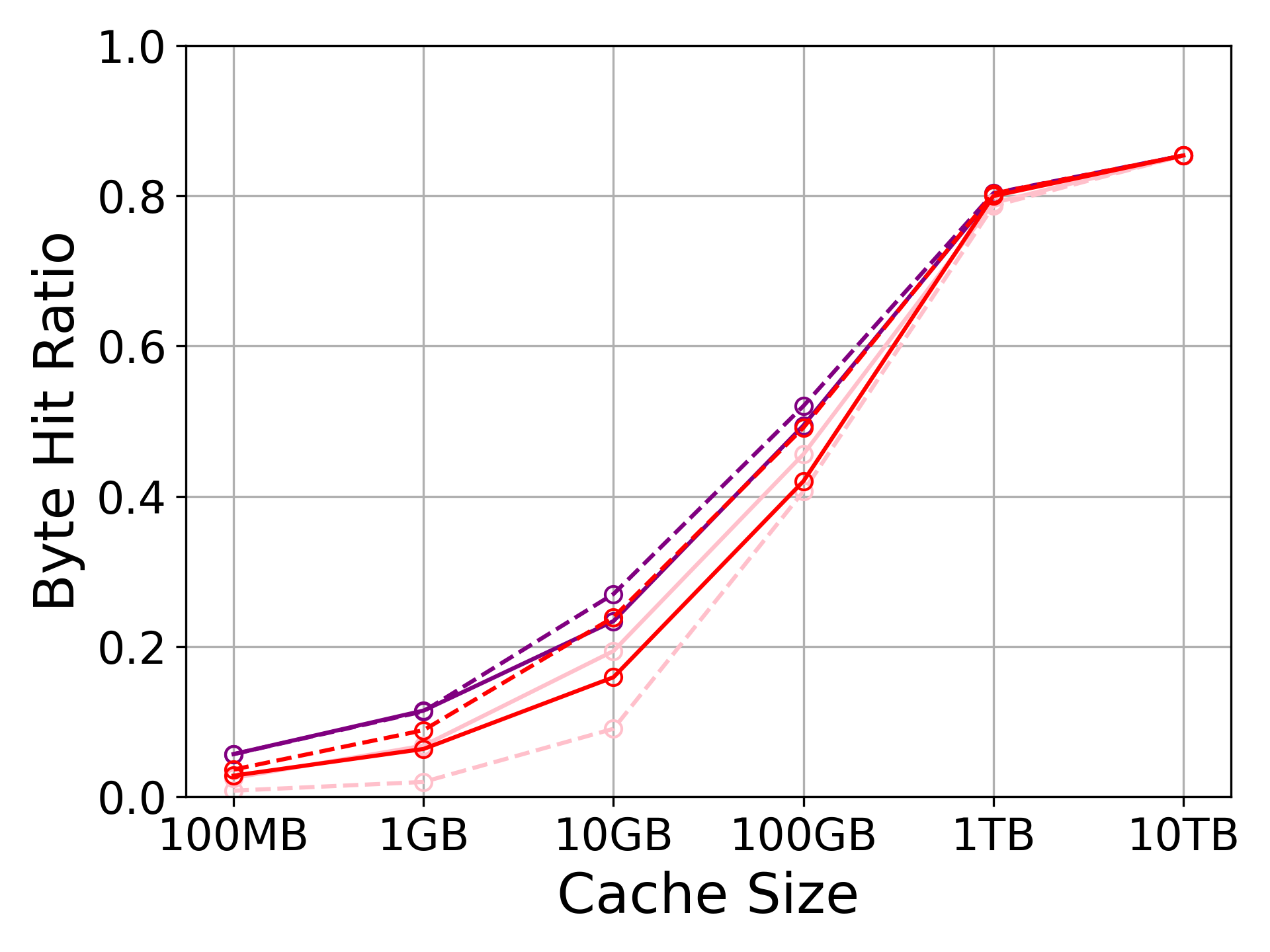}}
		\subfloat[TENCENT1\label{fig:wtinylfu:size:bhr:tencent1}]{\includegraphics[trim=30 0 0 0, clip, height = 3.0cm]{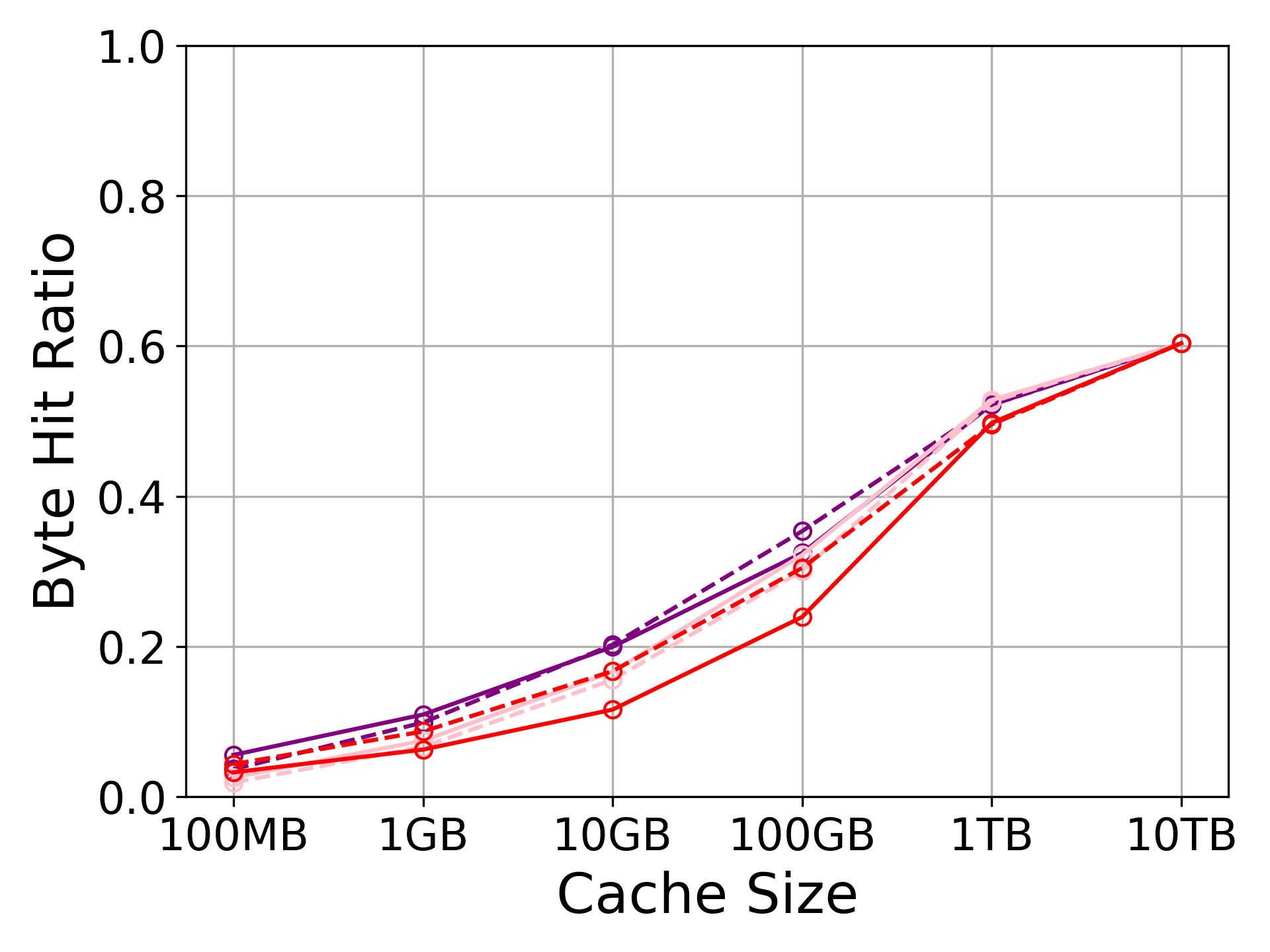}}
		\\
		\vspace{0.15cm}
		\includegraphics[trim=20 0 10 0, clip, height = 0.8cm]{Plots/wtinylfu/size-legend.png}
		\includegraphics[trim=20 0 20 0, clip, height = 0.8cm]{Plots/wtinylfu/size/legend.png}
		
	}
	\caption{Evaluation of byte hit-ratio for W-TinyLFU versions. For brevity, we present here only MSR2, SYSTOR2, TENCENT1, and CDN1 traces. Results for the rest of the traces are similar.}
	\label{fig:evaluation-between:bhr}
\end{figure*}

Another interesting aspect of these traces is their object size distributions.
This is depicted in Figure~\ref{fig:cdfs}.
As can be observed, all traces include objects of various sizes, ranging from less than 1KB to 0.5MB in the case of both storage systems, and up to 0.5GB or more in the case of the CDN.
We can also see that in MSR1 and MSR2, it is easy to divide the objects into a small number of ``size buckets'', suggesting that here slabbing might work well.
In particular, 3--4 slabs could suffice.
MSR3, TENCENT1 and SYSTORs traces are already more challenging, and the sizes of the objects are much less clustered.
Finally, in the CDNs, object sizes are spread in the entire range of up to 0.5GB, hinting that a true size aware policy is~needed.

\subsection{Finding the Best Hit-Ratio Filter Variant}
\label{sec:variants}

Figure~\ref{fig:evaluation-between:hr} compares our \PROTNAME{} with the IV approach (taken by Caffeine~\cite{CaffeineProject}) and the QV approach (used by Ristretto~\cite{RistrettoProject}). 
As illustrated, \PROTNAME{} is consistently better for all traces in the entire range.
Such improvement is explained by the fact that \PROTNAME{} more accurately captures the victim set's benefit compared with IV and QV.
Also, QV seems slightly better than~IV. 

Figure~\ref{fig:evaluation-between:bhr} repeats the above evaluation for byte hit ratio.
Here, QV seems to be the best admission policy.
Intuitively, AV's relative weakness on this metric is because it is more likely to admit smaller items than larger items. 
Specifically, smaller items are likely to be compared against few victims, while larger items are compared against many victims.
The byte hit ratio benefits more from serving a large item from the cache than from serving a small item.
The QV approach is indifferent to items' size and actively removes infrequent items from the cache even if it rejects the new item. 
These evictions free poorly utilized space, making QV more likely to admit large items to improve the byte hit ratio. 

\begin{figure*}[!ht]
	\center{
		\vspace{-0.3cm}
		\subfloat[MSR1\label{fig:others:hr:msr1}]{\includegraphics[trim=0 10 0 0, clip, height = 3.3cm]{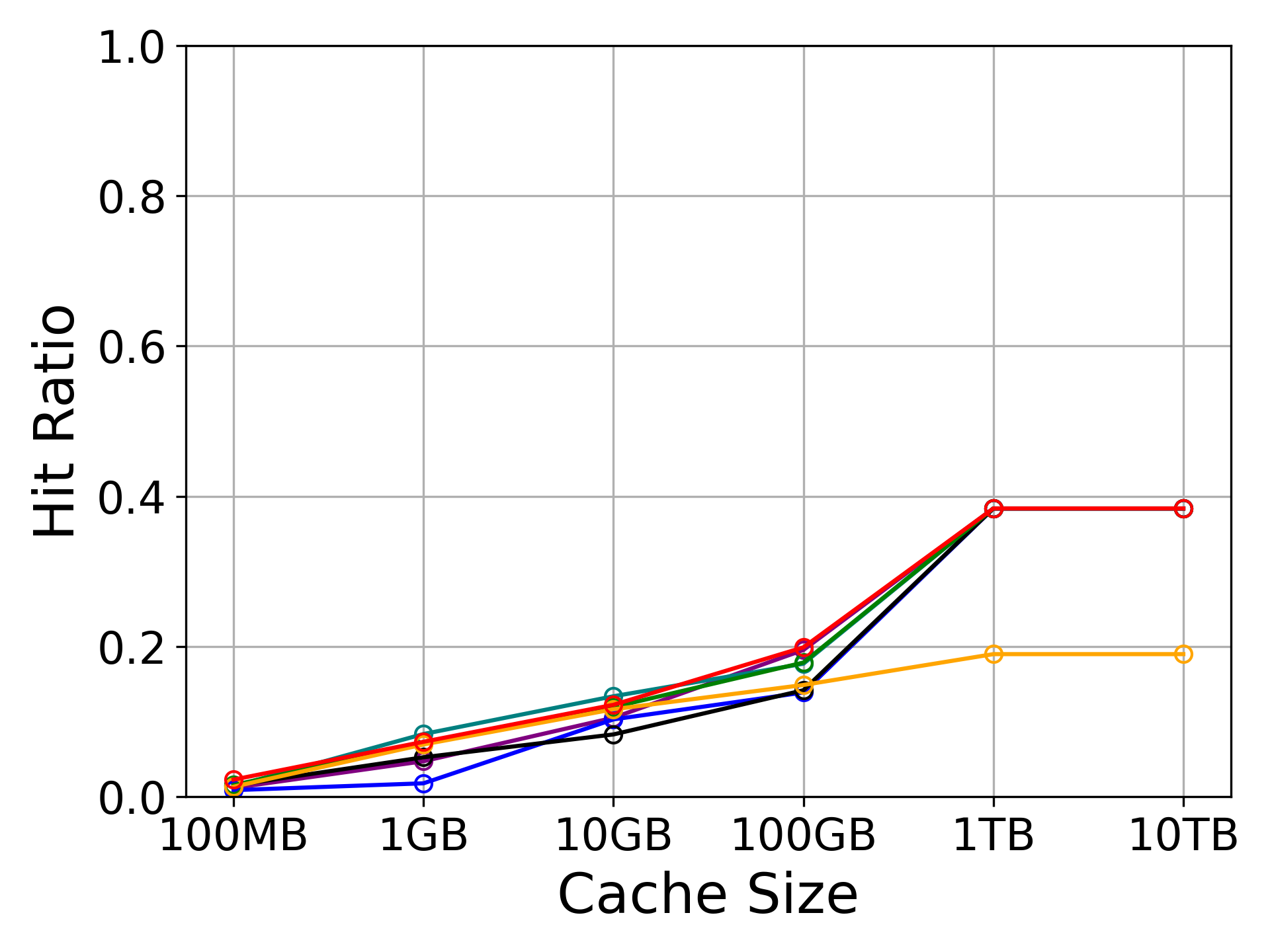}}
		\subfloat[MSR2\label{fig:others:hr:msr2}]{\includegraphics[trim=0 10 0 0, clip, height = 3.3cm]{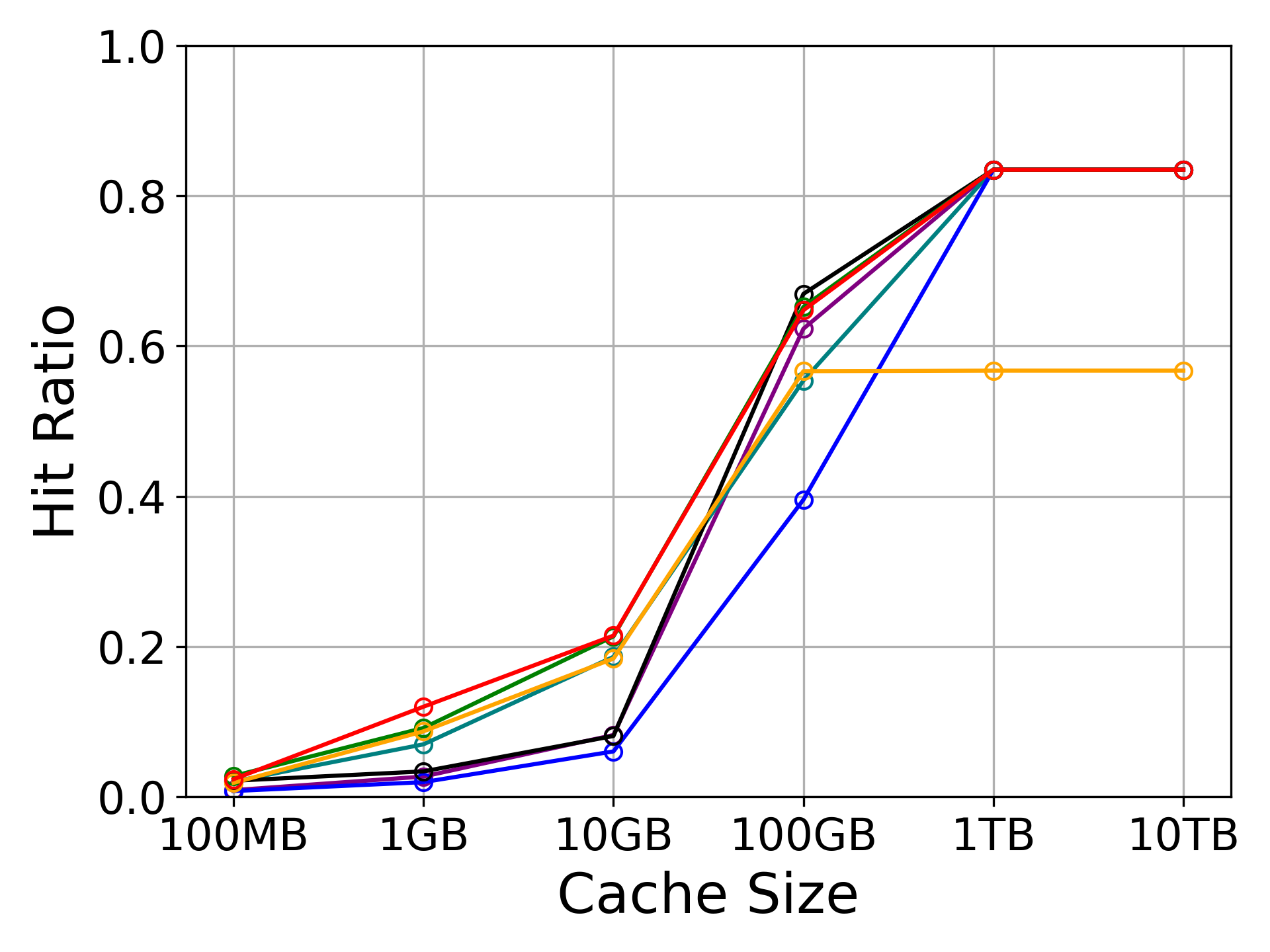}}
		\subfloat[MSR3\label{fig:others:hr:msr3}]{\includegraphics[trim=0 10 0 0, clip, height = 3.3cm]{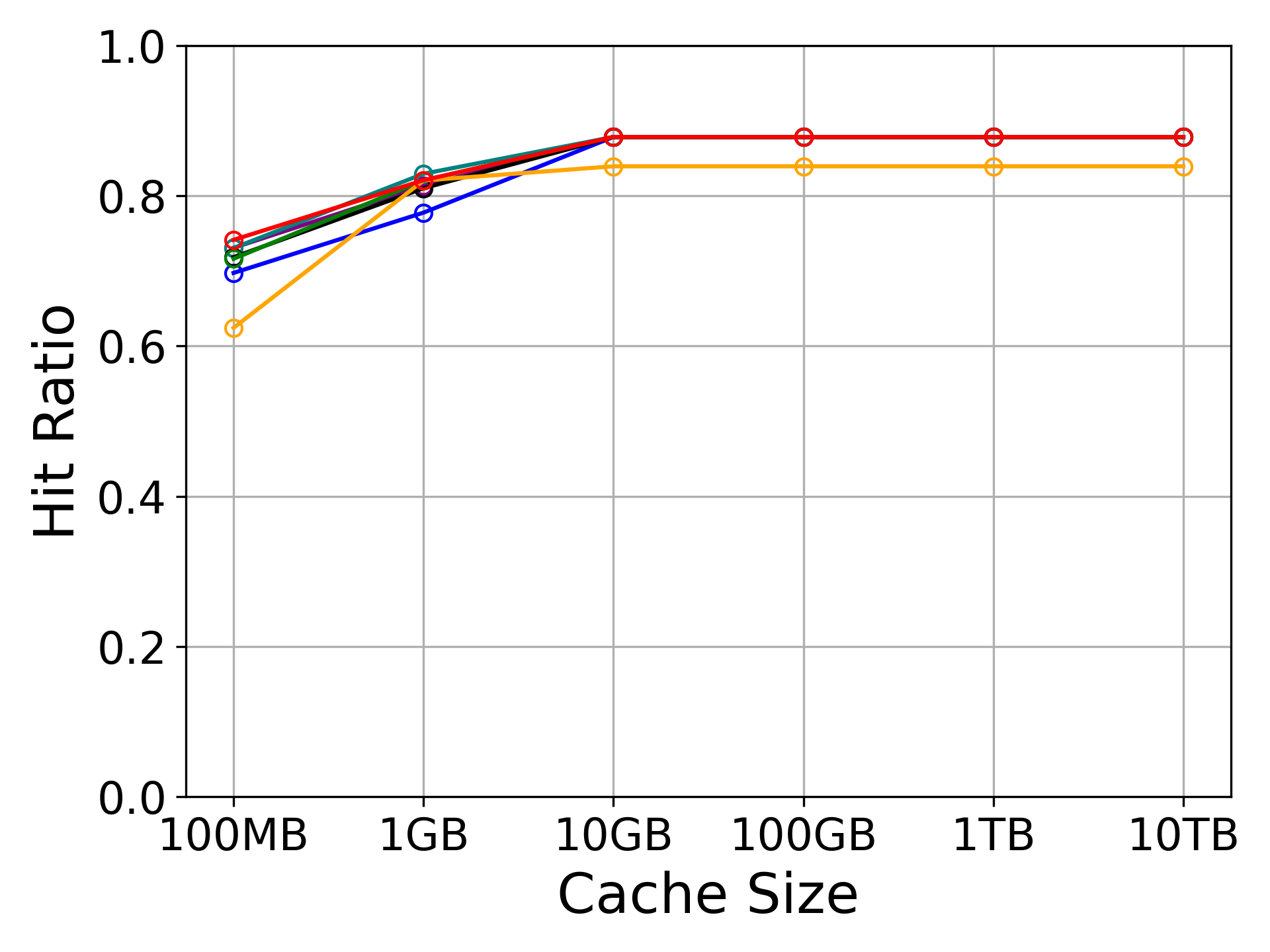}}\\
		\vspace{-0.3cm}
		\subfloat[SYSTOR1\label{fig:others:hr:systor1}]{\includegraphics[trim=0 10 0 0, clip, height = 3.3cm]{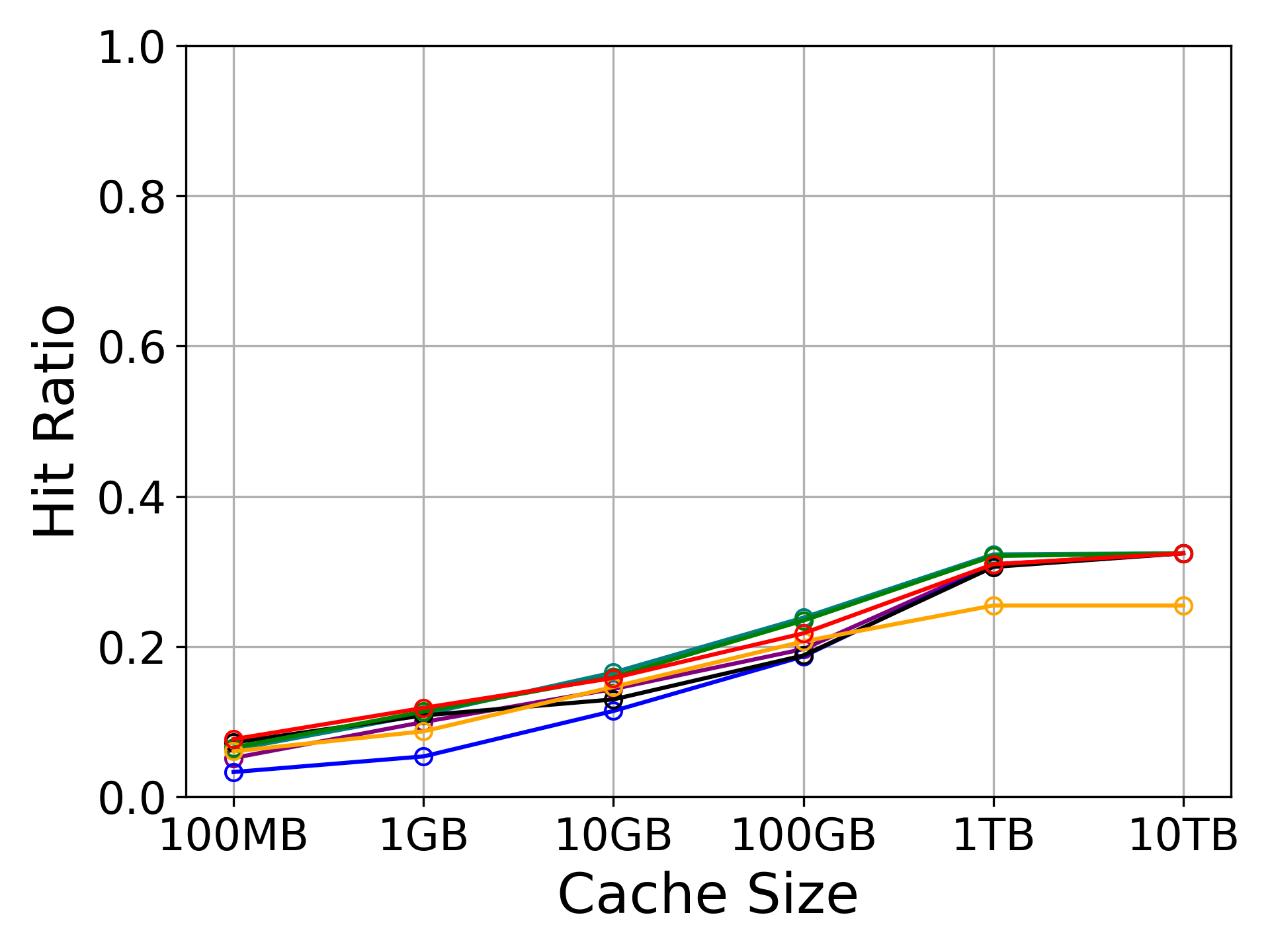}}
		\subfloat[SYSTOR2\label{fig:others:hr:systor2}]{\includegraphics[trim=0 10 0 0, clip, height = 3.3cm]{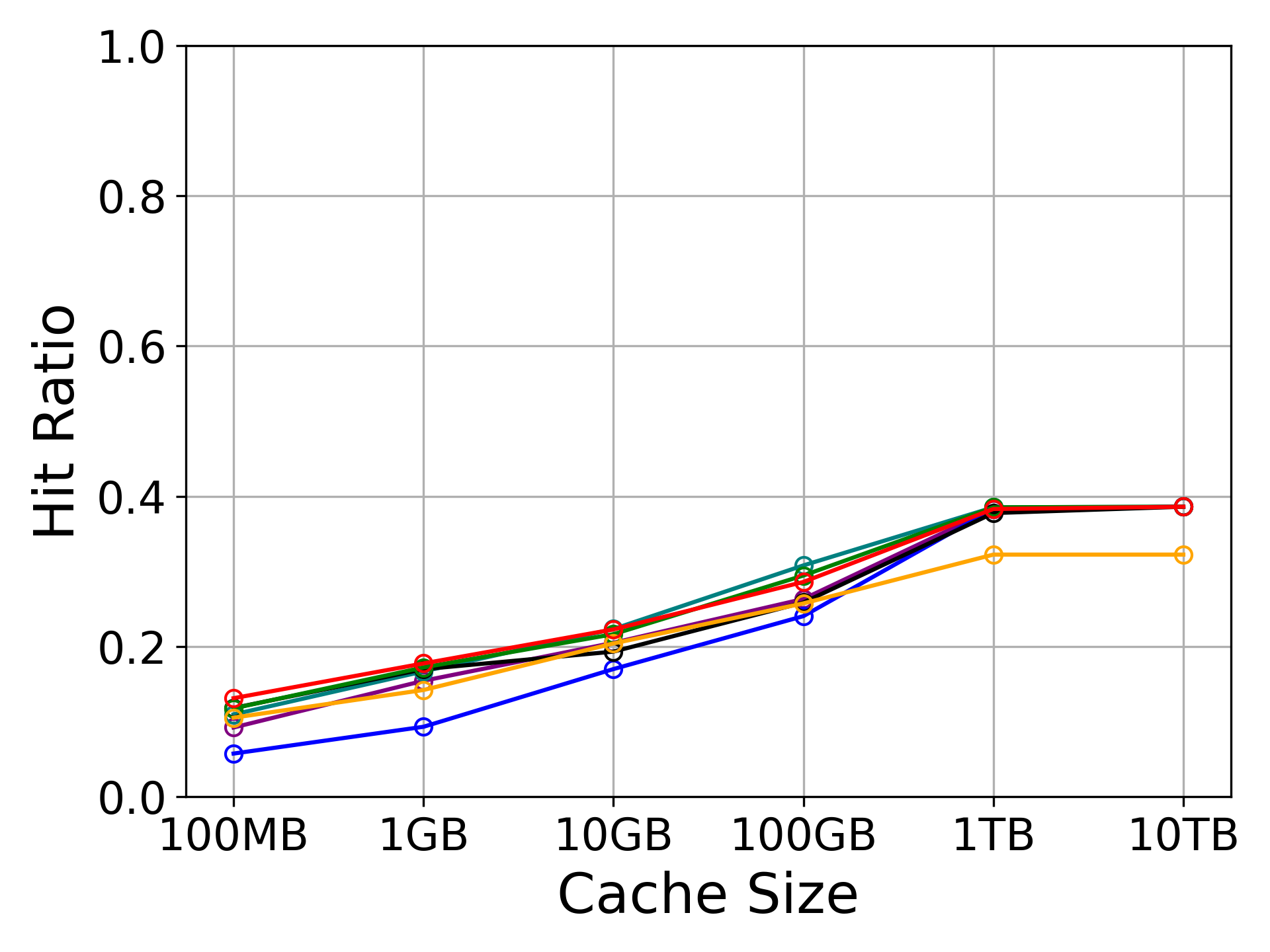}}
		\subfloat[SYSTOR3\label{fig:others:hr:systor3}]{\includegraphics[trim=0 10 0 0, clip, height = 3.3cm]{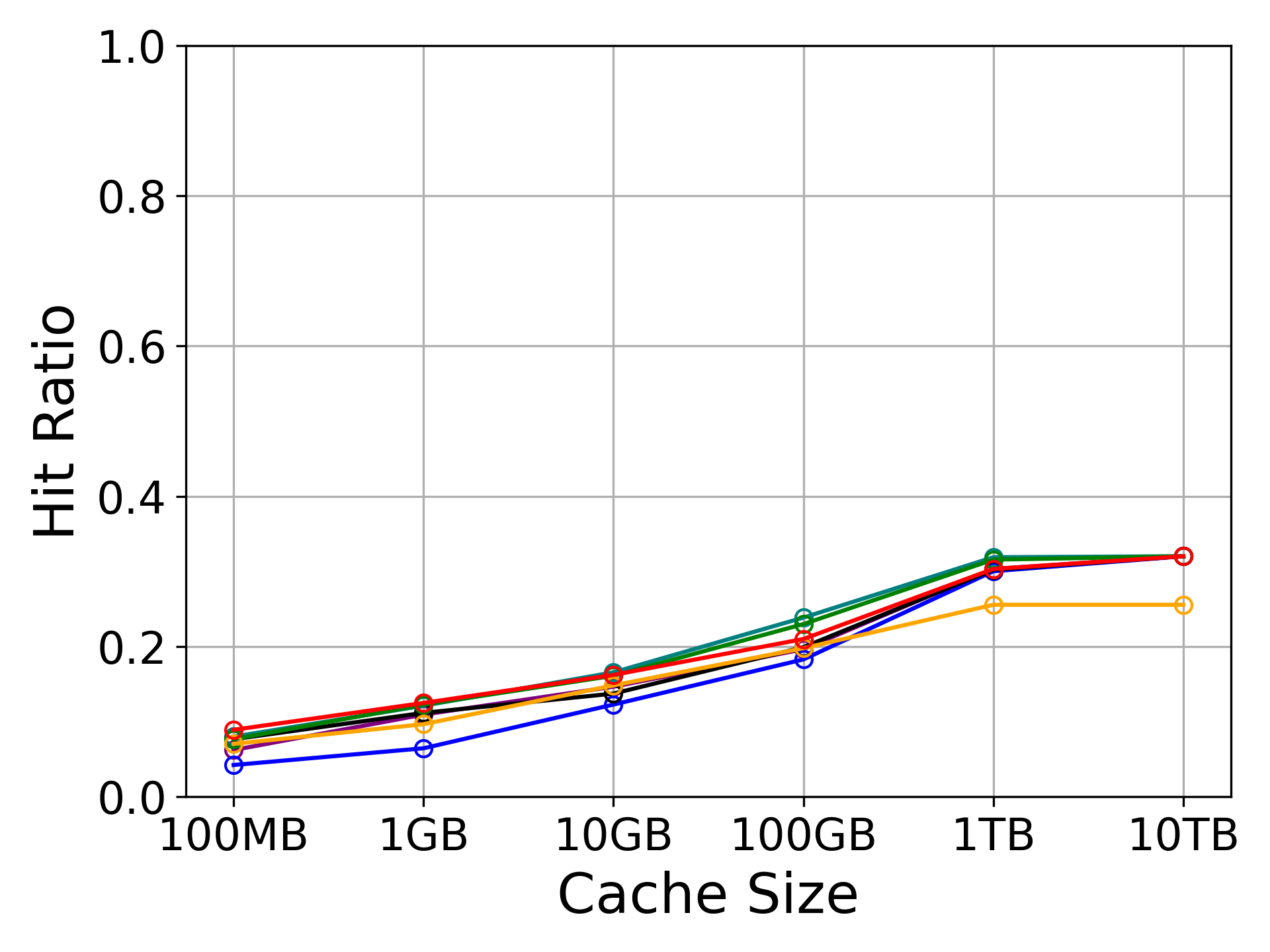}}\\
		\vspace{-0.3cm}
		\subfloat[CDN1\label{fig:others:hr:cdn1}]{\includegraphics[trim=0 10 0 0, clip, height = 3.3cm]{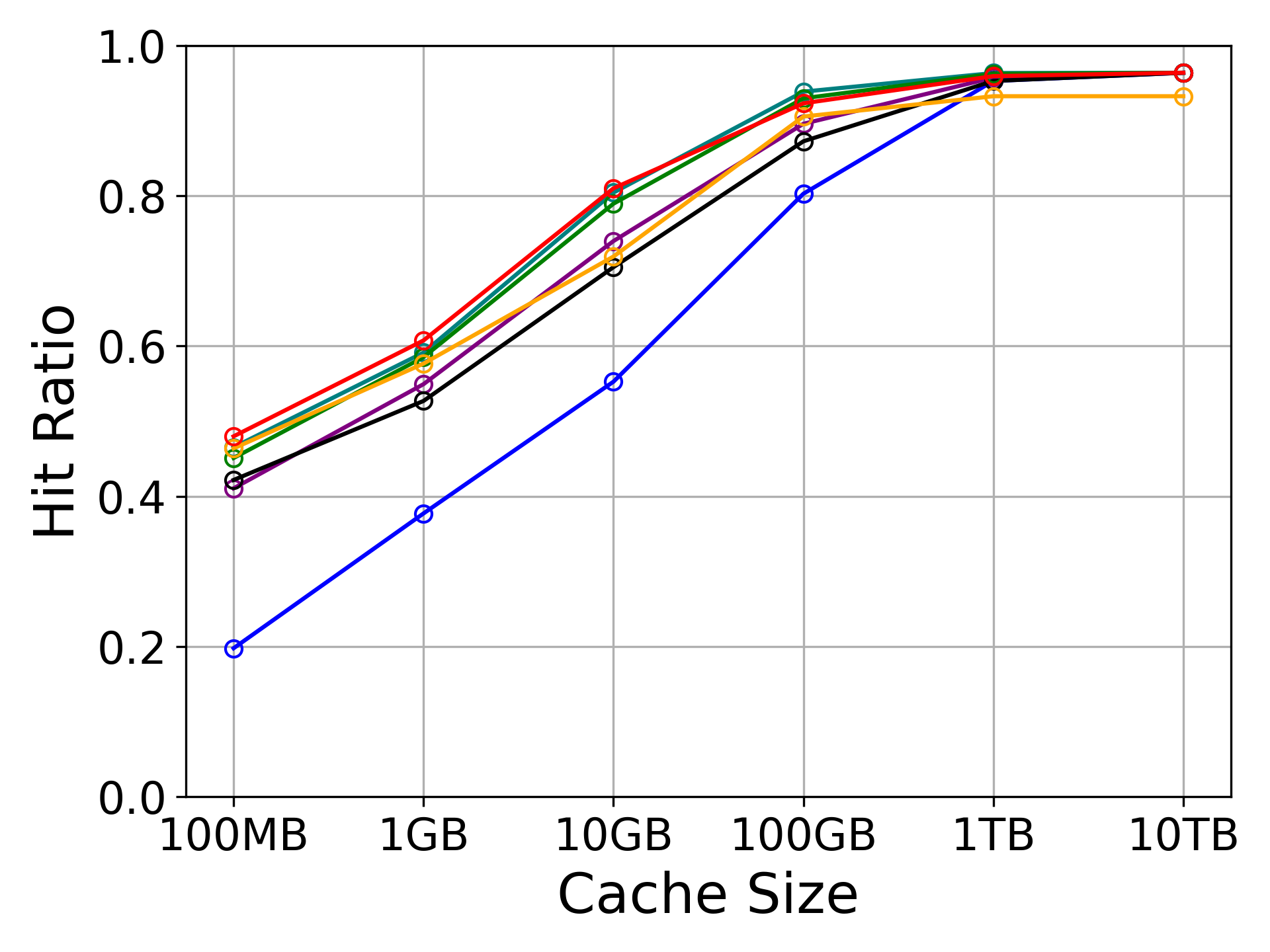}}
		\subfloat[CDN2\label{fig:others:hr:cdn2}]{\includegraphics[trim=0 10 0 0, clip, height = 3.3cm]{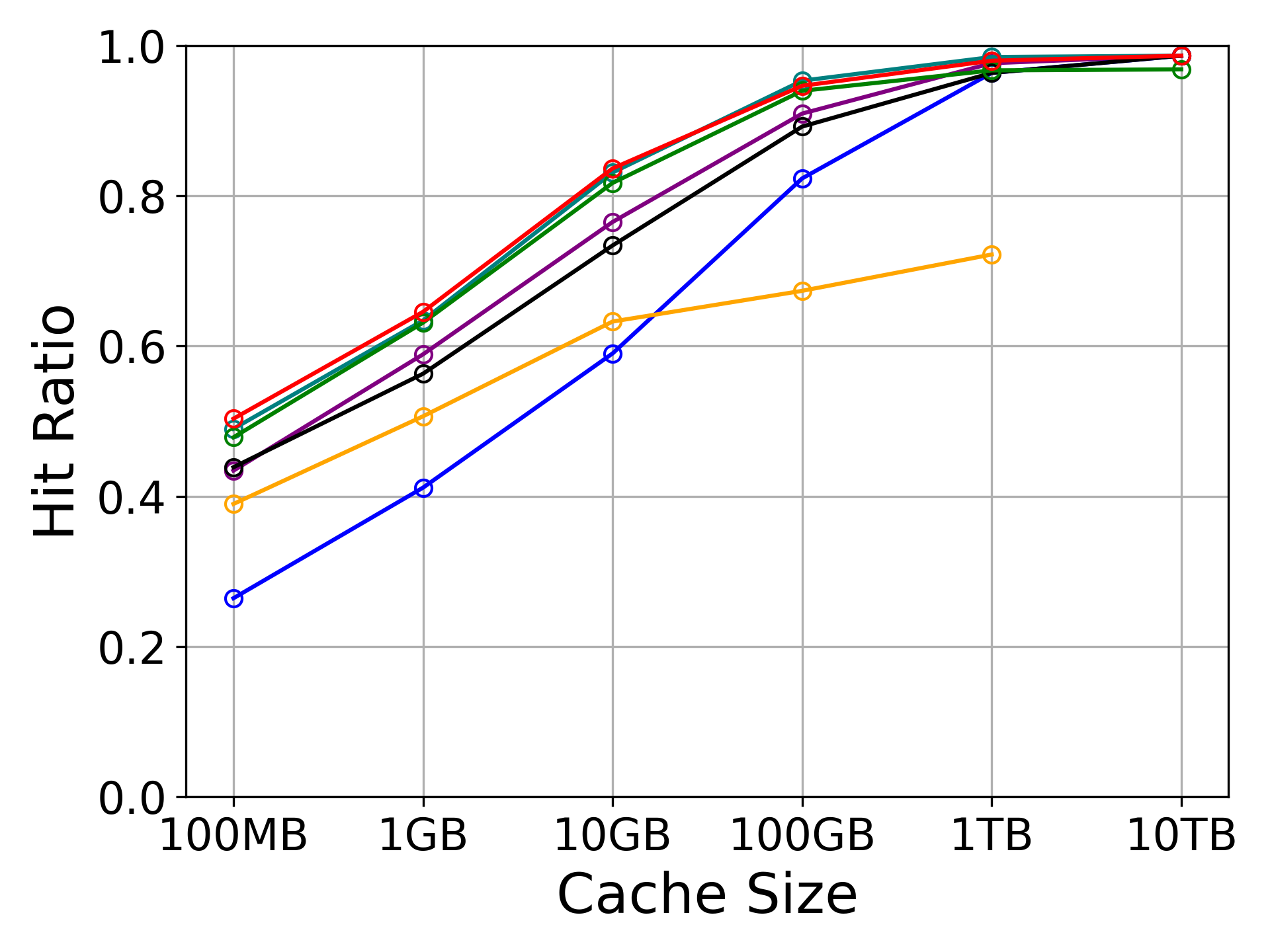}}
		\subfloat[CDN3\label{fig:others:hr:cdn3}]{\includegraphics[trim=0 10 0 0, clip, height = 3.3cm]{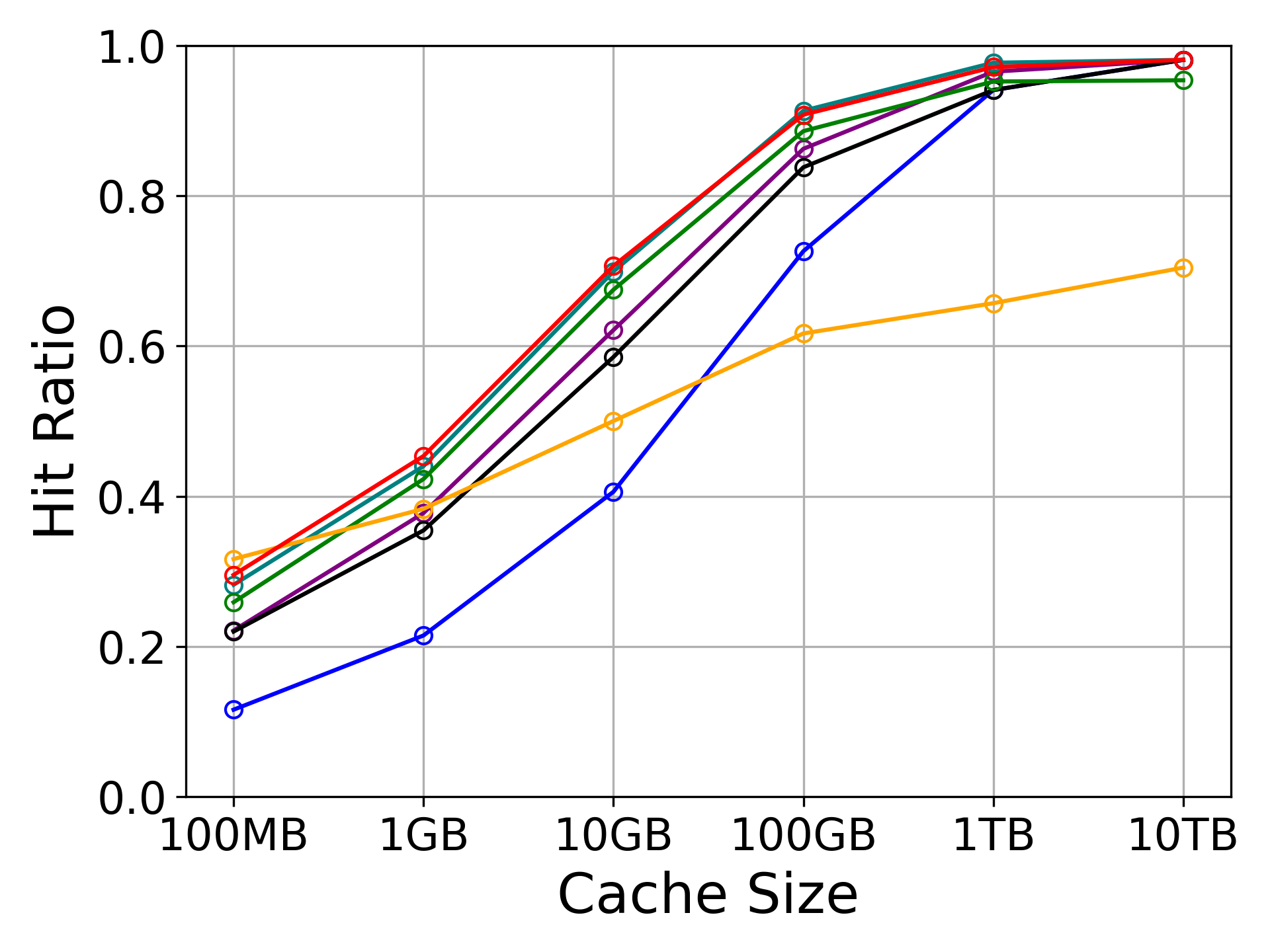}}\\
		\vspace{-0.3cm}
		\subfloat[TENCENT1\label{fig:others:hr:tencent1}]{\includegraphics[trim=0 10 0 0, clip, height = 3.3cm]{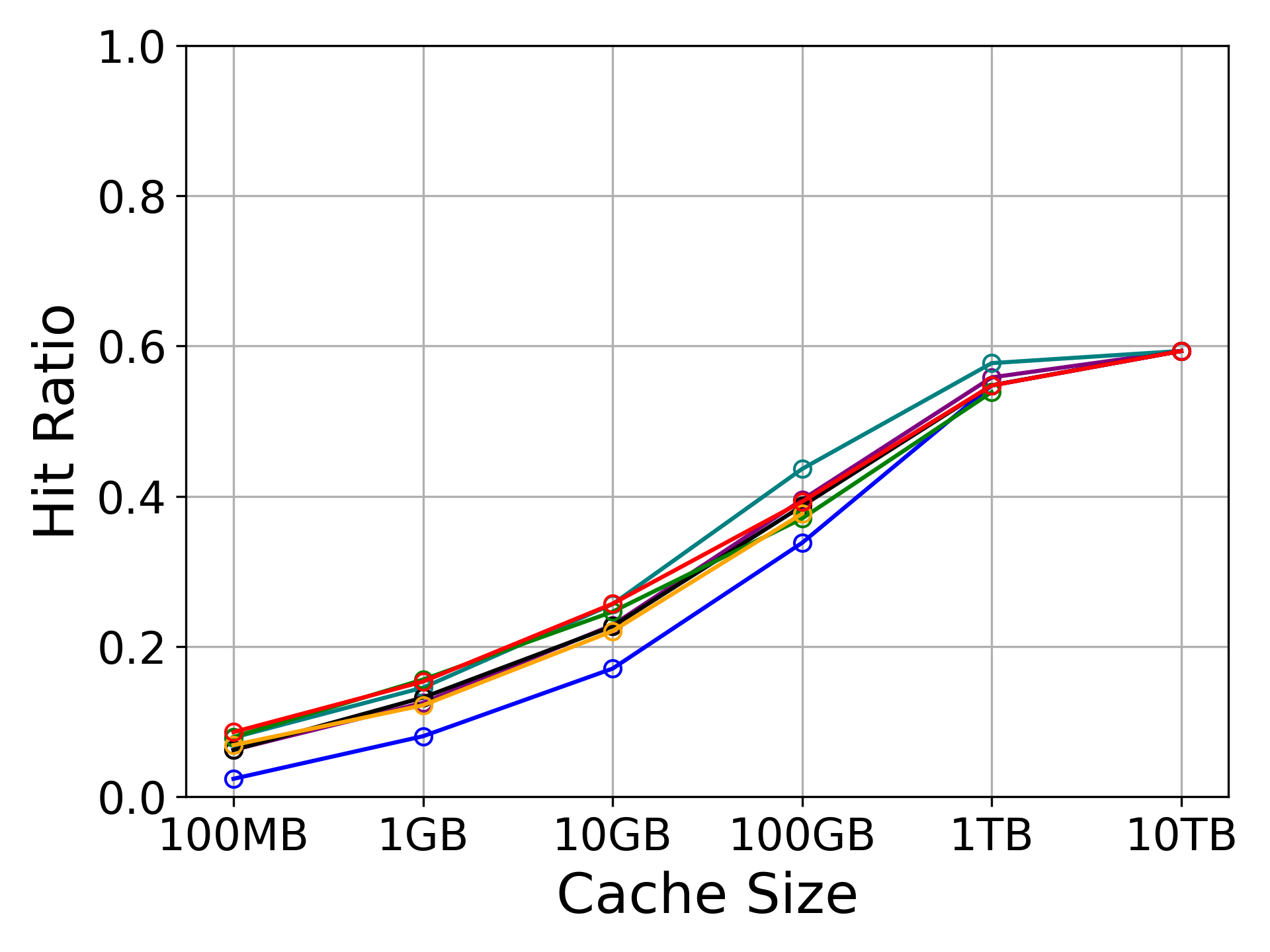}}
		\\	
		\vspace{0.15cm}
		\includegraphics[trim=0 0 0 0, clip, width = 11.5cm]{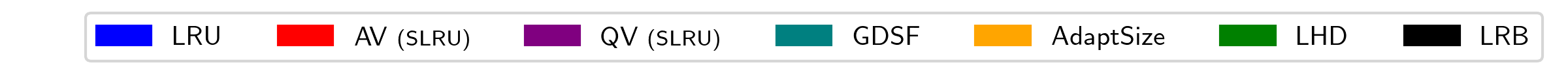}
		\vspace{-0.2cm}
	}
	\caption{Evaluation of hit-ratio for state-of-the-art policies (given the traces' characteristics in Table~\ref{tab:traces}, a 1TB cache is practically an unbounded cache for the MSR traces and an almost unbounded for the SYSTOR traces; a 10TB cache is effectively an unbounded cache for all the traces).}
	\label{fig:evaluation-others:hr}	
\end{figure*}

\ifdefined\SHOWHEATMAP
\begin{figure*}
		\center{
		\subfloat[\PROTNAME{}\label{fig:others:heatmap:hr:av}]{\includegraphics[trim=60 0 30 10, clip, width=0.5\columnwidth]{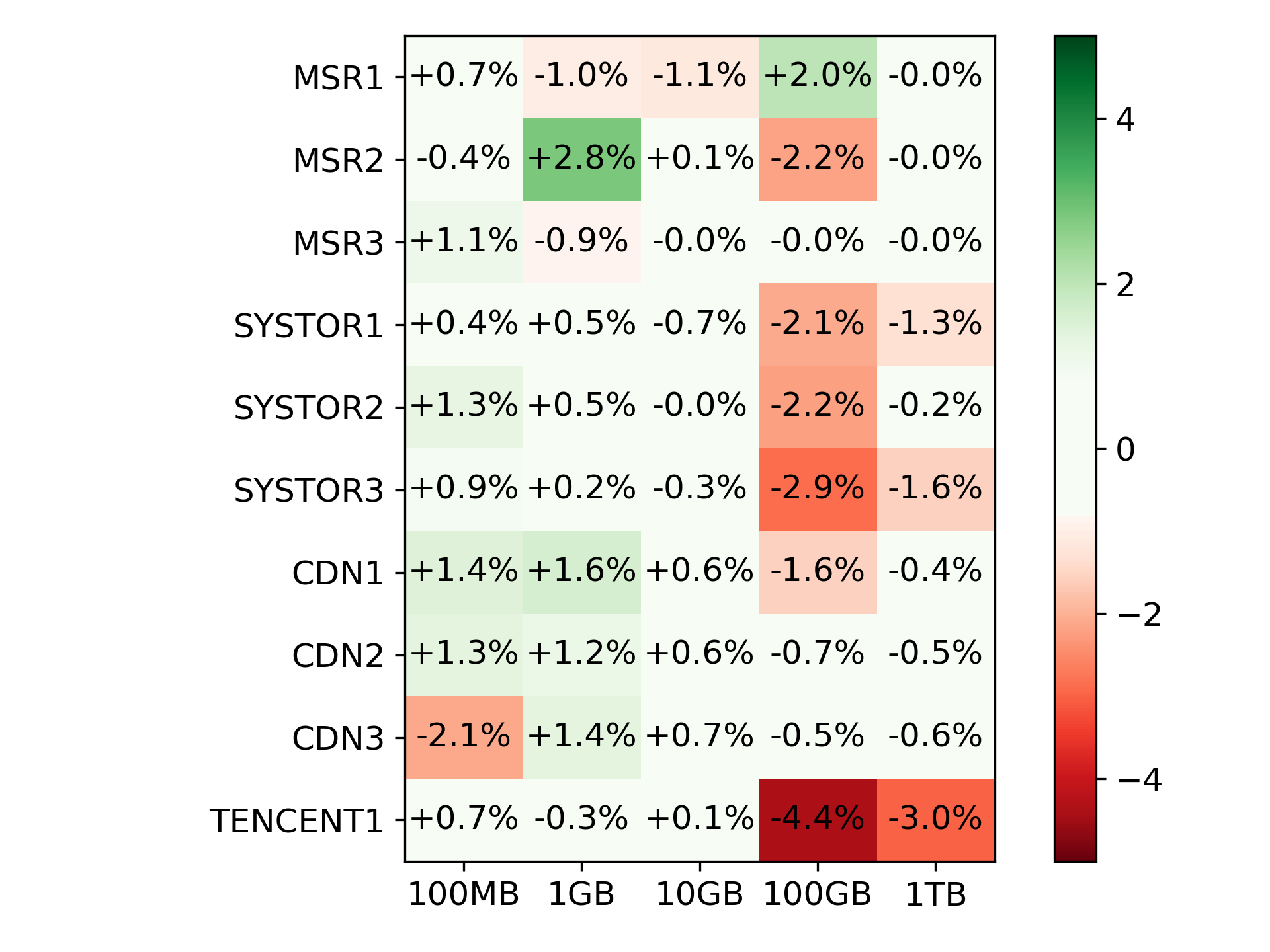}}
		\subfloat[QV\label{fig:others:heatmap:hr:qv}]{\includegraphics[trim=60 0 30 10, clip, width=0.5\columnwidth]{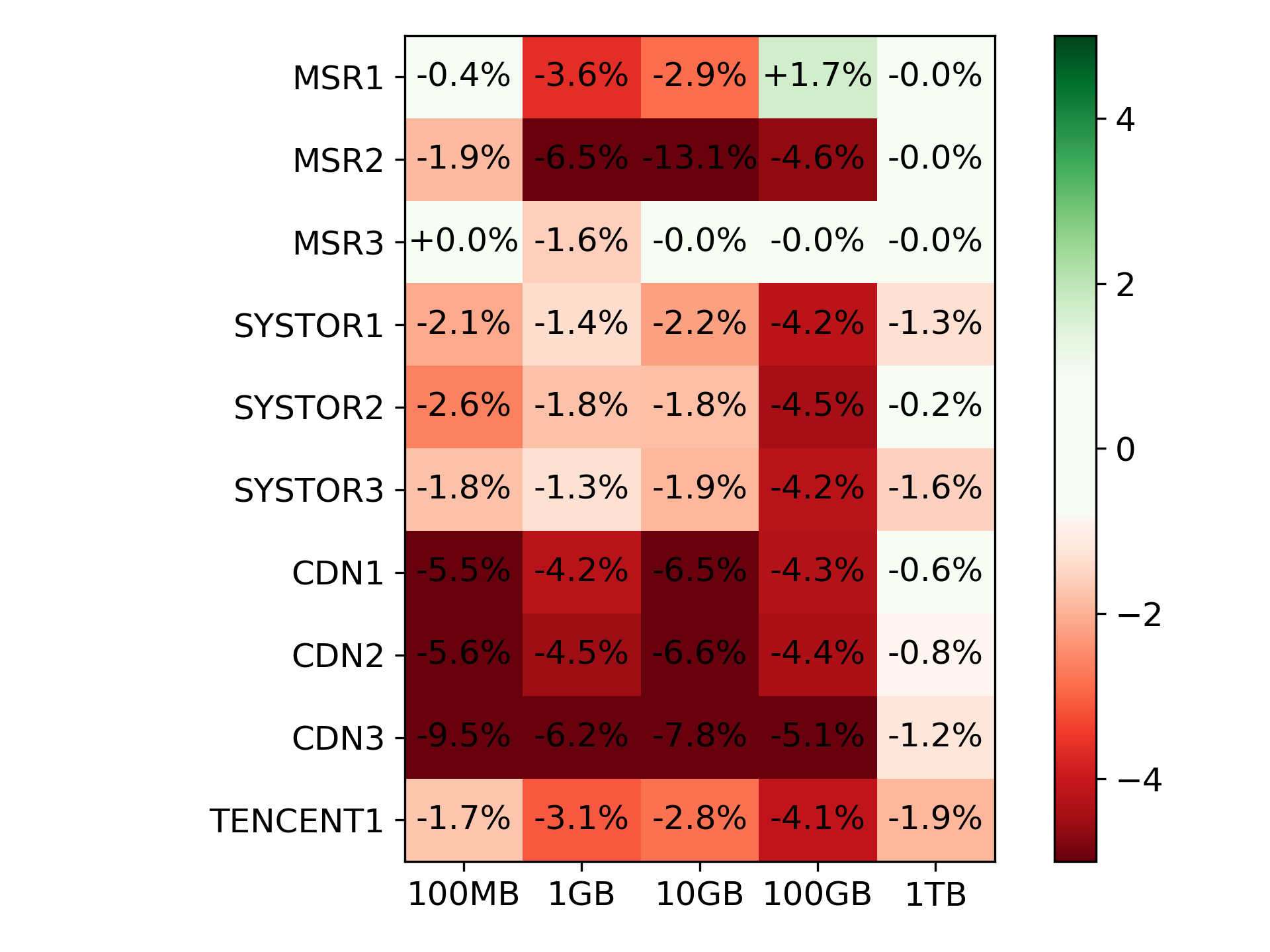}}
	}
	\caption{Hit-ratio of suggested policies vs. the best alternative.}
	\label{fig:heatmaps:hr}
\end{figure*}
\fi

\begin{figure*}[!ht]
	\center{
		\vspace{-0.3cm}
		\subfloat[MSR1\label{fig:others:bhr:msr1}]{\includegraphics[trim=0 10 0 0, clip, height = 3.3cm]{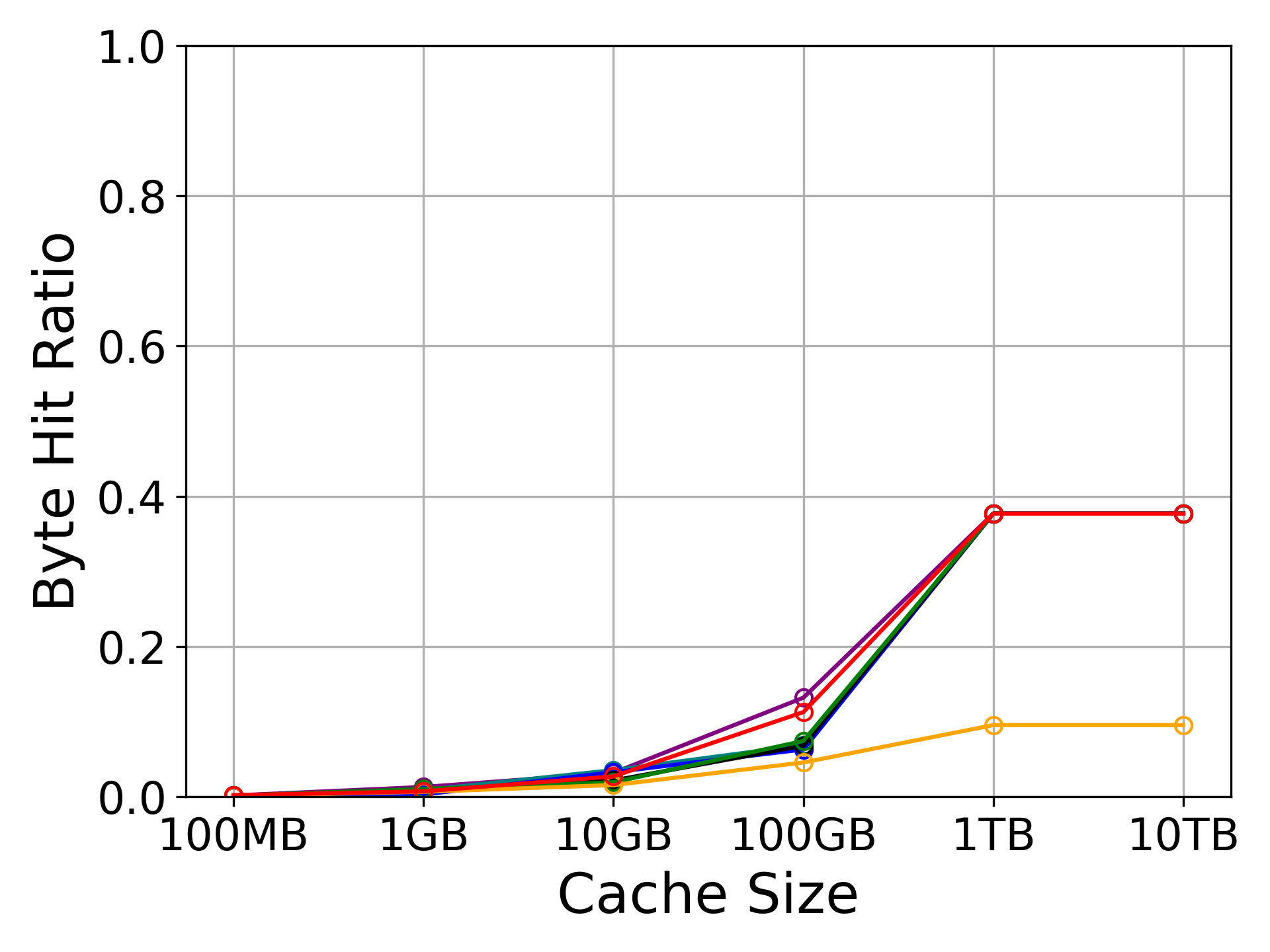}}
		\subfloat[MSR2\label{fig:others:bhr:msr2}]{\includegraphics[trim=0 10 0 0, clip, height = 3.3cm]{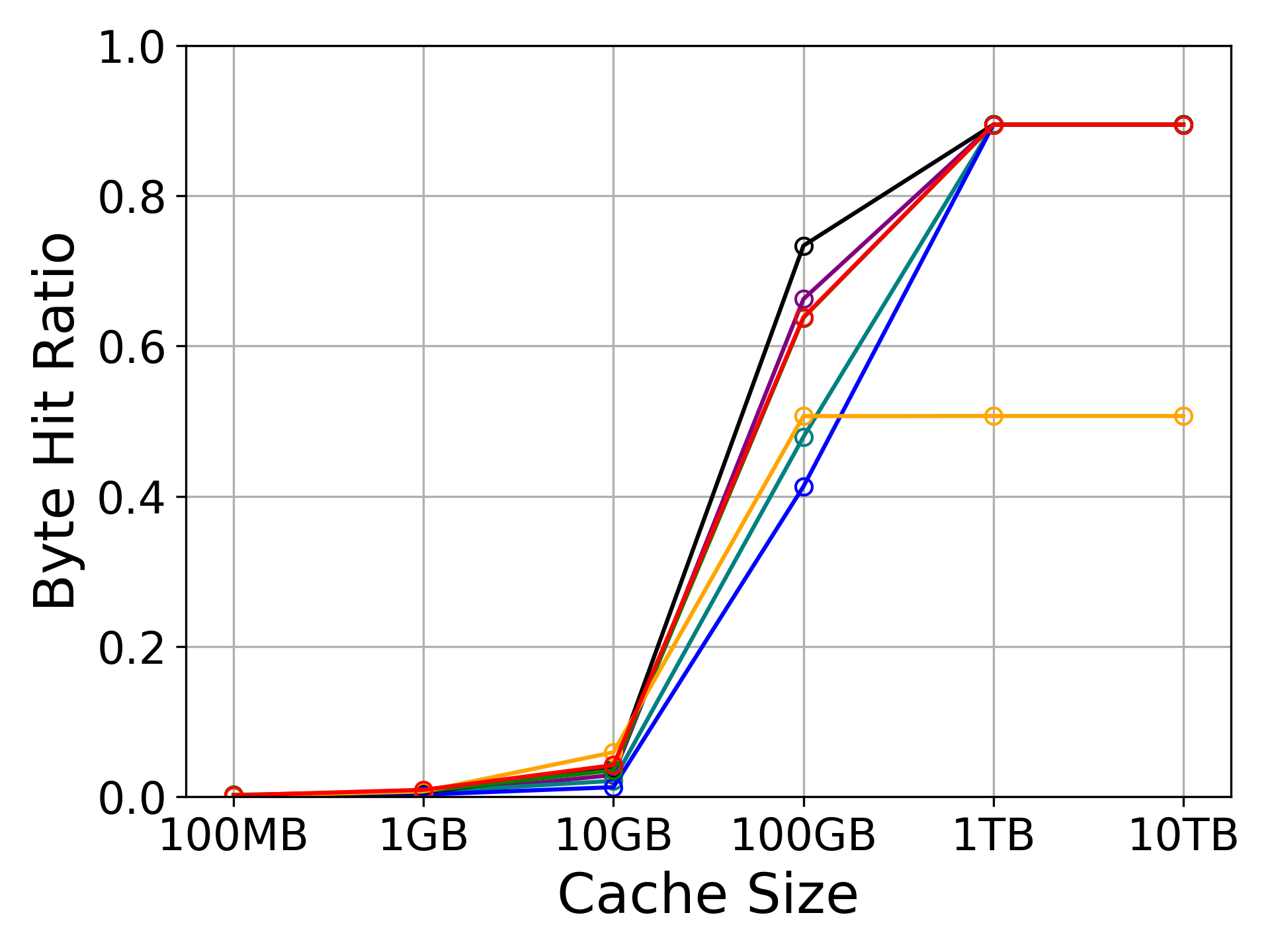}}
		\subfloat[MSR3\label{fig:others:bhr:msr3}]{\includegraphics[trim=0 10 0 0, clip, height = 3.3cm]{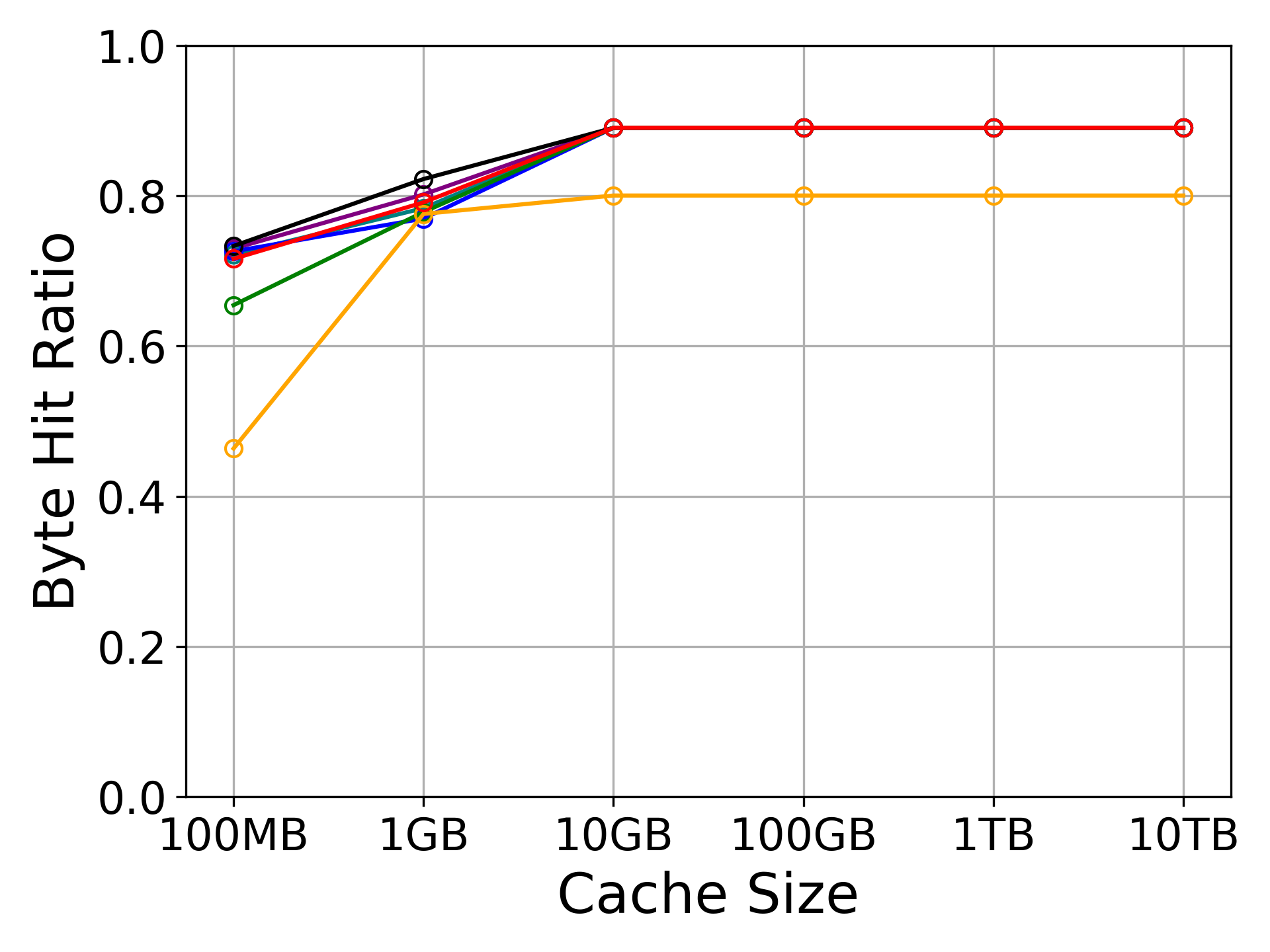}}\\
		\vspace{-0.3cm}
		\subfloat[SYSTOR1\label{fig:others:bhr:systor1}]{\includegraphics[trim=0 10 0 0, clip, height = 3.3cm]{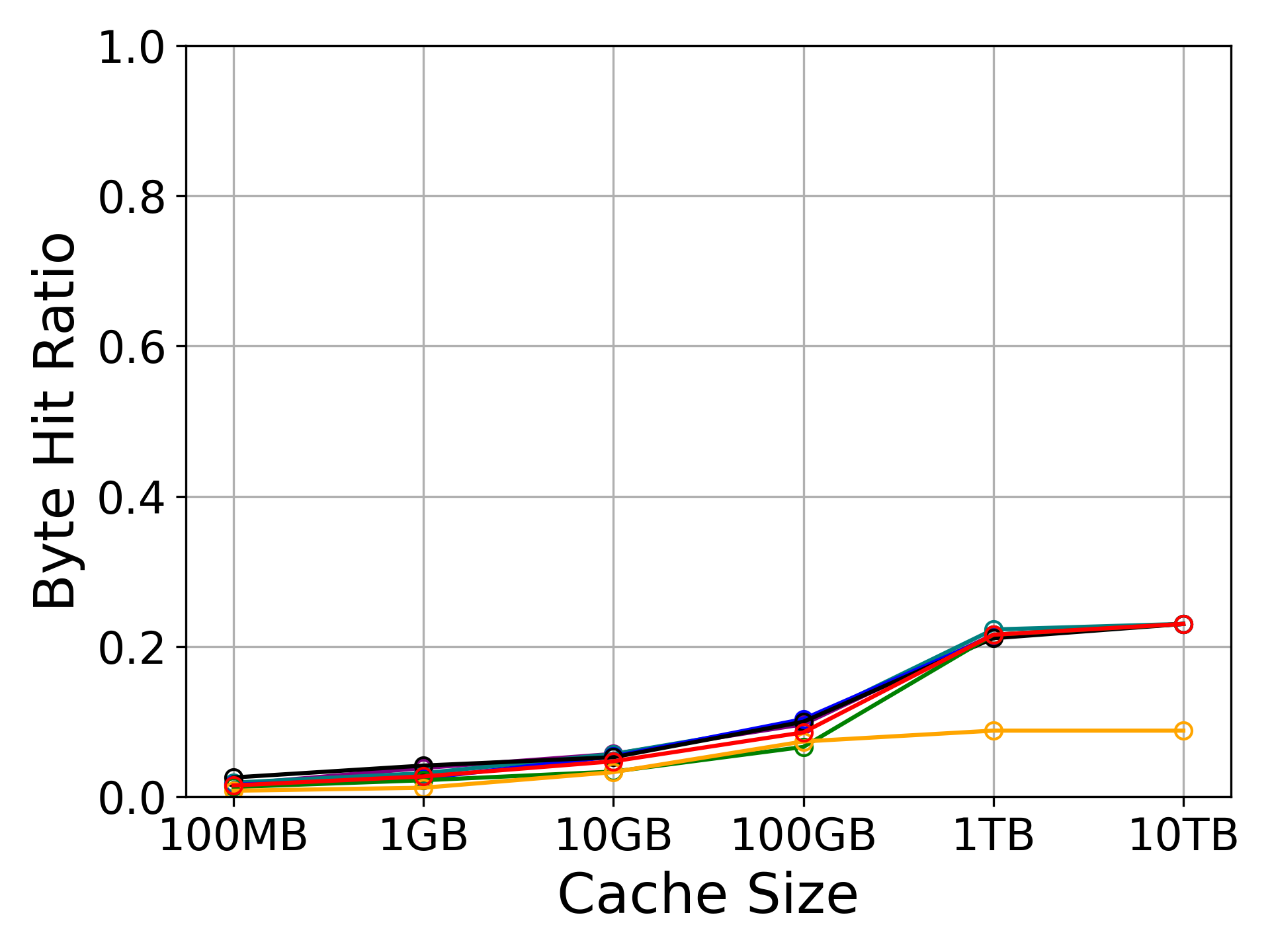}}
		\subfloat[SYSTOR2\label{fig:others:bhr:systor2}]{\includegraphics[trim=0 10 0 0, clip, height = 3.3cm]{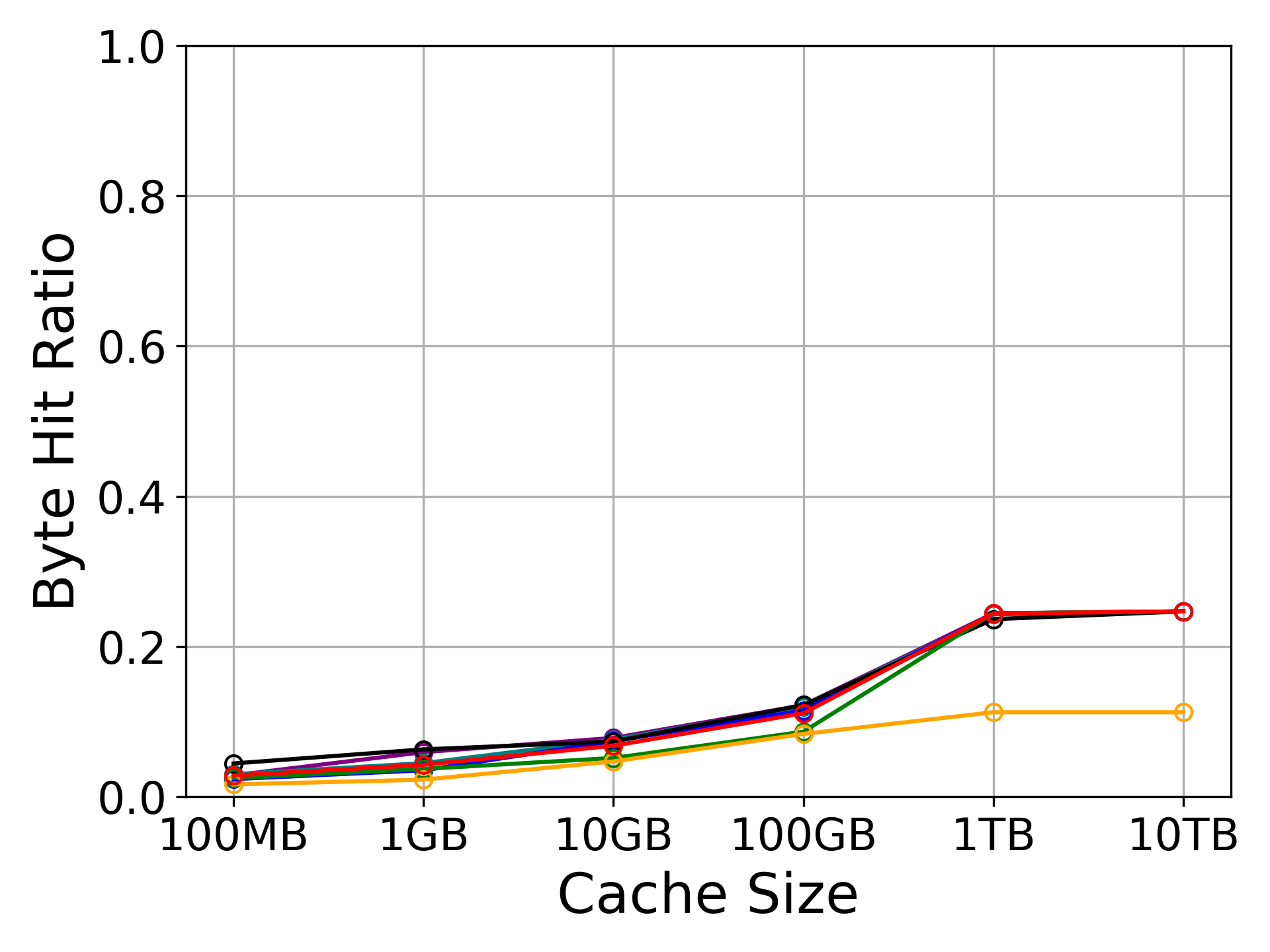}}
		\subfloat[SYSTOR3\label{fig:others:bhr:systor3}]{\includegraphics[trim=0 10 0 0, clip, height = 3.3cm]{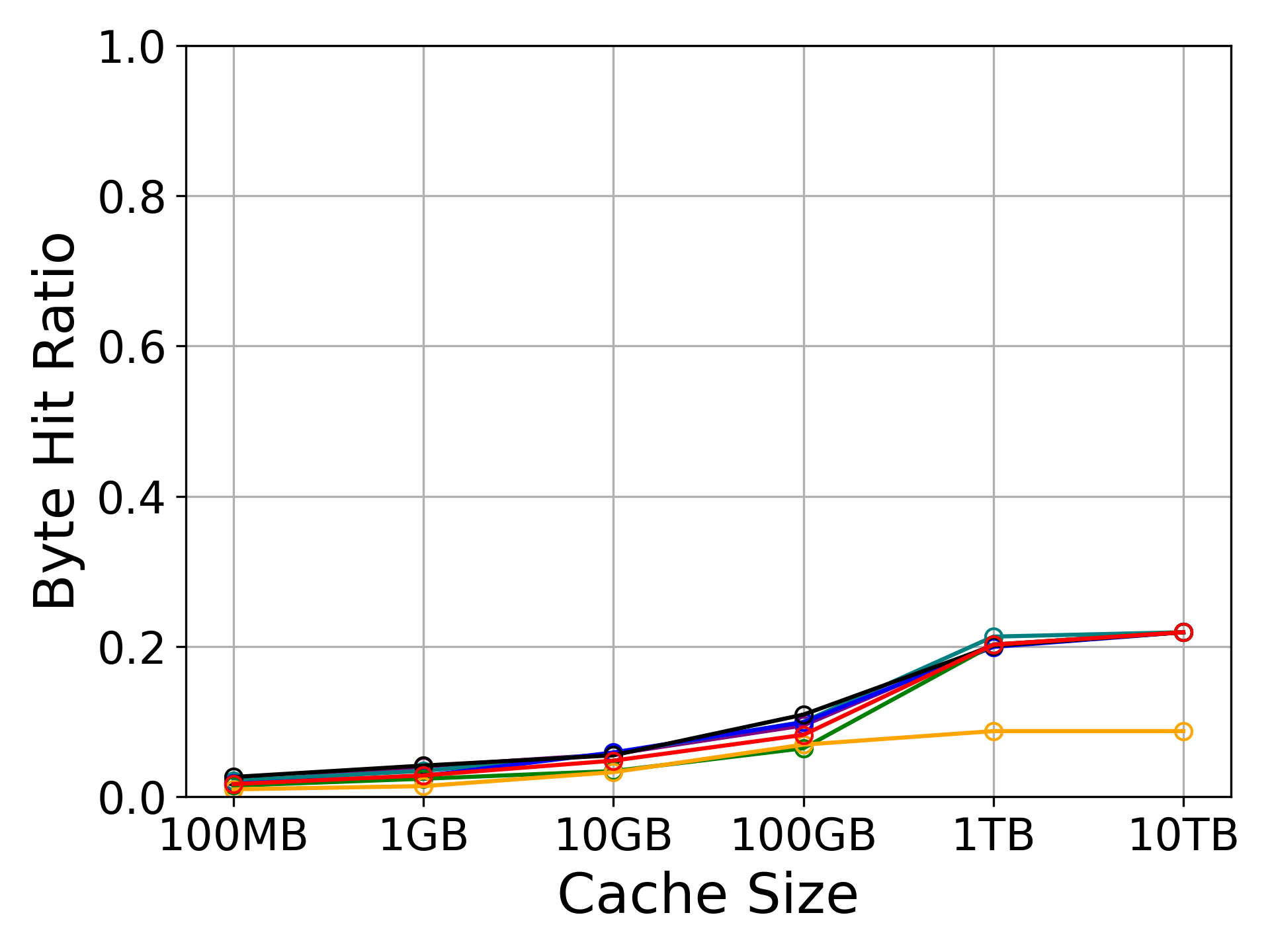}}\\
		\vspace{-0.3cm}
		\subfloat[CDN1\label{fig:others:bhr:cdn1}]{\includegraphics[trim=0 10 0 0, clip, height = 3.3cm]{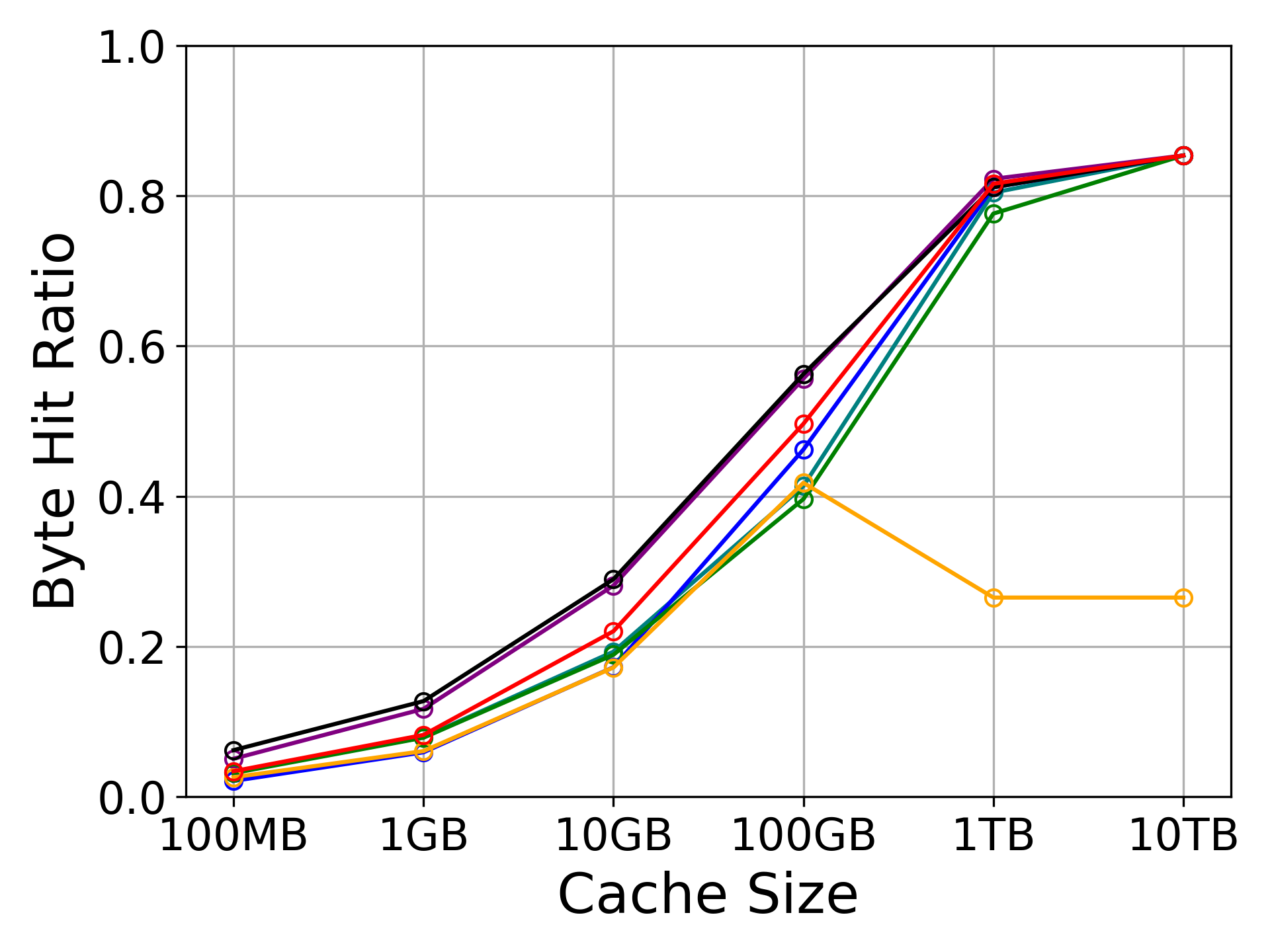}}
		\subfloat[CDN2\label{fig:others:bhr:cdn2}]{\includegraphics[trim=0 10 0 0, clip, height = 3.3cm]{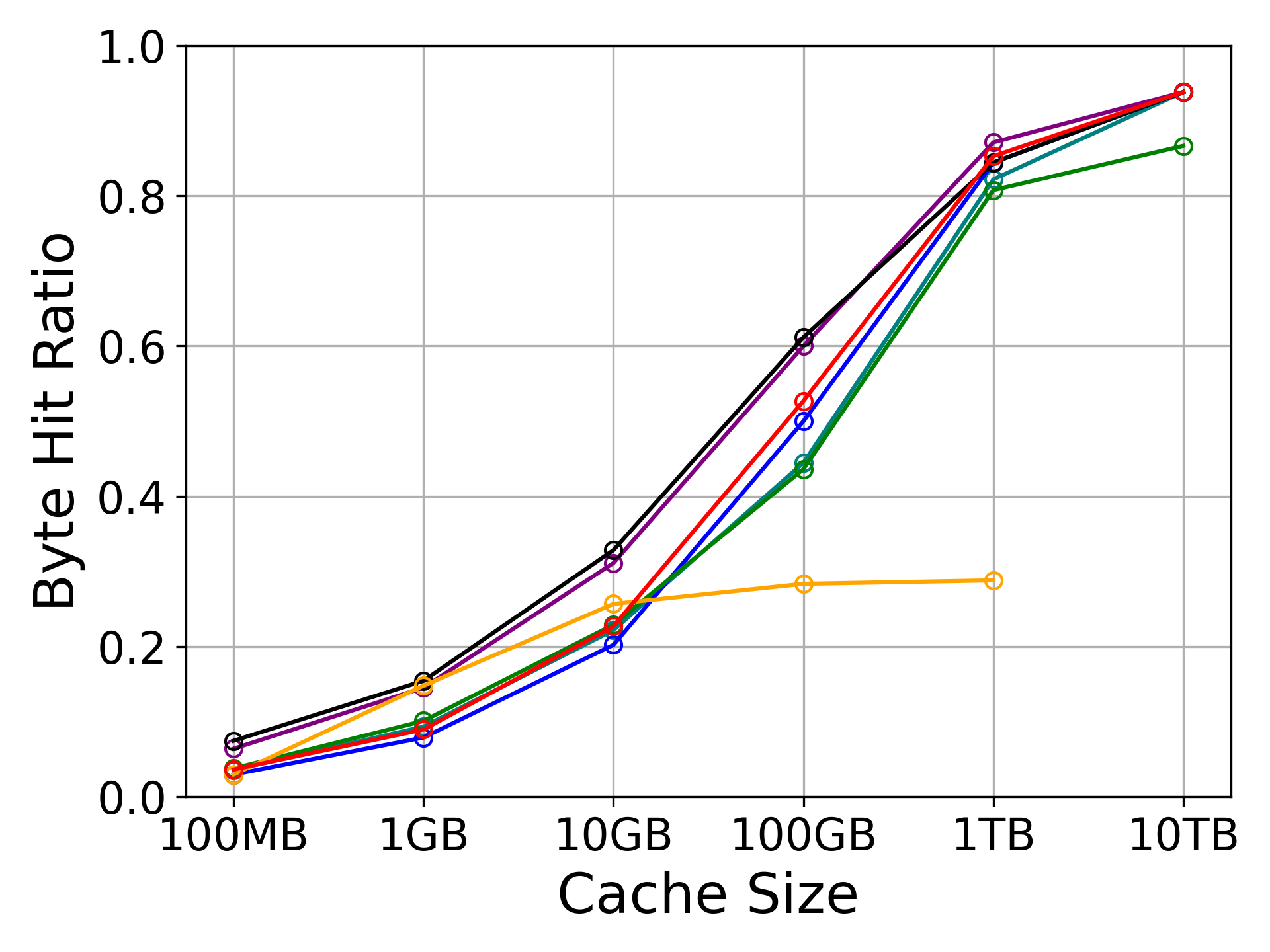}}
		\subfloat[CDN3\label{fig:others:bhr:cdn3}]{\includegraphics[trim=0 10 0 0, clip, height = 3.3cm]{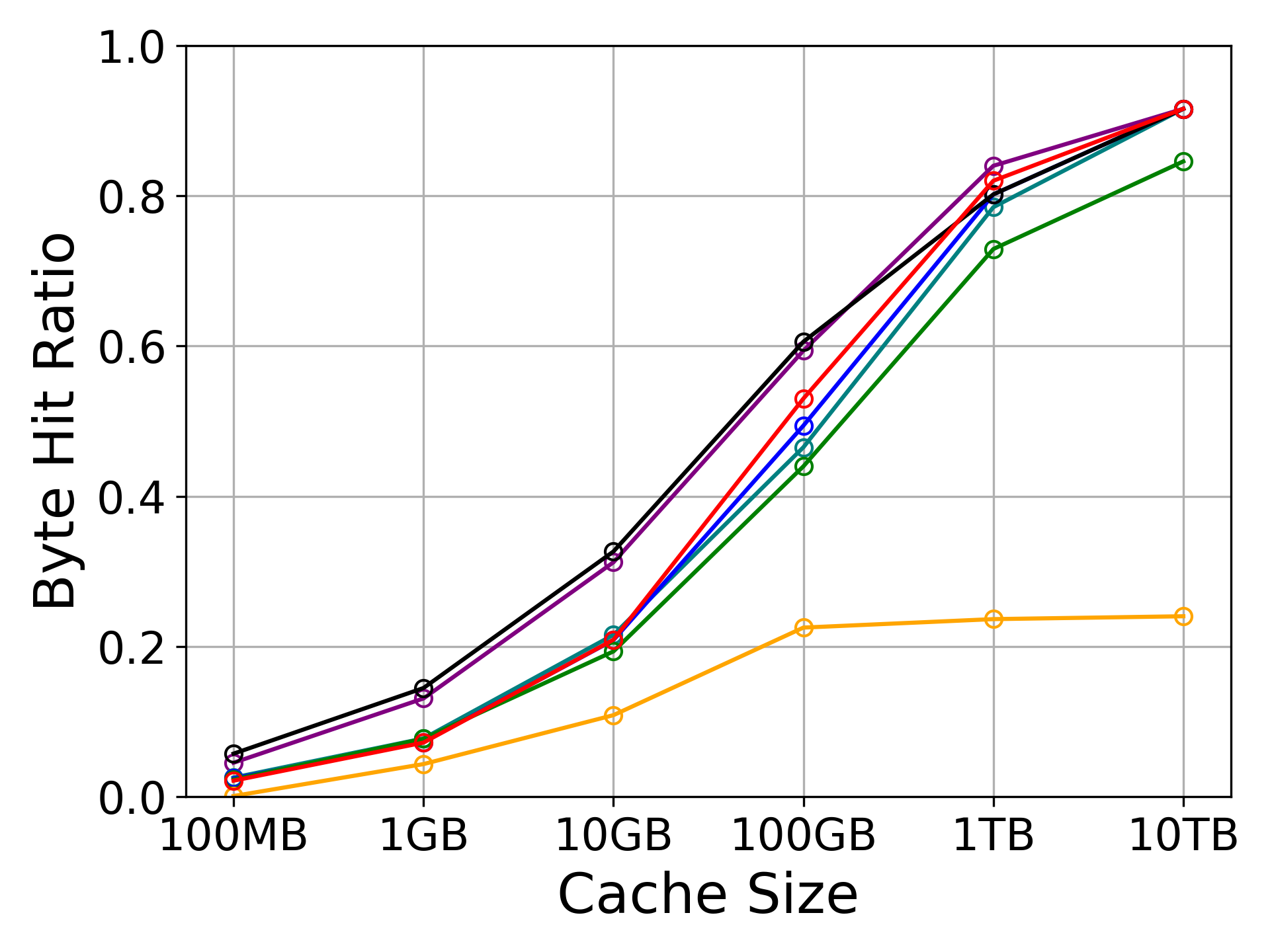}}\\
		\vspace{-0.3cm}
		\subfloat[TENCENT1\label{fig:others:bhr:tencent1}]{\includegraphics[trim=0 10 0 0, clip, height = 3.3cm]{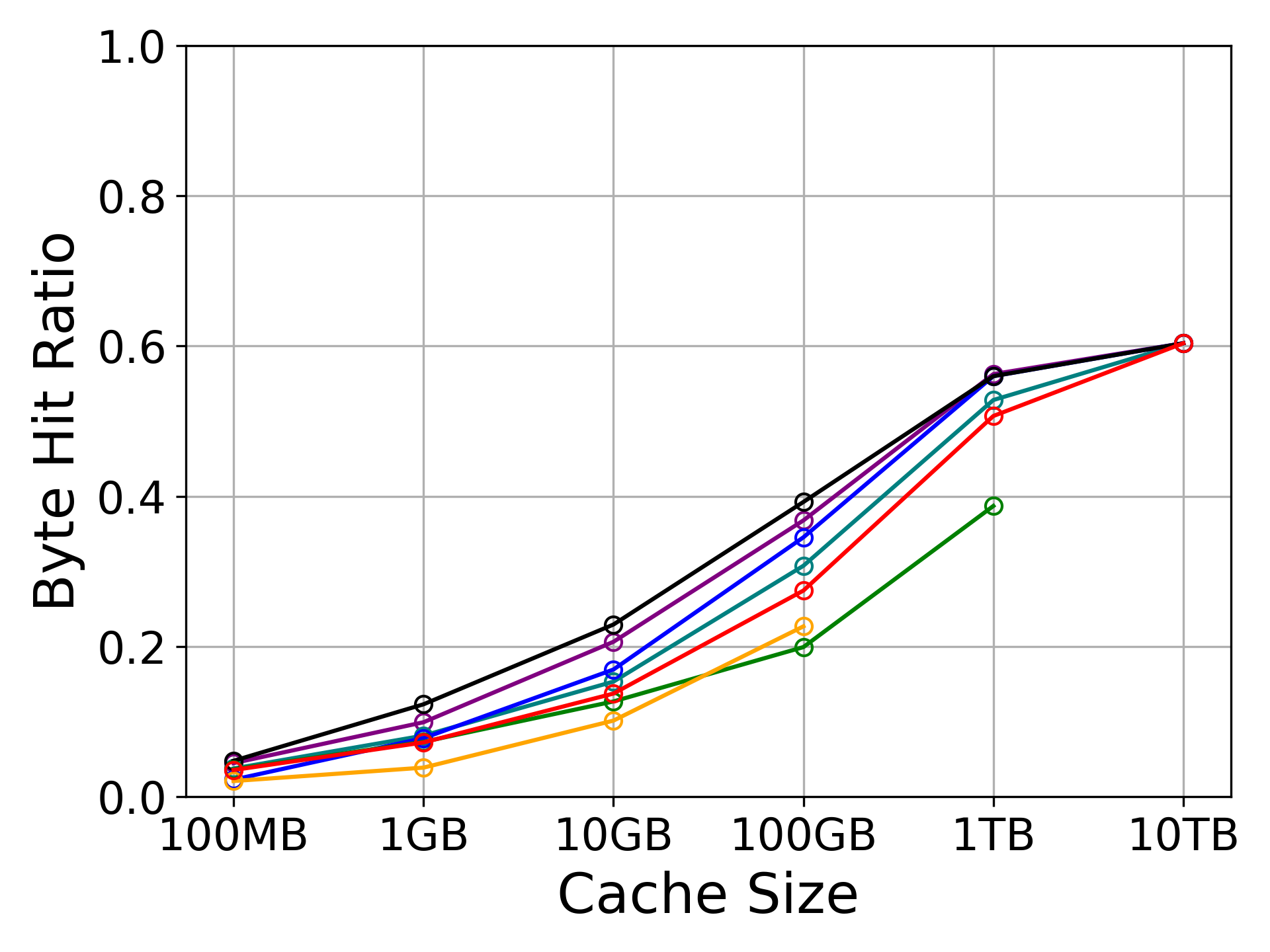}}
		\\
		\vspace{0.15cm}
		\includegraphics[trim=0 0 0 0, clip, width = 11.5cm]{Plots/others/legend.png}
		\vspace{-0.2cm}
	}
	\caption{Evaluation of byte hit-ratio for state-of-the-art policies (given the traces' characteristics in Table~\ref{tab:traces}, a 1TB cache is practically an unbounded cache for the MSR traces and an almost unbounded for the SYSTOR traces; a 10TB cache is effectively an unbounded cache for all the traces).}
	\label{fig:evaluation-others:bhr}
\end{figure*}

\ifdefined\SHOWHEATMAP
\begin{figure*}
	\center{
		\subfloat[\PROTNAME{}\label{fig:others:heatmap:bhr:av}]{\includegraphics[trim=60 0 30 10, clip, width=0.5\columnwidth]{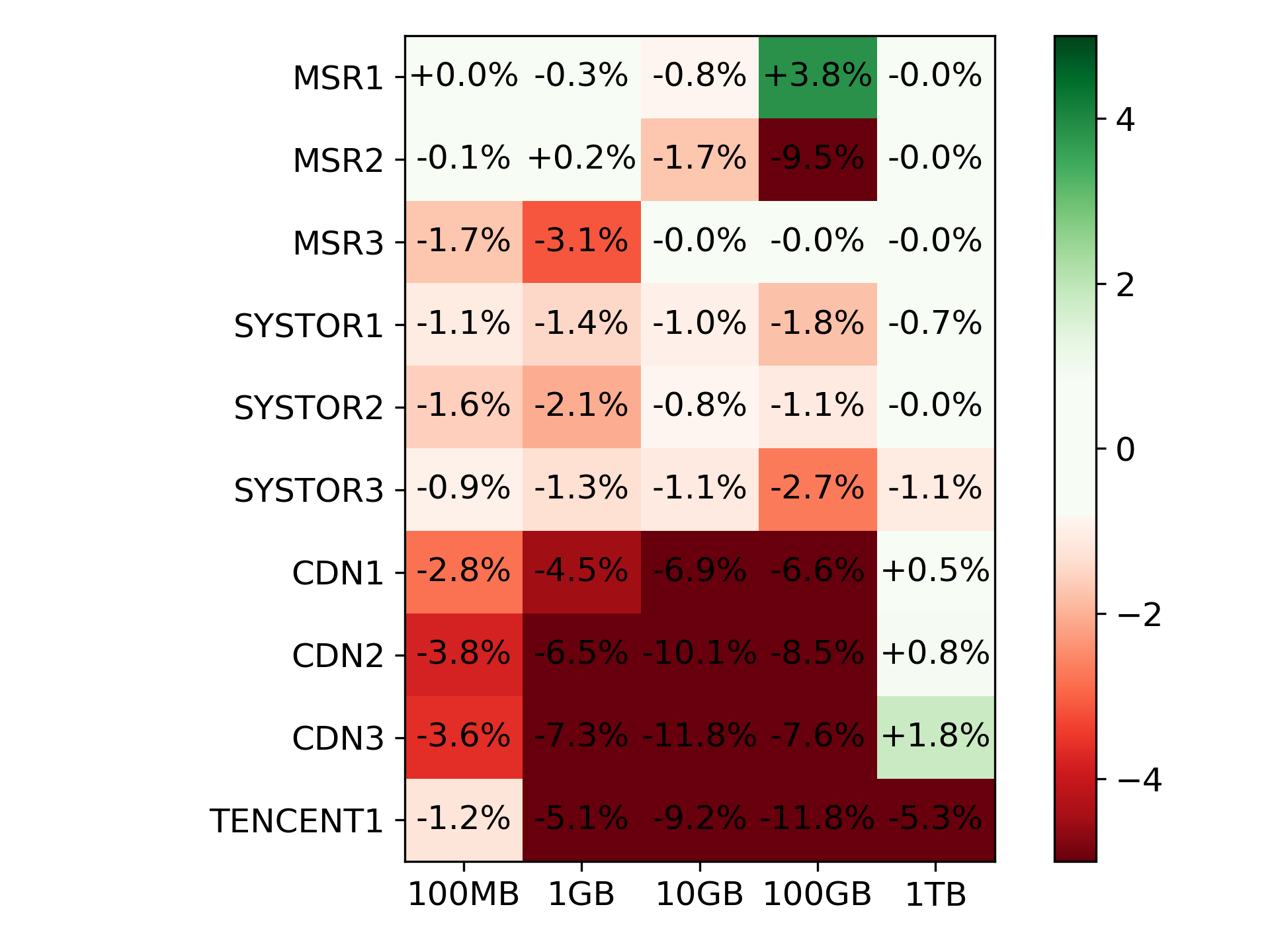}}
		\subfloat[QV\label{fig:others:heatmap:bhr:qv}]{\includegraphics[trim=60 0 30 10, clip, width=0.5\columnwidth]{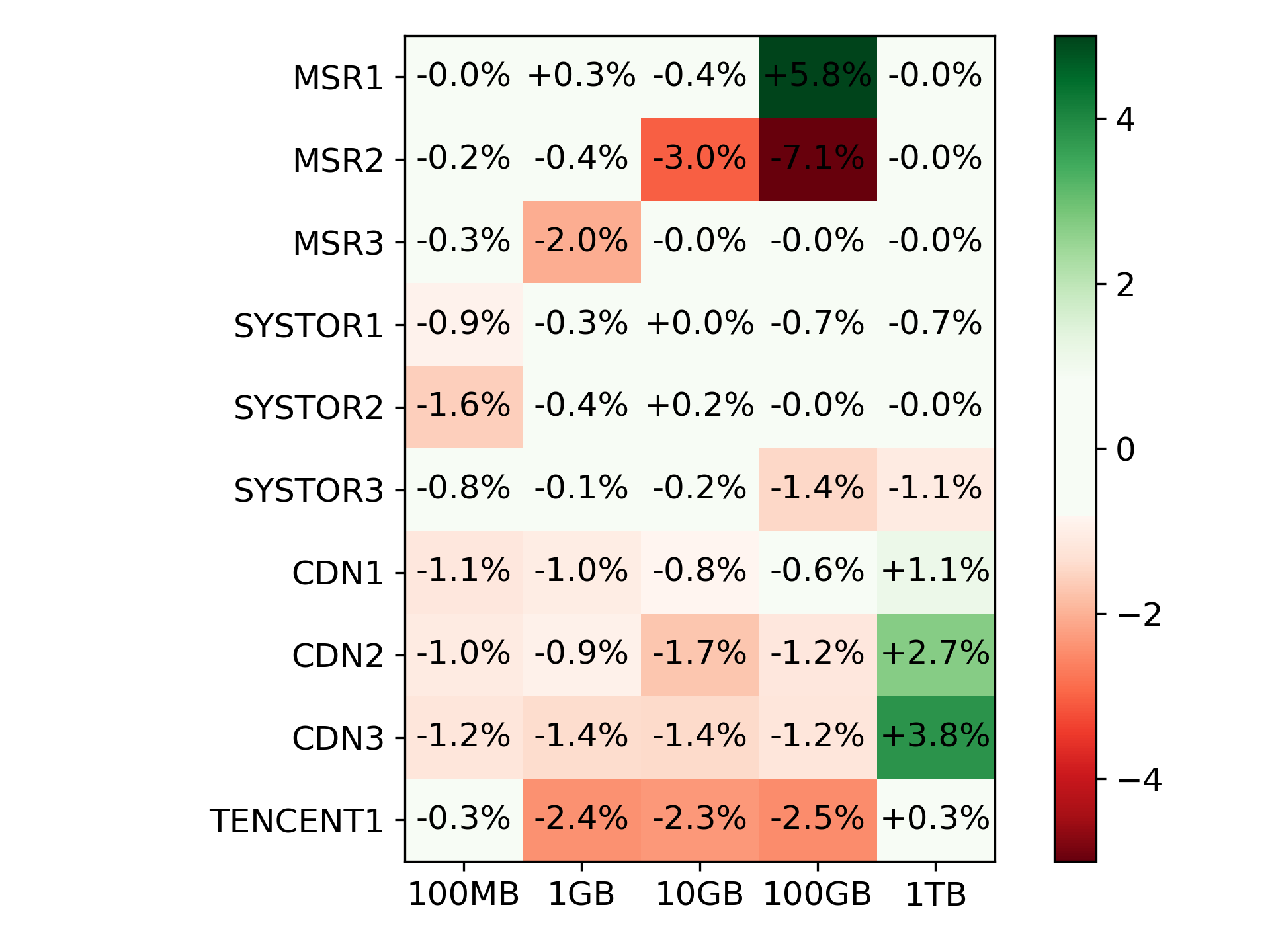}}
	}
	\caption{Byte hit-ratio of suggested policies vs. the best alternative.}
	\label{fig:heatmaps:bhr}
\end{figure*}
\fi

Notice that although there are clear differences between the admission approaches, the eviction policies make inconsistent difference, except for the Random policy that clearly lags. 
An interesting and expected observation is that Sampled Size is a little better than others for hit-rate due to the gain it gets from evicting large items, while it losses for byte-hit-rate. 
SLRU is decent for both measurements and more simple and efficient.
Hence we continue with SLRU for all subsequent~evaluations.

\subsection{Hit-Ratios of State-of-the-Art Approaches}

Next, we compare \PROTNAME{} (SLRU) and QV (SLRU) with the seminal GDSF~\cite{GDSF} as well as the recently proposed AdaptSize~\cite{AdaptSize}, and LHD~\cite{LHD}, and LRB~\cite{LRB} schemes. 
Figure~\ref{fig:evaluation-others:hr} shows the results for the hit ratio metric (important for user perceived reaction time).
First, observe that LHD lags for small cache sizes, which is expected as LHD does not use any metadata on non-cached items.
Intuitively, when the cache is large, LHD has enough information to reach sound decisions.
But, this approach is insufficient for small cache sizes. 
In comparison, \PROTNAME{} utilizes the TinyLFU data structure that collects metadata on a large sample of past accesses, which explains why \PROTNAME{} outperforms LHD for small caches.
We note that LRB is never the best policy despite its use of machine learning.
This may not be surprising, as it is trained to maximize the byte hit ratio~metric.

Next, observe that AdaptSize under-performs with large caches.
Specifically, observe the MSR traces, the SYSTOR traces, and the CDN traces and notice that the hit ratio of AdaptSize does not improve when moving from 100GB to 10TB (an x100~increase). 
We believe that this is intrinsic to Adaptsize's design.
Specifically, AdaptSize admits a new item to the cache with a probability that is inversely proportional to the item's size.
AdaptSize is, therefore, unlikely to admit a very large item even if the cache has enough space to accommodate this item without evictions.
In that sense, AdaptSize is too selective and fails to utilize large caches.

We argue that AdaptSize can be improved for large caches if it bases the admission probability on the victim set's size rather than on the cache candidate's size. 
Such a change would bring it closer to our approaches in terms of design philosophy. 
That is, in \PROTNAME{} the admission bar depends on the total frequency of the victim set. 
Similarly to AdaptSize, large items are less likely to enter the cache as they cause more evictions than small items do.
Unlike AdaptSize, \PROTNAME{} always admits an item if there is enough free space without evictions.
Consequently, we can see that \PROTNAME{} remains competitive across the entire range for all the traces.


The QV algorithm is not competitive in this metric since it evicts low-frequency items even when it fails to perform an admission.
Thus, it has lower space utilization compared to \PROTNAME.
Finally, observe that GDSF remains competitive with present-day algorithms even after all these years.
The main drawback of GDSF remains its logarithmic time complexity. 

Figure~\ref{fig:evaluation-others:bhr} shows the results for the byte hit-ratio metric (important for bandwidth conservation).
Notice that here, LRB is a very attractive suggestion, and is often one of the leading algorithms.
\PROTNAME{} is reasonable in the byte-hit-ratio metric as well but the QV algorithm excels in this metric.
It never lags by much after the best alternative, and in some cases is strictly better.  
QV's excellence in this metric is because QV actively evicts infrequent items even if the candidate is not admitted to the cache. 
Thus, it is more likely to pick up large and frequent items to increase the byte hit ratio. 

\subsection{Summary}
\ifdefined\SHOWHEATMAP
Figures~\ref{fig:others:heatmap:hr:av} and~\ref{fig:others:heatmap:bhr:av} summarize the relative results for \PROTNAME{} (SLRU). 
As illustrated, 
\else
As shown in our empirical study, 
\fi
\PROTNAME{} is competitive with the best alternative (which is not always the same algorithm) in most workloads, especially for small to midsize caches.
It is slightly better than the best alternatives in the byte hit-ratio metric.

\ifdefined\SHOWHEATMAP
Figures~\ref{fig:others:heatmap:hr:qv} and~\ref{fig:others:heatmap:bhr:qv} repeat the above for the QV (SLRU) algorithm.
As can be observed, 
\else
We can further conclude that 
\fi
QV is inferior in the hit-ratio metric but almost always superior in terms byte hit-ratio.
It is tough to excel in both hit ratio and byte hit ratio as there is some tension between them.
For example, hit ratio wise it is better to cache an item of size 0.5MB that is accessed six times, but for byte hit ratio, it may be better to cache an object of size 1MB that is accessed only five times. 

\begin{figure}[tp]
	\center{
		\subfloat[MSR1\label{fig:perf:msr}]{\includegraphics[width = 0.75\columnwidth]{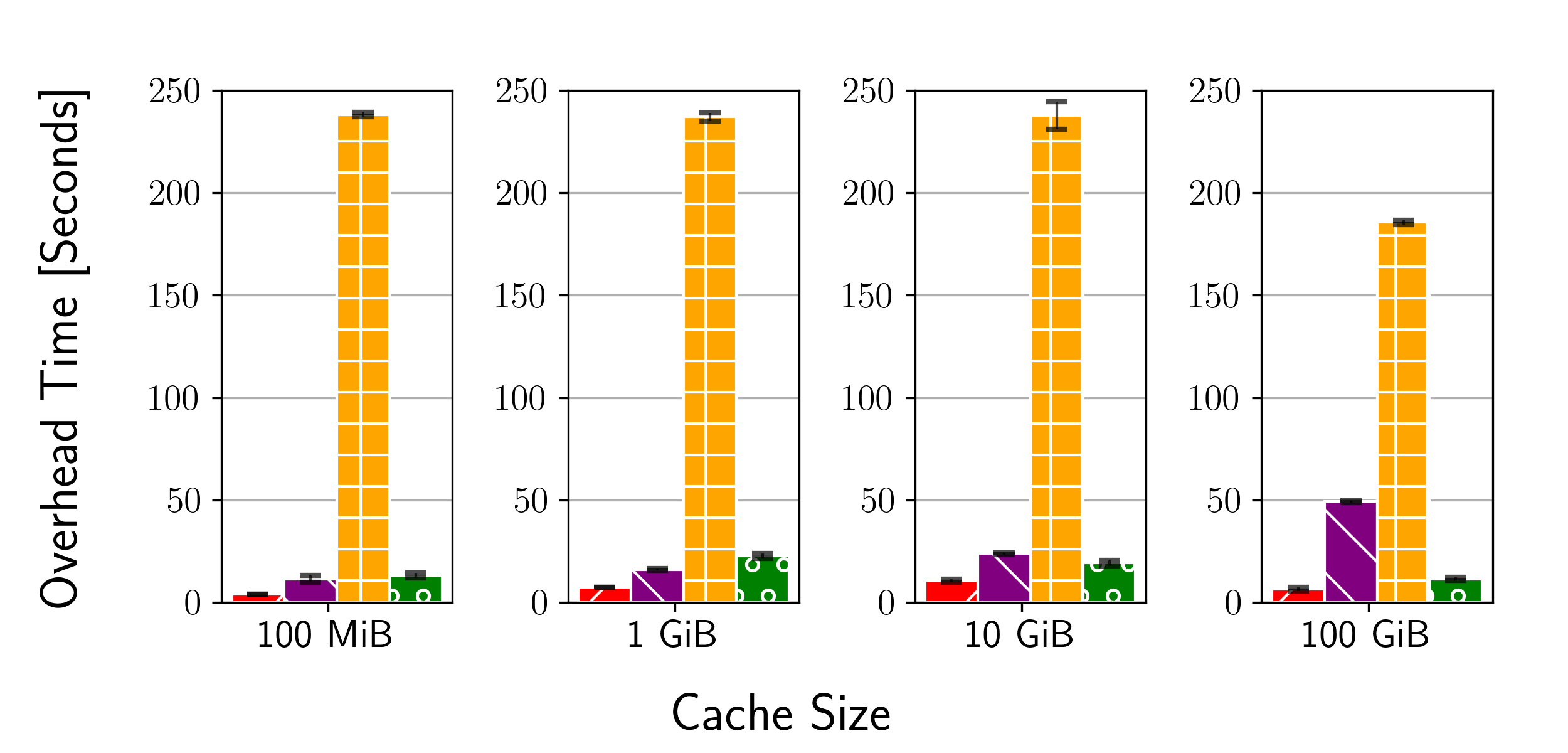}}\\
		\subfloat[SYSTOR1\label{fig:perf:systor}]{\includegraphics[width = 0.75\columnwidth]{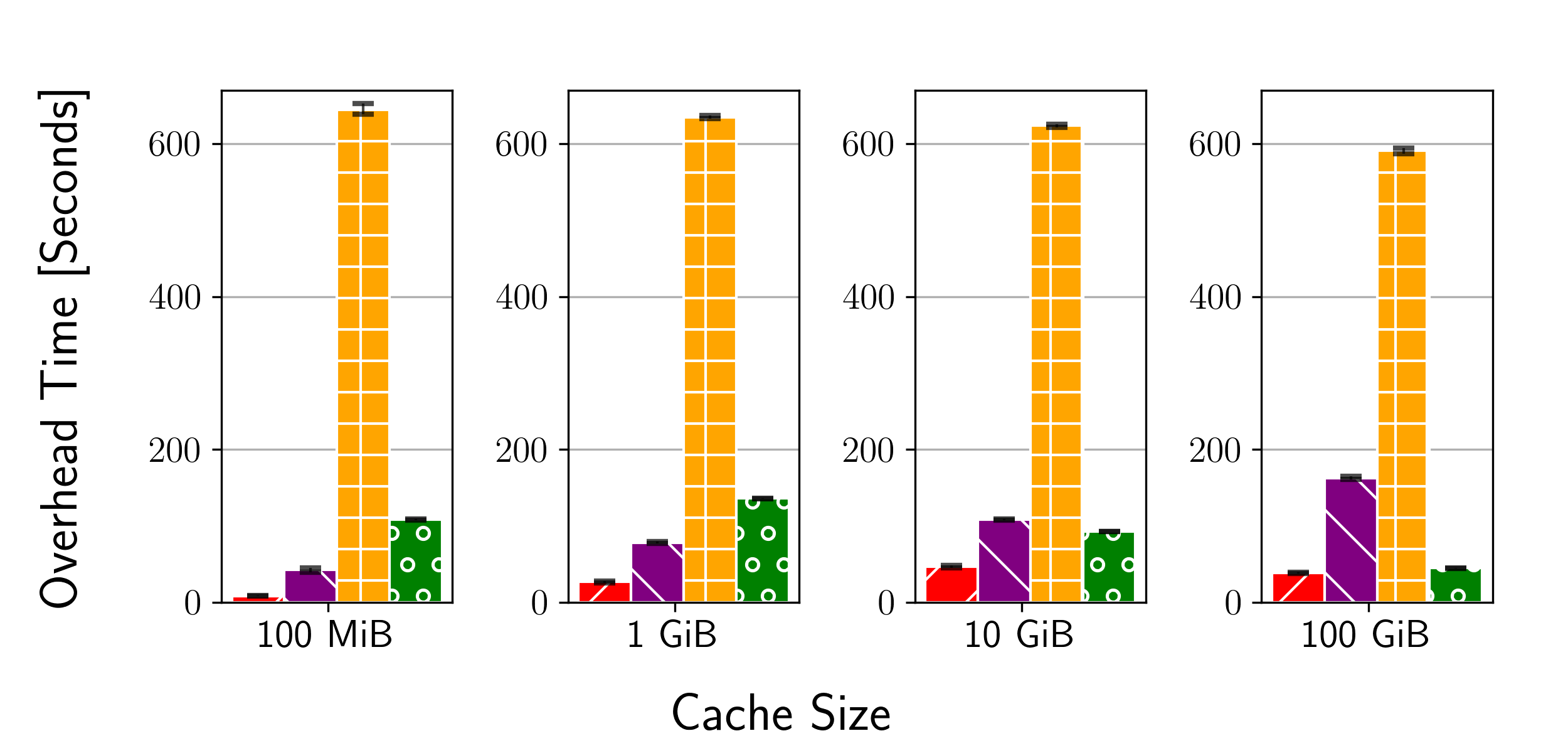}}\\
		\subfloat[CDN1\label{fig:perf:cdn}]{\includegraphics[width = 0.75\columnwidth]{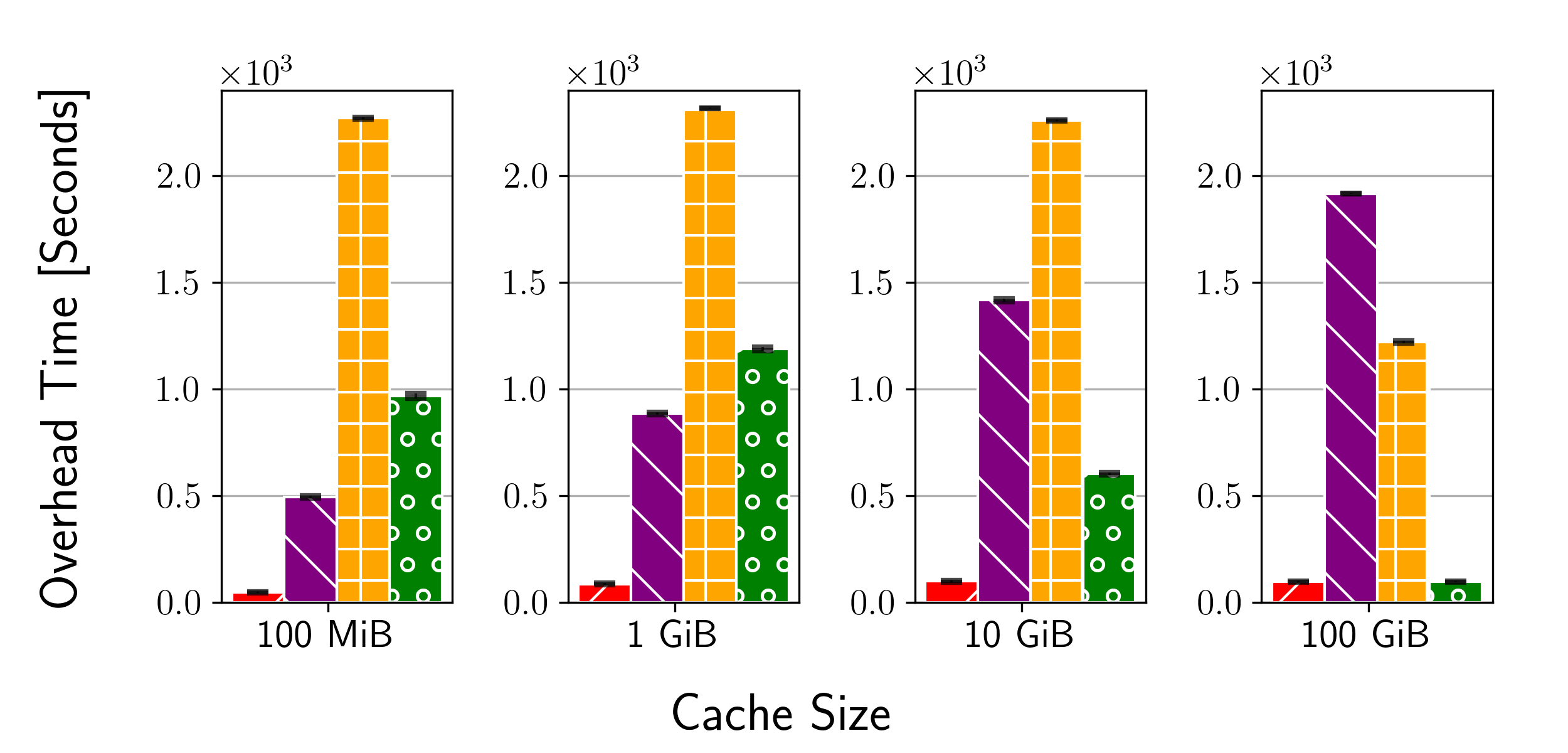}}\\		
		\vspace{0.2cm}
		\includegraphics[width = 8.5cm]{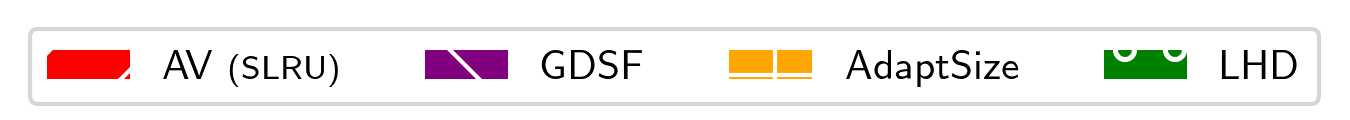}
	}
	\vspace*{-0.4cm}
	\caption{Overhead run-time of each policy for different traces and cache sizes. We ran GDSF and AV with our own Java implementation, while LHD and AdaptSize with their authors C++ implementation. To minimize the effect of using different frameworks and present the policy overhead only, we subtract the mean run-time of LRU in the matching implementation from the total run-time.} \label{fig:pref}
\end{figure}

\begin{table}[tp]
	\begin{center}
		\begin{tabular}{l|c|c|c} 
			\textbf{Trace} & \textbf{Cache Size} & \textbf{Overhead Time} \\
			& & $seconds$ \\
			\hline
			\multirow{4}{*}{MSR1} & 100 MB & 1,744 \\
			& 1 GB & 2,206 \\
			& 10 GB & 2,133 \\
			& 100 GB & 1,999 \\
			\hline
			\multirow{4}{*}{SYSTOR1} & 100 MB & 9,666 \\
			& 1 GB & 8,257 \\
			& 10 GB & 7,410 \\
			& 100 GB & 6,380 \\
			\hline
			\multirow{4}{*}{CDN1} & 100 MB & 26,744 \\
			& 1 GB & 25,725 \\ 
			& 10 GB & 16,712 \\ 
			& 100 GB & 14,769 \\
		\end{tabular}
	\end{center}
	\normalsize
	\vspace{0.4cm}
	\caption{Overhead run-time of LRB with the authors' C++ implementation. To minimize the effect of using different frameworks and present the policy overhead only, we subtract the mean run-time of LRU in the matching implementation from the total run-time.}
	\label{tab:lrb-run-time}		
\end{table}

\subsection{Computational Performance}
Finally, we complete the picture by measuring each policy's algorithmic computation time when varying the cache size. 
Figure~\ref{fig:pref} and Table~\ref{tab:lrb-run-time} illustrates these results. 
First, observe that \PROTNAME{} (SLRU) is the most lightweight size-aware policy by a large margin and that its calculation time is comparable to the LRU~policy. 

One of the critiques about GDSF is that its logarithmic complexity in the number of cached objects makes it impractical for large caches.
However, our evaluation shows that it is often considerably faster than both LHD and AdaptSize.
We did notice the impact of the logarithmic complexity of GDSF, as its running time increased with the cache size, despite the higher hit-ratios obtained by large caches.
This trend is evident in all traces, but is most acute for the CDN traces, which can be explained by their wide distribution of object sizes, as illustrated in Figure~\ref{fig:cdfs}.
Wider distributions of objects push the \emph{actual average} running times closer to the logarithmic \emph{upper bound} for GDSF's priority queue maintenance.
This is because when the size difference is very large, the distribution of objects' priorities is larger, so objects are moved further inside the priority queue, which involves more internal~operations.

Among the other algorithms, AdaptSize is the slowest one. Still, even LHD is at least x3 times slower than \PROTNAME{}, especially where the hit-ratio is not very high, and many evictions are needed.
Thus, we conclude that since neither of these algorithms yields a distinctive benefit in terms of hit-ratios, \PROTNAME{} offers the most attractive trade-off as it is very simple to implement, as shown in Algorithms~\ref{alg:wtinylfu} and~\ref{alg:our-alg}, and is considerably faster than the alternatives.

The run-time results for LRB (Table~\ref{tab:lrb-run-time}) are roughly an order of magnitude longer than the worst policy among the other schemes we tested (Figure~\ref{fig:pref}).
Yet, we notice that the performance improves when the cache size increases.
In fact, LRB's run-time is not directly dependent on the cache size.
Rather, LRB works much harder on each miss than on a hit, as whenever there is a miss, it has to sample 64 items and invoke the machine learning based mechanism on them.
Now, as the cache size increases, the hit ratio also improves, and therefore LRB does significantly less work.
This is somewhat similar to LHD's behavior, but on a grander scale.

\section{Discussion}
\label{sec:discussion}
Modern databases and data-stores need to cope with objects of significant size variability.
Therefore, traditional size oblivious cache management policies are often inadequate.
In this paper, we have studied size aware cache management policies for workloads in which objects' sizes vary.
In particular, we focused on three approaches to extending the size oblivious W-TinyLFU policy into this more challenging setting.
The first two, IV and QV, have been realized in two widely used open-source caching libraries~\cite{CaffeineProject,RistrettoProject}, but were never studied before, and the third is our new approach (\PROTNAME).
We also compared them against the state-of-the-art approaches GDSF~\cite{GDSF}, AdaptSize~\cite{AdaptSize}, LHD~\cite{LHD} and LRB~\cite{LRB}.
We showed that in terms of hit-ratio and byte hit-ratio, none of these policies always wins.
Yet, \PROTNAME{} is always competitive with the state-of-the-art and is considerably faster than previous approaches. 
On the other hand, QV is less competitive on the hit-ratio metric but offers very attractive byte hit-ratio.
In terms of CPU overhead, QV and \PROTNAME{} are consistently the fastest.

Being fast is vital since caching is a performance optimization, so quicker policies are applicable in more systems and have a smaller CPU footprint.  
Thus, we believe that \PROTNAME{} is a strong candidate as its hit-ratios are competitive, and it is considerably faster than the existing alternatives. 

However, in scenarios where byte hit-ratio is more beneficial than hit-ratio (e.g., to minimize network bandwidth consumption), QV is likely the best alternative. 
Further, \PROTNAME{} and QV are also very straightforward to implement. 
That is, their code is short, easy to explain, and requires no sophisticated math nor heavy machine learning mechanisms.

Our performance evaluation study has also revealed the following inherent drawback of AdaptSize~\cite{AdaptSize}, as a representative of policies that admit objects with a probability that is inversely proportional to the object's size~\cite{Treshold,LRUS,Treshold2}.
AdaptSize practically avoids inserting very large objects into the cache, even when it is far from full.
Consequently, it fails to utilize significant portions of large caches, which hurts AdaptSize's hit ratios, and even more significantly, its byte hit ratios when operating with very large~caches.

Let us note that slabbing is orthogonal to the management policy. 
Yet, intuitively, the more effective the management policy is inside a slab, the fewer slabs one needs, and therefore one can expect to obtain a better overall hit ratio.
Rigorously studying the tradeoffs of combining our policies with slabbing is left for future work.

\paragraph*{Code availability}
The full code of the algorithms suggested in this paper is available at \url{https://github.com/ohadeytan/caffeine/tree/arXiv_submission}.
The traces are available online and are pointed to from our code repository.

\clearpage
\bibliographystyle{abbrv}
\bibliography{refs}

\end{document}